%% file: SUS-19-008_temp.tex
\pdfoutput=1

\documentclass[11pt,twoside,a4paper,cmspaper,final,collab]{cms-tdr}
\def\svnVersion{4030dd1-D}\def\svnDate{2020/08/04}\def\cmsCernNoTag{CERN-EP-2020-001}\def\cmsCernDate{\today}\def\cmsMessage{"Published in the European Physical Journal C as \href{http://dx.doi.org/10.1140/epjc/s10052-020-8168-3}{\doi{10.1140/epjc/s10052-020-8168-3}.}"}

\begin{document}\cmsNoteHeader{SUS-19-008}

\hyphenation{had-ron-i-za-tion}
\hyphenation{cal-or-i-me-ter}
\hyphenation{de-vices}
\newlength\cmsTabSkip\setlength{\cmsTabSkip}{1.5ex}
\newcommand{\sslumiruntwo}{137\fbinv}
\newcommand{\MTmin}{\ensuremath{\mT^{\text{min}}}\xspace}
\newcommand{\Njets}{\ensuremath{N_\text{jets}}\xspace}
\newcommand{\Nbjets}{\ensuremath{N_{\PQb}}\xspace}
\newcommand{\gluino}{\PSg}
\newcommand{\susyq}{\PSQ}
\newcommand{\susytop}{\PSQt}
\newcommand{\susytopone}{\PSQt_{1}}
\newcommand{\susytoptwo}{\PSQt_{2}}
\newcommand{\ptcorr}{\ensuremath{\pt^\text{corr}}\xspace}
\newcommand{\nisrjet}{\ensuremath{N_J^\text{ISR}}\xspace}
\newcommand{\sbottomone}{\PSQb_{1}}
\newcommand{\lsp}{\PSGczDo}
\newcommand{\neutralinotwo}{\ensuremath{\PSGc_{2}^{0}}}
\newcommand{\chiminus}{\PSGcmDo}
\newcommand{\chiplus}{\PSGcpDo}
\newcommand{\chiplmin}{\PSGcpmDo}
\newcommand{\Totttt}{\ensuremath{\mathrm{T}1{\PQt\PQt\PQt\PQt}}\xspace}
\newcommand{\Tftttt}{\ensuremath{\mathrm{T}5{\PQt\PQt\PQt\PQt}}\xspace}
\newcommand{\Tfttcc}{\ensuremath{\mathrm{T}5{\PQt\PQt\PQc\PQc}}\xspace}
\newcommand{\TfqqqqWZ}{\ensuremath{\mathrm{T}5{\PQq\PQq\PQq\PQq}\PW\PZ}\xspace}
\newcommand{\TfqqqqWW}{\ensuremath{\mathrm{T}5{\PQq\PQq\PQq\PQq}\PW\PW}\xspace}
\newcommand{\TsttWW}{\ensuremath{\mathrm{T}6{\PQt\PQt}\PW\PW}\xspace}
\newcommand{\TsttHZ}{\ensuremath{\mathrm{T}6{\PQt\PQt}\PH\PZ}\xspace}
\newcommand{\TfttbbWW}{\ensuremath{\mathrm{T}5{\PQt\PQt\PQb\PQb}\PW\PW}\xspace}
\newcommand{\Totbs}{\ensuremath{\mathrm{T}1{\PQt\PQb\PQs}}\xspace}
\newcommand{\ToqqqqL}{\ensuremath{\mathrm{T}1{\PQq\PQq\PQq\PQq}\PL}\xspace}
\newcommand{\cmsTable}[1]{\resizebox{\textwidth}{!}{#1}}

\cmsNoteHeader{SUS-19-008}
\title{ Search for physics beyond the standard model in events with jets and two same-sign or at least three charged leptons in proton-proton collisions at $\sqrt{s}=13$\TeV }
\titlerunning{Search for physics beyond the standard model \ldots at $\sqrt{s}=13\TeV$}
\date{\today}

\abstract{
A data sample of events from proton-proton collisions with at least two jets,
and two isolated same-sign or three or more charged leptons, is studied in a
search for signatures of new physics phenomena. The data correspond to an integrated luminosity of 137\fbinv
at a center-of-mass energy of 13\TeV, collected in 2016--2018 by the CMS experiment at the LHC.
The search is performed using a total of 168 signal regions defined using several kinematic variables. The
properties of the events are found to be consistent with the expectations from standard
model processes. Exclusion limits at 95\% confidence level are set on cross
sections for the pair production of gluinos or squarks for various decay
scenarios in the context of supersymmetric models conserving or violating R
parity. The observed lower mass limits are as large as 2.1\TeV for
gluinos and 0.9\TeV for top and bottom squarks. To facilitate
reinterpretations, model-independent limits are provided in a set of
simplified signal regions.
}

\hypersetup{%
pdfauthor={CMS Collaboration},%
pdftitle={Search for physics beyond the standard model in events with jets and two same-sign or at least three charged leptons in proton-proton collisions at sqrt(s)=13 TeV},%
pdfsubject={CMS},%
pdfkeywords={CMS, physics, SUSY}}

\maketitle

\section{Introduction}
\label{sec:intro}

In the standard model (SM), the production of multiple jets
in conjunction with two same-sign (SS) or three or more charged leptons is a very rare process in proton-proton ($\Pp\Pp$) collisions.
These final states provide a promising starting point in the search for physics beyond the SM (BSM).
Many models attempting to address the shortcomings of the SM lead to such signatures.
Examples include the production of supersymmetric (SUSY) particles~\cite{Barnett:1993ea,Guchait:1994zk},
SS top quark pairs~\cite{Bai:2008sk,Berger:2011ua},
scalar gluons (sgluons)~\cite{Plehn:2008ae,Calvet:2012rk}, heavy scalar bosons of extended Higgs sectors~\cite{Gaemers:1984sj,Branco:2011iw},
Majorana neutrinos~\cite{Almeida:1997em}, and vector-like quarks~\cite{Contino:2008hi}.

{\tolerance=5000
In SUSY models~\cite{Ramond:1971gb,Golfand:1971iw,Neveu:1971rx,Volkov:1972jx,Wess:1973kz,Wess:1974tw,Fayet:1974pd,Nilles:1983ge,Martin:1997ns},
the decay chain of pair-produced gluinos or squarks can contain multiple \PW or \PZ bosons, with the potential to have at least one pair of SS \PW bosons. 
Such a decay chain is realized, for example in gluino pair production, when a gluino decays into a top quark-antiquark pair and a neutralino, or into a pair of quarks and a chargino that subsequently decays into a \PW\ boson and a neutralino.
In R parity~\cite{Farrar:1978xj} conserving (RPC) scenarios, the lightest SUSY particle is neutral and stable and escapes detection, leading to an imbalance in the measured
transverse momentum. The magnitude of the missing transverse momentum strongly depends on the details of the model, and
in particular on the mass spectrum of the particles involved.
Scenarios with R parity violation (RPV)~\cite{Nikolidakis:2007fc,Csaki:2011ge} additionally allow decays of SUSY particles into SM particles only, leading in many cases
to signatures with little or no missing transverse momentum.
For many SUSY models, the SS and multilepton signatures provide complementarity with searches in the zero- or one-lepton final states,
and they are particularly suitable for probing
compressed mass spectra and other scenarios involving low-momentum leptons or low missing transverse momentum.
Both the ATLAS~\cite{Aaboud:2017dmy} and CMS~\cite{SUS-16-035,SUS-16-041} Collaborations have carried out searches in these channels using LHC data collected up to and including 2016. 
The ATLAS Collaboration has also recently released a search with the full data set recorded between 2015 and 2018~\cite{ATLAS-SUSY-2018-09}.
\par}

In this paper, we extend and refine the searches described in Refs.~\cite{SUS-16-035,SUS-16-041} using a larger data set of $\Pp\Pp$ collisions at $\sqrt{s} = 13\TeV$ recorded by the CMS detector at the CERN LHC in 2016--2018,
corresponding to an integrated luminosity of~\sslumiruntwo.
We base our search on an initial selection of events with at least two hadronic jets and
two SS or three or more light leptons (electrons and muons), including those from leptonic decays of $\tau$ leptons.  Several signal regions (SRs) are then
constructed with requirements on variables such as the number of leptons, the number of jets (possibly identified as originating from \PQb~quarks), and the magnitude
of missing transverse momentum.
A simultaneous comparison of the observed and SM plus BSM expected event yields in all SRs is performed to constrain the BSM models described in Section~\ref{sec:samples}.
After a brief description of the CMS experiment in Section~\ref{sec:cms}, we present the details of the search strategy and event selection in Section~\ref{sec:selection} and discuss the various relevant backgrounds from SM processes in Section~\ref{sec:backgrounds}. The systematic uncertainties considered in the analysis are presented in Section~\ref{sec:systematics}. In Section~\ref{sec:results}, the observed yields are compared to the background expectation and the results are interpreted to constrain the various BSM models introduced earlier. Model independent limits are also derived. Finally, the main results are summarized in Section~\ref{sec:summary}.

\section{Background and signal simulation}
\label{sec:samples}

Monte Carlo (MC) simulations are used to study the SM backgrounds and to estimate the event selection efficiency of the BSM signals under
consideration.
Three sets of simulated events for each process are used in order to match the different data taking conditions in 2016, 2017, and 2018.

{\tolerance=5000
The hard scattering process of the dominant backgrounds estimated from simulation (including the $\ttbar \PW$, $\ttbar \PZ$ and $\PW\PZ$ contributions) is simulated with the \MGvATNLO 2.2.2 (2.4.2)~\cite{MADGRAPH5,Alwall:2007fs,Frederix:2012ps} generator for 2016 (2017 and 2018) conditions. 
An exception is the $\PW\PZ$ process for the 2016 conditions that, as with a few subdominant backgrounds, is simulated using the \POWHEG~v2~\cite{Nason:2004rx,Frixione:2007vw,Alioli:2010xd,Melia:2011tj,Nason:2013ydw} next-to-leading order (NLO) generator.
Samples of signal events, as well as of SS \PW boson pairs and other very rare SM processes,
are generated at leading order (LO) accuracy with \MGvATNLO, with up to two additional partons in the matrix element calculations.
The set of parton distribution functions (PDFs) used was NNPDF3.0~\cite{Ball:2014uwa} for the 2016 simulation and NNPDF3.1~\cite{Ball:2017nwa} for the 2017 and 2018 simulations.
\par}

Parton showering and hadronization, as well as the double parton scattering production of $\PW^{\pm} \PW^{\pm}$,
are described using the \PYTHIA~8.230 generator~\cite{Sjostrand:2014zea}
with the CUETP8M1 (CP5) underlying event tune for 2016 (2017 and 2018) simulation~\cite{Skands:2014pea,Khachatryan:2015pea,Sirunyan:2019dfx}.
The response of the CMS detector is modeled using the \GEANTfour program~\cite{Geant} for SM
background samples, while the CMS fast simulation package~\cite{Abdullin:2011zz,Giammanco:2014bza} is used for signal samples.

To improve the \MGvATNLO modeling of the
multiplicity of additional jets from initial-state radiation (ISR),
2016 MC events are reweighted according to the
number of ISR jets (\nisrjet).  The reweighting factors
are extracted from a study
of the light-flavor jet
multiplicity in dilepton \ttbar events.  They
vary between 0.92 and 0.77 for
\nisrjet between 1 and 4, with one half of the deviation
from unity taken as the systematic uncertainty.
This reweighting is not necessary for the 2017 and 2018 MC samples that are generated
using an updated \PYTHIA tune.

\begin{figure*}[htb!]
\centering
\includegraphics[width=1.0\textwidth]{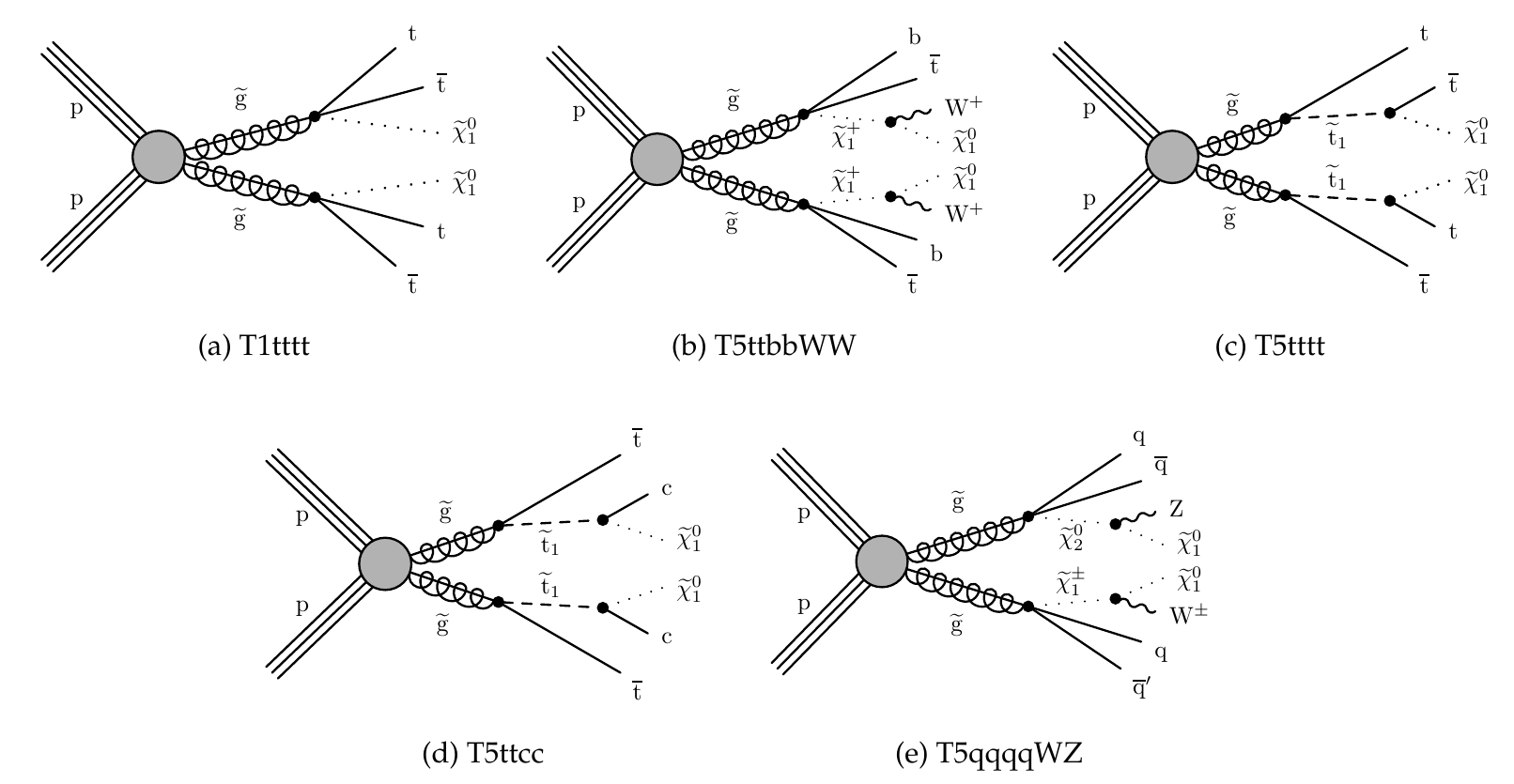}
\\
\caption{Diagrams illustrating the simplified RPC SUSY models with gluino production considered in this analysis.}
\label{fig:diagrams1}
\end{figure*}

\begin{figure*}[htb!]
\centering
\includegraphics[width=0.65\textwidth]{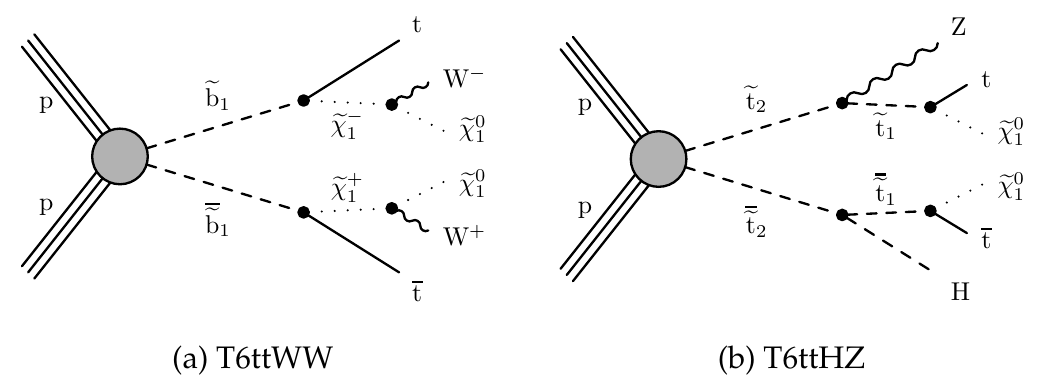}
\\
\caption{Diagrams illustrating the simplified RPC SUSY models with squark production considered in this analysis.}
\label{fig:diagrams2}
\end{figure*}

\begin{figure*}[htb!]
\centering
\includegraphics[width=0.7\textwidth]{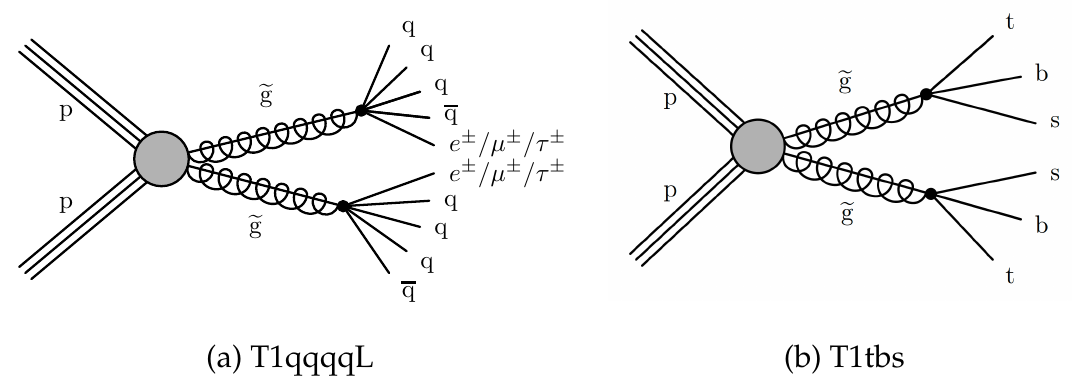}
\\
\caption{Diagrams illustrating the two simplified RPV SUSY models considered in this analysis.}
\label{fig:diagramsRPV}
\end{figure*}

{\tolerance=19200
The phenomenology of a given SUSY model strongly depends on its underlying details such as the masses of the SUSY particles and their couplings with the SM particles and each other, many of which can be free parameters. 
The signal models used by this search are
simplified SUSY models~\cite{Alves:2011wf,Chatrchyan:2013sza}
of either
gluino or squark pair production, followed by a variety of
RPC (Figs.~\ref{fig:diagrams1} and~\ref{fig:diagrams2}) or
RPV (Fig.~\ref{fig:diagramsRPV}) decays and where several leptons can arise in the final state. 
Production cross sections are calculated at approximate next-to-next-to-leading order plus
next-to-next-to-leading logarithmic (NNLO+NNLL)
accuracy~\cite{bib-nnll,bib-nlo-nll-01,bib-nlo-nll-02,bib-nlo-nll-03,bib-nlo-nll-04,bib-nlo-nll-05,bib-nlo-nll-06,bib-nlo-nll-07,bib-nlo-nll-08,bib-nlo-nll-09,bib-nlo-nll-10,bib-nlo-nll-11,Borschensky:2014cia}.
The branching fractions for the decays shown are assumed to be 100\%, unless otherwise
specified, and all decays are assumed to be prompt.
\par}

Gluino pair production models giving rise to signatures with up to four \cPqb\ quarks and up to four \PW\ bosons are
shown in Fig.~\ref{fig:diagrams1}.
In these models, the gluino decays to the lightest squark ($\gluino \to \susyq \cPq$),
which in turn decays to same-flavor ($\susyq \to \cPq \lsp$)
or different-flavor ($\susyq \to \cPq' \chiplmin$) quarks.
The chargino ($\chiplmin$) decays to a \PW\ boson and a neutralino ($\lsp$) via $\chiplmin \to \PW^{\pm} \lsp$,
where the \lsp is taken to be the lightest stable SUSY particle and escapes detection.

The first scenario, denoted by \Totttt and
displayed in Fig.~\ref{fig:diagrams1}a,
includes an off-shell top squark ($\susytop$) leading to the three-body decay of the gluino,
$\gluino \to \ttbar\lsp$,
resulting in events with four \PW\ bosons and four \cPqb\ quarks.
Figure~\ref{fig:diagrams1}b presents a similar model (\TfttbbWW)
where the gluino decay results
in a chargino that further decays into a neutralino and a \PW\ boson.
The model shown in Fig.~\ref{fig:diagrams1}c (\Tftttt)
is the same as \Totttt except that the intermediate top squark
is on-shell.
The mass splitting between
the $\susytop$ and the \lsp
is taken to be $m_{\susytop} - m_{\lsp} = m_{\cPqt}$, where
$m_{\cPqt}$ is the top quark mass.  This choice maximizes
the kinematic differences between this model and \Totttt, and also
corresponds to one of the most challenging regions of parameter
space for the
observation of the $\susytop \to \cPqt \lsp$ decay since the neutralino is produced at rest in the top squark rest frame.
The decay chain of Fig.~\ref{fig:diagrams1}d (\Tfttcc)
is identical to that of \Tftttt
except that the $\susytop$ decay involves a \PQc quark.
In Fig.~\ref{fig:diagrams1}e, the decay process includes a virtual light-flavor squark,
leading to three-body decays of $\gluino \to \cPq \cPq' \chiplmin$ or $\gluino \to \cPq \cPq' \neutralinotwo$, with a resulting signature
of two \PW\ bosons, two \PZ\ bosons, or one of each (the case shown in Fig.~\ref{fig:diagrams2}e), and four light-flavor jets.
This model, \TfqqqqWZ, with a resulting signature of one \PW\ boson and one \PZ\ boson,
is studied with two different assumptions for the chargino mass: $m_{\chiplmin} = 0.5(m_{\gluino} +  m_{\lsp})$,
and $m_{\chiplmin} = m_{\lsp}+20 \GeV$, producing on- and off-shell bosons, respectively.
The model is also considered with the assumption of decays to two \PW\ bosons exclusively (\TfqqqqWW).

Figure~\ref{fig:diagrams2}a shows a model of bottom squark production with subsequent
decay of $\sbottomone \to \cPqt \chiplmin$, yielding two \cPqb\ quarks and four \PW\ bosons.
This model, \TsttWW, is considered as a function of the the lightest bottom squark, $\sbottomone$, and \chiplmin masses.
The \lsp mass is fixed to be 50\GeV, causing two of the \PW\ bosons to be produced off-shell
when the \chiplmin mass is less than approximately 130\GeV.
Figure~\ref{fig:diagrams2}b displays a model similar to \TsttWW, but with top squark pair production and
a subsequent decay of $\susytoptwo\to\susytopone\PH/\PZ$, with $\susytopone \to \cPqt\lsp$, producing signatures with two \PH\ bosons, two \PZ\ bosons, or one of each.
In this model, \TsttHZ, the \lsp mass is fixed such that $m(\susytopone)-m(\lsp)=m_{\cPqt}$.

The R parity violating decays considered in this analysis are \ToqqqqL (Fig.~\ref{fig:diagramsRPV}a)
and \Totbs (Fig.~\ref{fig:diagramsRPV}b).
In \ToqqqqL, the gluino decays to the lightest squark ($\gluino \to \susyq \cPq$),
which in turn decays to a quark ($\susyq \to \cPq \lsp$), but decays with the $\lsp$ off shell (violating R parity) into two quarks and a charged lepton,
giving rise to a prompt 5-body decay of the gluino.
In \Totbs, each gluino decays into three different SM quarks (a top, a bottom, and a strange quark).

\section{ The CMS detector and event reconstruction}
\label{sec:cms}

The central feature of the CMS detector is a superconducting solenoid of 6\unit{m} internal diameter, providing a magnetic field of 3.8\unit{T}. Within the solenoid volume are a silicon pixel and strip tracker, a lead tungstate crystal electromagnetic calorimeter (ECAL), and a brass and scintillator hadron calorimeter (HCAL), each composed of a barrel and two endcap sections. Forward calorimeters extend the pseudorapidity ($\eta$) coverage provided by the barrel and endcap detectors. Muons are detected in gas-ionization chambers embedded in the steel flux-return yoke outside the solenoid.
A more detailed description of the CMS detector, together with a definition of the coordinate system used and the relevant kinematic variables, can be found in Ref.~\cite{Chatrchyan:2008zzk}.

Events of interest are selected using a two-tiered trigger system~\cite{Khachatryan:2016bia}. The first level, composed of custom hardware processors, uses information from the calorimeters and muon detectors to select events at a rate of around 100\unit{kHz} within a fixed time interval of less than 4\mus. The second level, known as the high-level trigger, consists of a farm of processors running a version of the full event reconstruction software optimized for fast processing, and reduces the event rate to around 1\unit{kHz} before data storage.

{\tolerance=800
The reconstructed vertex with the largest value of summed physics-object
squared-transverse-momentum is taken to be the primary $\Pp\Pp$ interaction vertex. The physics
objects are the jets, clustered using the jet finding
algorithm of Refs.~\cite{Cacciari:2008gp,Cacciari:2011ma} with the tracks assigned to
the vertex as inputs, and the associated missing transverse momentum, taken as
the negative vector sum of the transverse momentum (\pt) of those jets.
\par}

The particle-flow (PF) algorithm~\cite{CMS-PRF-14-001} aims to reconstruct and identify each individual particle in an event, with an optimized combination of information from the various elements of the CMS detector.  The energy of photons is directly obtained from the ECAL measurement. The energy of electrons is determined from a combination of the electron momentum at the primary interaction vertex as determined by the tracker, the energy of the corresponding ECAL cluster, and the energy sum of all bremsstrahlung photons spatially compatible with the electron track~\cite{Khachatryan:2015hwa}. The momentum of muons is obtained from the curvature of the corresponding track, combining information from the silicon tracker and the muon system~\cite{Sirunyan:2018fpa}. The energy of charged hadrons is determined from a combination of their momentum measured in the tracker and the matching ECAL and HCAL energy deposits, corrected
for the response function of the calorimeters to hadronic showers. The energy of neutral hadrons is obtained from the corresponding corrected ECAL and HCAL energies.

Hadronic jets are clustered from charged PF candidates associated with the primary vertex and from all neutral PF candidates using the anti-\kt algorithm~\cite{Cacciari:2008gp, Cacciari:2011ma} with a distance parameter of 0.4.
The jet momentum is determined as the vectorial sum of all PF candidate momenta in the jet.
An offset correction is applied to jet energies to take into account the contribution from pileup~\cite{cacciari-2008-659}.
Additional jet energy corrections are derived from simulation to bring the detector response to unity, and are improved with in situ measurements of the energy balance in dijet, multijet, photon+jets, and leptonically decaying $\PZ\text{+jets}$ events~\cite{Khachatryan:2016kdb, CMS-PAS-JME-16-003}. 
Additional selection criteria are applied to each jet to remove jets potentially dominated by instrumental effects or reconstruction failures.
Jets originating from \cPqb\ quarks are identified as \PQb-tagged jets using a deep neural network algorithm, DeepCSV~\cite{Sirunyan:2017ezt},
with a working point chosen such that the efficiency to identify a \PQb~jet is  55--70\% for a jet \pt between 20 and 400\GeV.
The misidentification rate for a light-flavor jet is 1--2\% in the same jet \pt range.

The vector \ptvecmiss is defined as the projection onto the plane perpendicular
to the beams of the negative vector sum of the momenta of all reconstructed PF
candidates in an event~\cite{Sirunyan:2019kia}. Its magnitude, called missing
transverse momentum, is referred to as \ptmiss. The scalar \pt sum of all jets
in an event is referred to as \HT.

\section{Search strategy and event selection}
\label{sec:selection}

The search strategy is similar to the one adopted in Refs.~\cite{SUS-16-035,SUS-16-041}.
The event selection requires the presence of at least two hadronic jets and at least two leptons, among which is an SS pair, as described below.
Each selected event is assigned to an SR, based on its content.
Maximum likelihood fits of the background (or signal plus background) predictions to the data in all SRs are then performed.
Such a strategy ensures sensitivity to a broad range of possible signatures of new physics, even beyond the signal benchmarks considered in this analysis. 

The kinematic requirements applied to leptons and jets are presented in Table~\ref{tab:objetselection}. 
The analysis requires at least two jets with $\pt>40\GeV$ and two light SS leptons with $\pt>15\GeV$ ($10\GeV$) for electrons (muons).
Electrons are identified based on a discriminant using shower shape and track
quality variables, while the muon identification relies on the quality of the
geometrical matching between the tracker and muon system measurements.  In
order to reject leptons from the decay of heavy flavor hadrons, the tracks are
required to have an impact parameter compatible with the position of the
primary vertex.  Several isolation criteria are also applied, based on the
scalar sum of hadron and photon \pt within a cone centered on the lepton
direction and whose radius decreases with its \pt, the ratio of the \pt of the
lepton to that of the closest jet, and the relative \pt of the lepton to that of
the closest jet after lepton momentum subtraction.  These criteria are designed to
mitigate the loss of lepton efficiency caused by lepton-jet overlaps that
occurs frequently in events with significant hadronic activity.  A more
detailed description of the set of identification and isolation variables used
in the lepton selection can be found in Ref.~\cite{SUS-15-008}.

The lepton reconstruction and identification efficiency is in the range of 45--70\% (70--90\%)  for electrons (muons),
with $\pt>25\GeV$, increasing as a function of \pt and reaching the maximum value for $\pt>60\GeV$.
In the low-momentum regime, $15<\pt<25\GeV$ for electrons and $10<\pt<25\GeV$ for muons,
the efficiencies are approximately 40\% for electrons and 55\% for muons.
The lepton trigger efficiency for electrons is in the range of 90--98\%, converging to the maximum
value for $\pt>30 \GeV$, and it is around 92\% for muons.

\begin{table}[!hbtp]
\centering
\topcaption{Transverse momentum and pseudorapidity requirements for leptons and jets.
    Note that the $\pt$ thresholds to count jets and \PQb-tagged jets are different;
    the jet multiplicity \Njets includes \PQb-tagged jets if their \pt exceeds $40\GeV$.}
\label{tab:objetselection}
\begin{tabular}{ccc}
    Object        & $\pt$ (\GeVns{})           & $\abs{\eta}$ \\
\hline
Electrons     & $>$15 & $<$2.5 \\
Muons         & $>$10 & $<$2.4 \\
Jets          & $>$40 & $<$2.4 \\
\PQb-tagged jets & $>$25 & $<$2.4 \\
\end{tabular}

\end{table}

In order to reduce backgrounds from the decays of \PQc- and \PQb-hadrons or from the Drell--Yan process,
we reject events with same-flavor lepton pairs with invariant mass ($m_{\ell\ell}$) less than 12\GeV,
where leptons are reconstructed with a looser set of requirements compared to the nominal selection.
Furthermore, events containing a lepton pair with $m_{\ell\ell}< 8 \GeV$, regardless of charge or flavor,
are rejected in order to emulate a similar condition applied at the trigger level.
Events are then separated according to the $\pt$ of the leptons forming the SS pair: high-high if both have $\pt > 25\GeV$, low-low if both have $\pt < 25\GeV$, and high-low otherwise.

Two sets of trigger algorithms are used to select the events: pure dilepton triggers, which require the presence of two isolated leptons with $\pt$ thresholds
on the leading (subleading) lepton in the 17--23 (8--12)\GeV range,
and dilepton triggers with no isolation requirements, a lower $\pt$
threshold of 8\GeV, an invariant mass condition $m_{\ell\ell}> 8 \GeV$ to reject low mass resonances, and with a minimum $\HT$ in the range of 300--350\GeV.
The ranges listed here reflect the varying trigger conditions during the
data taking periods.
The pure dilepton triggers are
used to select high-high and high-low pairs, while low-low pairs are selected using the
triggers with $\HT$ requirements.

Six exclusive categories are then defined as follows:
\begin{itemize}
\item High-High SS pair, significant \ptmiss (HH): exactly 2 leptons, both with $\pt>25 \GeV$, and $\ptmiss>50 \GeV$;
\item High-Low SS pair, significant \ptmiss (HL): exactly 2 leptons, one with $\pt>25 \GeV$, one with $\pt<25 \GeV$, and $\ptmiss>50 \GeV$;
\item Low-Low SS pair, significant \ptmiss (LL): exactly 2 leptons, both with $\pt<25 \GeV$ and $\ptmiss>50 \GeV$;
\item Low \ptmiss (LM): exactly 2 leptons, both with $\pt>25 \GeV$, and $\ptmiss<50 \GeV$; and
\item Multilepton with an on-shell Z boson (on-\PZ ML): $\geq$3 leptons, at least one with $\pt>25 \GeV$, $\ptmiss>50 \GeV$, $\geq$ \PZ boson candidate formed by a pair of opposite-sign (OS), same-flavor leptons with $76 < m_{\ell\ell}< 106 \GeV$. 
\item Multilepton without an on-shell Z boson (off-\PZ ML): same as on-\PZ ML but without a \PZ boson candidate.
\end{itemize}

The categories are typically sensitive to different new physics scenarios and enriched in different SM backgrounds. 
For example the HH category drives the sensitivity for most of the RPC scenarios (\Totttt, \TfttbbWW, \Tftttt, \Totttt, \TfqqqqWW) with a large mass splitting between the gluino and the lightest neutralino. The HL and LL categories become relevant for a lower mass splitting when one or both leptons tend to be soft. Scenarios resulting in the presence of one or multiple \PZ bosons in the final state such as \TfqqqqWZ and \TsttHZ will typically be primarily constrained by the on-Z or off-Z category, also depending on the considered SUSY mass spectrum. Finally the LM category enhances the analysis sensitivity for RPV scenarios, in particular for \ToqqqqL where no genuine \ptmiss is expected.

Various SRs are constructed based on the jet multiplicity \Njets,
the \PQb-tagged jet multiplicity \Nbjets, \HT,  \ptmiss, the charge of the SS pair, and \MTmin, which is defined below.
The \MTmin variable, introduced in Ref.~\cite{SUS-15-008}, is defined as the minimum of the transverse masses calculated from each of the leptons forming the SS pair and \ptvecmiss,
except for the on-\PZ ML category where we only consider the transverse mass computed using the leptons not forming the \PZ candidate.
It is characterized by a kinematic cutoff for events where \ptmiss only arises from the leptonic decay of a single \PW\  boson and is effective at discriminating signal and background signatures.

A subset of SRs is split by the charge of the leptons in an SS pair which is used to
take advantage of the charge asymmetry in most of the background processes, such as $\PW\PZ$, $\ttbar \PW$ or SS $\PW\PW$.
The SRs corresponding to each category, HH, HL, LL, LM, on-\PZ ML, and off-\PZ ML, are summarized in Tables~\ref{tab:SRDefHH}, \ref{tab:SRDefHL}, \ref{tab:SRDefLL}, \ref{tab:SRDefLM}, \ref{tab:SRDefMLonZ} and \ref{tab:SRDefMLoffZ}, respectively. 
The binning ranges are chosen to maximize the sensitivity to a variety of SUSY benchmark points
and are such that the expected SM yield in any SR has relative statistical uncertainties typically smaller than unity.

\begin{table*}[htb!]
    \centering
      \topcaption{\label{tab:SRDefHH} The SR definitions for the HH category. Charge-split regions are indicated with (++) and (-$\,$-).
        The three highest $\HT$ regions are split only by $\Njets$, resulting in 62 regions in total.
        Quantities are specified in units of \GeV where applicable.}
        \cmsTable{
\begin{tabular}{|c|c|c|c|c|c|c|c|c|}
                \hline
                $\Nbjets$                             & $\MTmin$              & $\ptmiss$              & $\Njets$            & $\HT < 300$                          & $\HT\in[300, 1125]$ & $\HT\in[1125, 1300]$                  & $\HT\in[1300, 1600]$                  & $\HT > 1600$ \\ \hline \hline
                \multirow{8}{*}{0}                & \multirow{4}{*}{$<$120}    & \multirow{2}{*}{ 50--200}  & 2--4                 & SR1                                      & SR2                     & \multirow{10}{*}{\shortstack{SR54  \\ $\Njets < 5$ }}& \multirow{10}{*}{\shortstack{SR55  \\ $\Njets < 5$ }}& \multirow{10}{*}{\shortstack{SR56  \\ $\Njets < 5$ }}\\ \cline{4-6}
                                                  &                            &                             & $\geq$5                  & \multirow{7}{*}{SR3}                     & SR4                     &                                           &                                           & \\ \cline{3-4} \cline{6-6}
                                                  &                            & \multirow{2}{*}{200--300}  & 2--4                 &                                          & SR5 (++) / SR6 (-$\,$-)   &                                           &                                           & \\ \cline{4-4} \cline{6-6}
                                                  &                            &                             & $\geq$5                  &                                          & SR7                     &                                           &                                           & \\ \cline{2-4} \cline{6-6}
                                                  & \multirow{4}{*}{$>$120}    & \multirow{2}{*}{ 50--200}  & 2--4                 &                                          & SR8 (++) / SR9 (-$\,$-)   &                                           &                                           & \\
\hline
        \end{tabular}
        }

\end{table*}

\begin{table*}[htb!]
    \centering
    \topcaption{\label{tab:SRDefHL} The SR definitions for the HL category. Charge-split regions are indicated with (++) and (-$\,$-).
        There are 43 regions in total.
        Quantities are specified in units of \GeV where applicable.}
        \cmsTable{
            \begin{tabular}{|c|c|c|c|c|c|c|c|}
                \hline
                $\Nbjets$                   & $\MTmin$              & $\ptmiss$              & $\Njets$             & $\HT < 300$                          & $\HT\in[300, 1125]$                                                   & $\HT\in[1125, 1300]$                  & $\HT > 1300$ \\ \hline \hline
                \multirow{4}{*}{0}          & \multirow{4}{*}{$<$120}    & \multirow{2}{*}{50--200}   & 2--4                  & SR1                                      & SR2                                                                       & \multirow{17}{*}{\shortstack{SR40 (++) / \\ SR41 (-$\,$-)}} & \multirow{17}{*}{\shortstack{SR42 (++) / \\ SR43 (-$\,$-)}} \\ \cline{4-6}
                                            &                            &                             & $\geq$5                   & \multirow{3}{*}{SR3}                     & SR4                                                                       &                                           & \\ \cline{3-4} \cline{6-6}
                                            &                            & \multirow{2}{*}{200--300}  & 2--4                  &                                          & SR5 (++) / SR6 (-$\,$-)                                                     &                                           & \\ \cline{4-4} \cline{6-6}
                                            &                            &                             & $\geq$5                   &                                          & SR7                                                                       &                                           & \\ \cline{1-6}
                \multirow{4}{*}{1}          & \multirow{4}{*}{$<$120}    & \multirow{2}{*}{ 50--200}  & 2--4                  & SR8                                      & SR9                                                                       &                                           & \\
\hline
        \end{tabular}
        }

\end{table*}

\begin{table*}[htb!]
    \centering
    \topcaption{\label{tab:SRDefLL} The SR definitions for the LL category.
        All SRs in this category require $\Njets \geq 2$.
        There are 8 regions in total.
        Quantities are specified in units of \GeV where applicable.}
            \begin{tabular}{|c|c|c|c|c|}
                \hline
                $\Nbjets$ & $\MTmin$           & $\HT$             & $\ptmiss\in[50, 200]$ & $\ptmiss> 200$ \\
                \hline
                0         & \multirow{4}{*}{$<$120} &\multirow{5}{*}{$>$400} & SR1                       & SR2 \\
                \cline{1-1} \cline{4-5}
                1         &                         &                        & SR3                       & SR4 \\
                \cline{1-1} \cline{4-5}
                2         &                         &                        & SR5                       & SR6 \\
                \cline{1-1} \cline{4-5}
                $\geq$3  &                         &                        & \multicolumn{2}{c|}{SR7} \\
                \cline{1-2} \cline{4-5}
                Inclusive & $>$120                  &                        & \multicolumn{2}{c|}{SR8} \\
                \hline
        \end{tabular}

\end{table*}

\begin{table*}[htb!]
    \centering
        \topcaption{\label{tab:SRDefLM}  The SR definitions for the LM category. All SRs in this category require $\ptmiss < 50\GeV$ and $\HT>300\GeV$.
        The two high-$\HT$ regions are split only by $\Njets$, resulting in 11 regions in total.
        Quantities are specified in units of \GeV where applicable.}
            \begin{tabular}{|c|c|c|c|c|}
                \hline
                $\Nbjets$                      & $\Njets$ & $\HT\in[300,1125]$ & $\HT\in[1125,1300]$                 & $\HT> 1300$ \\ \hline
                \multirow{2}{*}{0}             & 2--4      & SR1                    & \multirow{3}{*}{SR8 ($\Njets<5$)}       & \multirow{3}{*}{SR10 ($\Njets<5$)}       \\ \cline{2-3}
                                               & $\geq$5       & SR2                    &                                         &                                           \\ \cline{1-3}
                \multirow{2}{*}{1}             & 2--4      & SR3                    &                                         &                                                                                     \\ \cline{2-3}
                                               & $\geq$5       & SR4                    & \multirow{3}{*}{SR9 ($\Njets\geq 5$)}       & \multirow{3}{*}{SR11 ($\Njets\geq 5$)}        \\ \cline{1-3}
                \multirow{2}{*}{2}             & 2--4      & SR5                    &                                         &                                                                                      \\ \cline{2-3}
                                               & $\geq$5       & SR6                    &                                         &                                                                                       \\ \cline{1-3}
                $\geq$3                        & $\geq$2       & SR7                    &                                         &                                           \\ \hline
        \end{tabular}

\end{table*}

\begin{table*}[htb!]
    \centering
    \topcaption{\label{tab:SRDefMLonZ} The SR definitions for the on-\PZ ML category. All SRs in these categories require $\Njets \geq 2$.
        Regions marked with ${}^\dagger$ are split by $\MTmin=120\GeV$, with the high-$\MTmin$ region specified by the second SR label.
        There are 23 regions in total.
        Quantities are specified in units of \GeV where applicable.}
            \begin{tabular}{|c|c|c|c|c|}
            \hline
                $\Nbjets$               & $\HT$  & $\ptmiss\in[50,150]$   & $\ptmiss\in[150,300]$ & $\ptmiss\geq300$ \\ \hline
                \multirow{2}{*}{0}      & $<$400      & SR1/SR2${}^\dagger$      & SR3/SR4${}^\dagger$     & \multirow{8}{*}{SR22/SR23${}^\dagger$}\\ \cline{2-2}
                                        & 400--600     & SR5/SR6${}^\dagger$      & SR7/SR8${}^\dagger$     & \\ \cline{1-2}
                \multirow{2}{*}{1}      & $<$400      & SR9                       & SR10                      & \\ \cline{2-2}
                                        & 400--600     & SR11                       & SR12                      & \\ \cline{1-2}
                \multirow{2}{*}{2}      & $<$400      & SR13                       & SR14                      & \\ \cline{2-2}
                                        & 400--600     & SR15                       & SR16                      & \\ \cline{1-4}
                $\geq$3                 & $<$600      & \multicolumn{2}{c|}{SR17}              & \\ \cline{1-4} 
                Inclusive               & $\geq$600   & SR18/SR19${}^\dagger$      & SR20/SR21${}^\dagger$     & \\ \hline
        \end{tabular} 
    
\end{table*}

\begin{table*}[htb!]
    \centering
    \topcaption{\label{tab:SRDefMLoffZ} The SR definitions for the off-\PZ category. All SRs in these categories require $\Njets \geq 2$.
        Regions marked with ${}^\dagger$ are split by $\MTmin=120\GeV$, with the high-$\MTmin$ region specified by the second SR label.
        There are 21 regions in total.
        Quantities are specified in units of \GeV where applicable.}
            \begin{tabular}{|c|c|c|c|c|}
                \hline
                $\Nbjets$               & $\HT$  & $\ptmiss\in[50,150]$    & $\ptmiss\in[150,300]$  & $\ptmiss\geq300$ \\ \hline
                \multirow{2}{*}{0}      & $<$400      & SR1/SR2${}^\dagger$         & SR3/SR4${}^\dagger$    & \multirow{8}{*}{SR20/SR21${}^\dagger$}              \\ \cline{2-2}
                                        & 400--600     & SR5                         & SR6                    & \\ \cline{1-2}
                \multirow{2}{*}{1}      & $<$400      & SR7                         & SR8                    & \\ \cline{2-2}
                                        & 400--600     & SR9                         & SR10                   & \\ \cline{1-2}
                \multirow{2}{*}{2}      & $<$400      & SR11                        & SR12                   & \\ \cline{2-2}
                                        & 400--600     & SR13                        & SR14                   & \\ \cline{1-4}
                $\geq$3                 & $<$600      & \multicolumn{2}{c|}{SR15}                 & \\ \cline{1-4}
                Inclusive               & $\geq$600   & SR16/SR17${}^\dagger$       & SR18/SR19${}^\dagger$  & \\ \hline
        \end{tabular}

\end{table*}

\clearpage

\section{Backgrounds}
\label{sec:backgrounds}

Several SM processes can lead to the signatures studied in this analysis.
There are three background categories, depending on the lepton content of the event:
\begin{itemize}
\item Events with two or more prompt leptons, including an SS pair;
\item Events with at least one nonprompt lepton (defined below); and
\item Events with a pair of OS leptons, one of which is reconstructed with the wrong charge.
\end{itemize}

The first category includes a variety of low cross section processes where multiple electroweak bosons are produced, possibly in the decay of top quarks, which then
decay leptonically leading to an SS lepton pair.
This category usually dominates the background yields in SRs with large \ptmiss or \HT and in most of the ML SRs with a \PZ candidate.
The main contributions arise from the production of a $\PW\PZ$ or an SS \PW pair, or of a $\ttbar $ pair in association with a \PW, \PZ or \PH boson. The event yields for these processes are estimated individually.
In contrast, the expected event yields from other rare processes (including $\PZ\PZ$, triple boson production, $\cPqt\PW\PZ$, $\cPqt\PZ\cPq$, $\ttbar\ttbar$, and double parton scattering)
are summed up into a single contribution denoted as ``Rare''.
Processes including a genuine photon, such as $\PW\Pgg$, $\PZ\Pgg$, $\ttbar \Pgg$, and $\cPqt\Pgg$, are also considered and grouped together.
They are referred to as ``X\Pgg''.
All contributions from this category are estimated using simulated samples. Correction factors are applied to take into account small differences between data and simulation,
including trigger, lepton selection, and \PQb tagging efficiencies, with associated systematic uncertainties listed in Section~\ref{sec:systematics}.

The second category consists of events where one of the selected leptons, generically denoted as ``nonprompt lepton'', is either a decay product of a heavy flavor hadron or, more rarely, a misidentified hadron. 
This category is typically the dominant one in SRs with moderate or low \ptmiss or low \MTmin (except for the on-\PZ ML SRs).
This background is estimated directly from data using the ``tight-to-loose''
method~\cite{SUS-16-035,SUS-16-041}.
This method is based on the probability for a nonprompt lepton passing loose selection criteria to also
satisfy the tighter lepton selection used in the analysis.
The number of events in an SR
with $N$ leptons, including at least one nonprompt lepton,
can be estimated by applying this probability to a corresponding control region (CR) of
events with $N$ loose leptons where at least one of them
fails the tight selection.

The measurement of the tight-to-loose ratio is performed in a sample enriched in dijet events with exactly one loose lepton, low \ptmiss, and low \MTmin.
This sample is contaminated by prompt leptons from \PW\ boson decays.
The contamination is estimated from the \MTmin distribution, and it is subtracted before calculating
the ratio.
The tight-to-loose ratio is computed separately for electrons and muons, and is parameterized as a function of the lepton $\eta$ and \ptcorr.
The \ptcorr variable is defined as the sum of the lepton \pt and the energy in the isolation cone exceeding the isolation threshold value applied to tight leptons.
This parametrization improves the stability of the tight-to-loose ratio with respect to variations in the \pt of the partons
from which the leptons originate.

The performance of the tight-to-loose ratio was assessed in a MC closure test.  A tight-to-loose ratio was extracted from a MC sample of QCD events.  This ratio was then used to predict the number of events with one prompt and one nonprompt SS dileptons in MC \ttbar and $\PW\text{+jets}$ events.  The predicted and observed rates of SS dileptons were compared as a function of kinematic properties and found to agree within 30\%.
The data driven estimate was also compared to a direct prediction from simulation and a similar level of agreement was reached.

The final category is a subdominant background in all SRs and corresponds to events where the charge of a lepton is incorrectly measured.
Charge misidentification primarily occurs when an electron undergoes brems\-strah\-lung in the tracker
material or in the beam pipe.
Similarly to the tight-to-loose method, the number of SS lepton pairs where one of the leptons has its charge misidentified
can be determined using the number of OS pairs and the knowledge of the charge misidentification rate.
We use simulation to parameterize this rate as a function of \pt and $\eta$ for electrons and find values varying between
$10^{-5}$ (central electrons with \pt$\approx 20\GeV$) and $5\times 10^{-3}$ (forward electrons with \pt$\approx 200\GeV$).  To calibrate the charge misidentification rate, we exploit the fact that
charge misidentification only has a small effect on the electron energy measurement in the calorimeter.
As a result, electron pairs from $\PZ$ boson decays yield a sharp peak near the $\PZ$ mass even when
one of the electrons has a misidentified charge.
The SS dielectron invariant mass distributions in data and MC can then be used to derive
a correction factor to the MC charge misidentification rate.
Good agreement between data and MC is found in 2016, while the charge misidentification rate in simulation corresponding to 2017 and 2018 data
needs to be scaled up by a factor of 1.4.
Muon charge misidentification arises from a relatively large uncertainty in the transverse momentum at high momentum or from a poor quality track.
The various criteria applied in this analysis on the quality of the muon reconstruction lead to a misidentification rate at least one order of magnitude smaller than for electrons according to simulation. The muon charge misassignment has also been studied using cosmic ray muons with \pt up to several hundred \GeV, confirming the predictions from simulation~\cite{Sirunyan:2019yvv}. It is therefore neglected.
Correction factors are however applied to the simulation to account for a possible difference in the selection efficiency related to these criteria.

\section{Systematic uncertainties}
\label{sec:systematics}

The predicted yields of signal and background processes are affected by several sources of uncertainty, summarized in Table~\ref{tab:systSummary}.
Depending on their source, they are treated as fully correlated or uncorrelated between the three years of data taking.
Signal and background contributions estimated from simulation are affected by experimental uncertainties in the efficiency of the trigger,
lepton reconstruction and identification~\cite{Khachatryan:2015hwa,Chatrchyan:2012xi}, the efficiency of \PQb~tagging~\cite{Sirunyan:2017ezt},
the jet energy scale~\cite{Khachatryan:2016kdb},  the integrated luminosity~\cite{cmsLumi2016,cmsLumi2017,cmsLumi2018}.
An uncertainty is also assigned to the value of the inelastic cross section, which affects the pileup rate~\cite{Sirunyan:2018nqx} and that can impact the description of the jet multiplicity or the \ptmiss resolution. 
Simulation is also affected by theoretical uncertainties, which are evaluated by varying the factorization and renormalization scales up and down by a factor of two,
and by using different PDFs within the NNPDF3.0 and 3.1 PDF sets~\cite{Ball:2014uwa,Ball:2017nwa,Kalogeropoulos:2018cke}.
These uncertainties can affect both the overall yield (normalization) and the relative population (shape) across the SRs.
Background normalization uncertainties are increased to 30\%, either to account for the additional hadronic activity required (for $\PW\PZ$ and $\PW^{\pm} \PW^{\pm}$)
or to take into consideration recent measurements (for $\ttbar\PW$, $\ttbar\PZ$)~\cite{ATLAS:ttV,CMS:ttV}.
The Rare and X\Pgg backgrounds, which are less well understood experimentally and theoretically, are assigned a 50\% uncertainty.

To account for possible mismodeling of the flavor of additional jets, an additional 70\% uncertainty is applied to
$\ttbar \PW$, $\ttbar \PZ$, and $\ttbar \PH$ events produced in association with a pair of \PQb~jets, reflecting the
measured ratio of $\ttbar \bbbar / \ttbar \text{jj}$ cross sections reported in Ref.~\cite{CMSttbb}.

As discussed in Section~\ref{sec:backgrounds}, the nonprompt lepton
and charge misidentification backgrounds are estimated from CRs.  The associated uncertainties include the
statistical uncertainties in the CR yields, as well as the systematic
uncertainties in the extrapolations from the CRs to the SRs, as described below.
In the case of the nonprompt lepton background, we include a 30\% uncertainty
from studies of the closure of the method in simulation.
Furthermore, the uncertainty in the measurement of the tight-to-loose ratio, because of the prompt lepton contamination,
results in a 1--30\% additional uncertainty in the background yields.
The charge misidentification background is assigned a 20\% uncertainty based on a comparison of the kinematic properties of simulated and data events
in the $\PZ\to\Pep\Pem$ CR with one electron or positron having a misidentified charge.

In general, the systematic uncertainties with the largest impact on the expected limits defined below are related to the lepton identification and isolation scale factors, the cross section of the rare processes, and the $\PW\PZ$ background normalization.

\begin{table*}[!hbtp]
\centering
    \topcaption{
   Summary of the sources of systematic uncertainty and their effect on the yields of different processes in the SRs.
    The first two groups list experimental and theoretical uncertainties assigned to processes estimated using simulation,
    while the last group lists uncertainties assigned to processes whose yield is estimated from the data.
    The uncertainties in the first group also apply to signal samples.
    Reported values are representative for the most relevant signal regions.
    }
    \label{tab:systSummary}
\begin{tabular}{lcc}
    \hline
Source          & Typical uncertainty (\%) & Correlation across years\\
\hline
Integrated luminosity & 2.3--2.5 &Uncorrelated \\
Lepton selection & 2--10 &Uncorrelated \\
Trigger efficiency & 2--7  &Uncorrelated \\
Pileup & 0--6  &Uncorrelated\\
Jet energy scale & 1--15 &Uncorrelated\\
$\cPqb$ tagging & 1--10  &Uncorrelated\\
Simulated sample size & 1--20 &Uncorrelated \\
[\cmsTabSkip]
Scale and PDF variations & 10--20 &Correlated\\
Theoretical background cross sections & 30--50& Correlated\\
[\cmsTabSkip]
Nonprompt leptons & 30 & Correlated \\
Charge misidentification & 20  & Uncorrelated\\
\nisrjet & 1--30  & Uncorrelated\\
    \hline
\end{tabular}

\end{table*}

\section{Results and interpretation}
\label{sec:results}
\begin{figure*}[!hbtp]
\centering
\includegraphics[width=.48\textwidth]{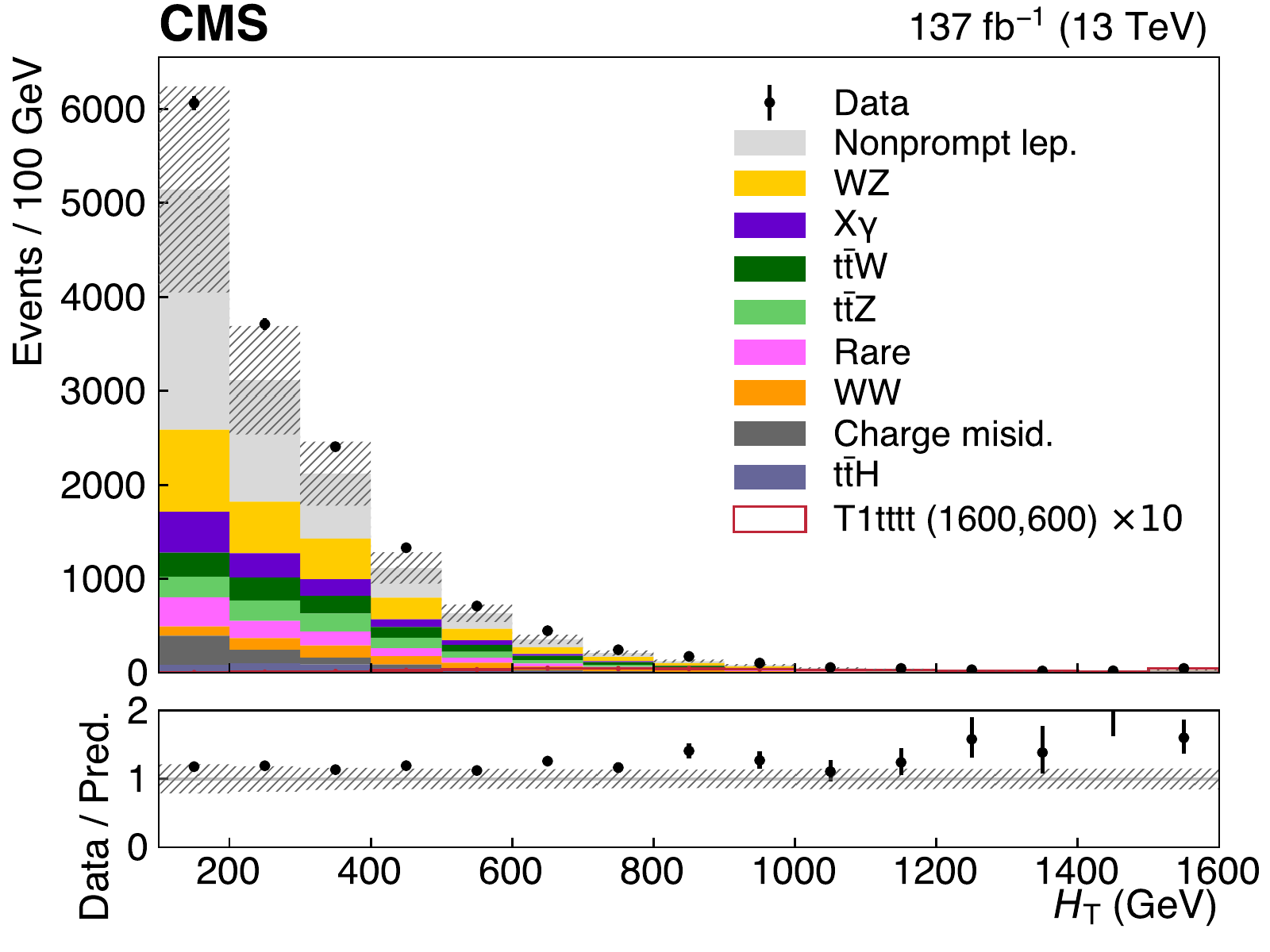}
\includegraphics[width=.48\textwidth]{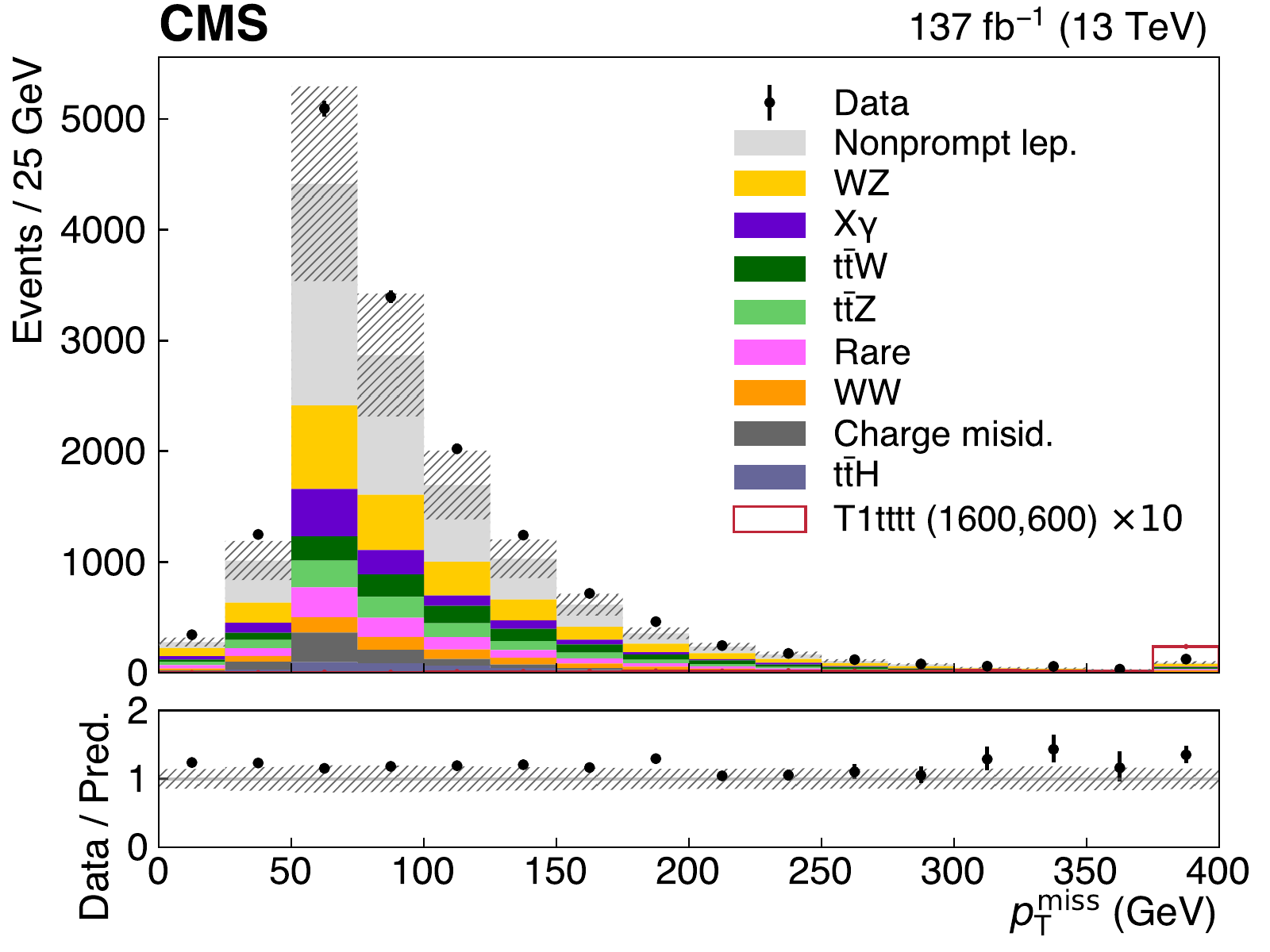}
\includegraphics[width=.48\textwidth]{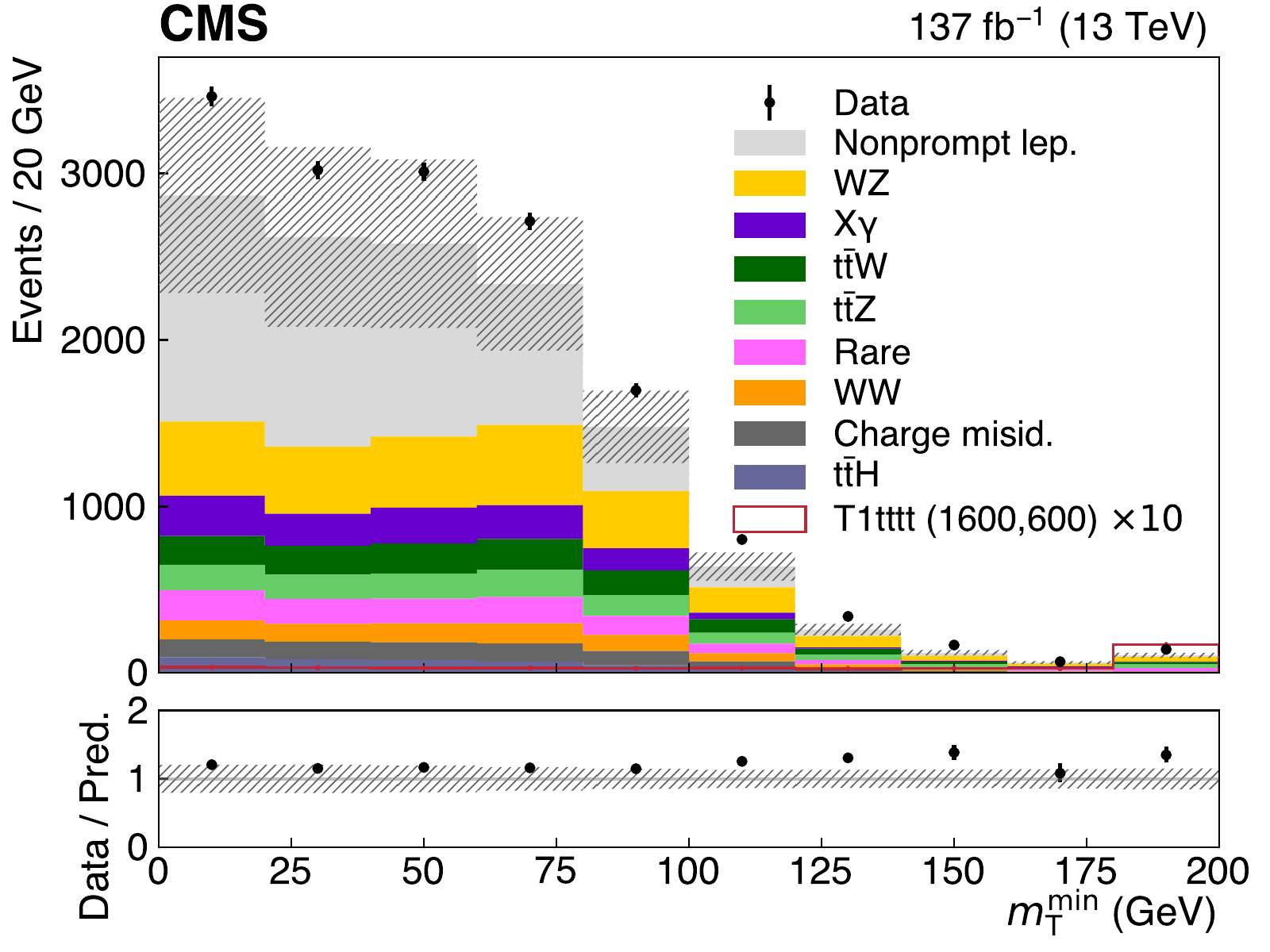}
\includegraphics[width=.48\textwidth]{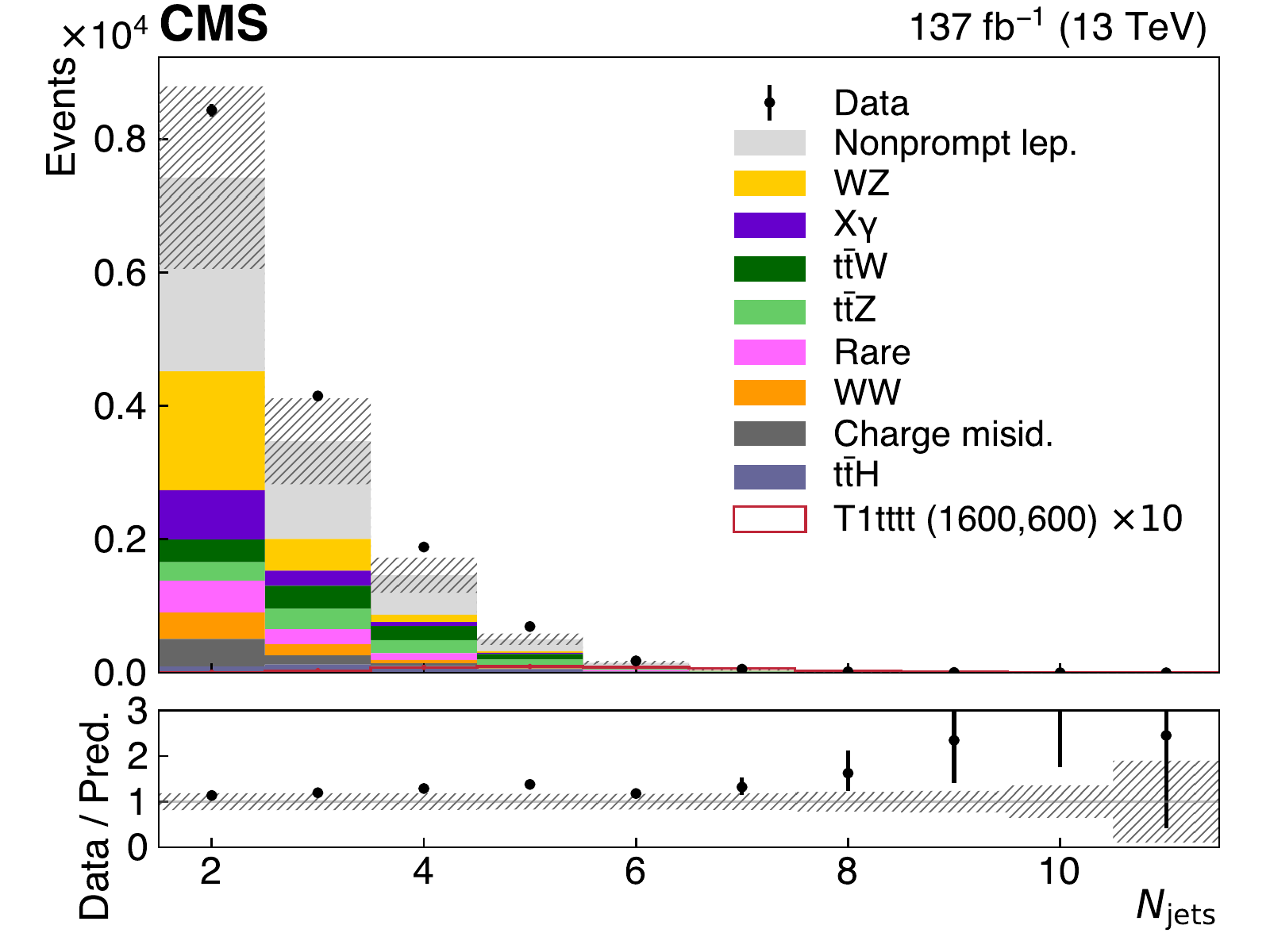}
\includegraphics[width=.48\textwidth]{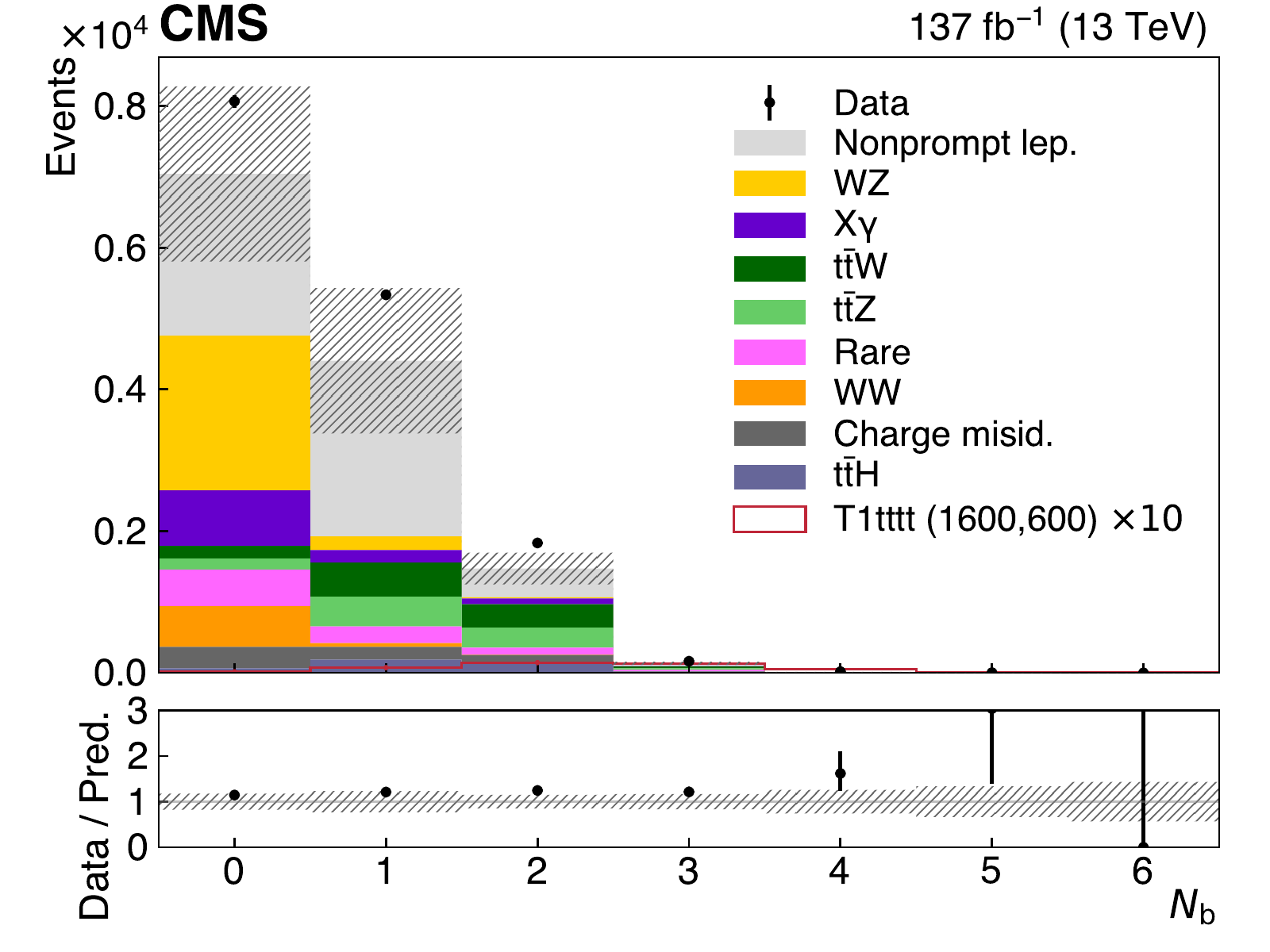}
\includegraphics[width=.48\textwidth]{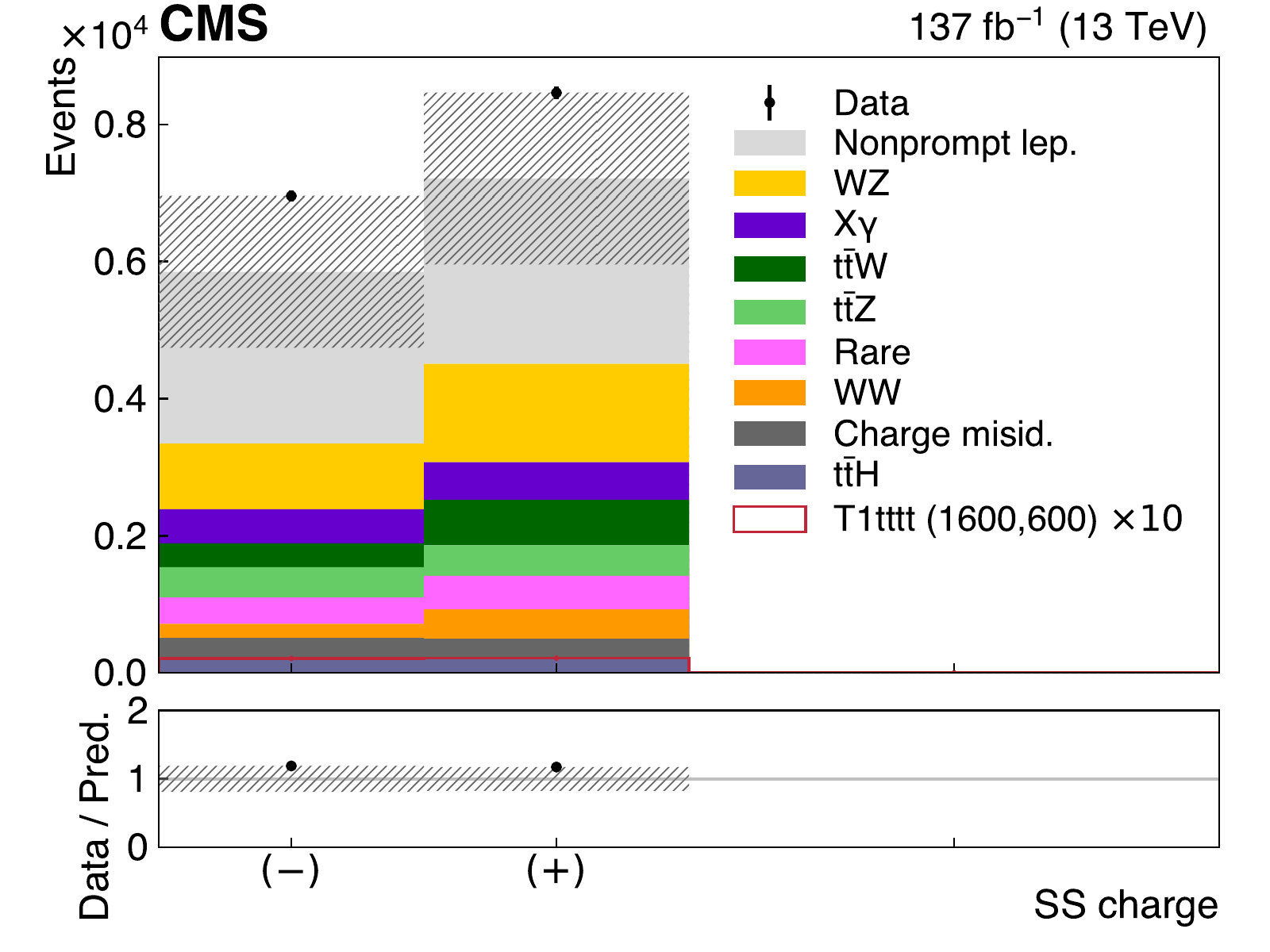}
\\
\caption{Distributions of the main analysis variables after the event selection:
\HT, \ptmiss, \MTmin, \Njets, \Nbjets, and the charge of the SS pair,
    where the last bin includes the overflow (where applicable). The hatched area represents the total statistical and systematic uncertainty in the background prediction.
The lower panels show the ratio of the observed event yield to the background prediction.
The prediction for the SUSY model \Totttt with $m_{\gluino}=1600\GeV$ and $m_{\lsp}=600\GeV$ is overlaid.
}
\label{fig:kinem}
\end{figure*}

\begin{figure*}[!hbtp]
\centering
\includegraphics[width=.60\textwidth]{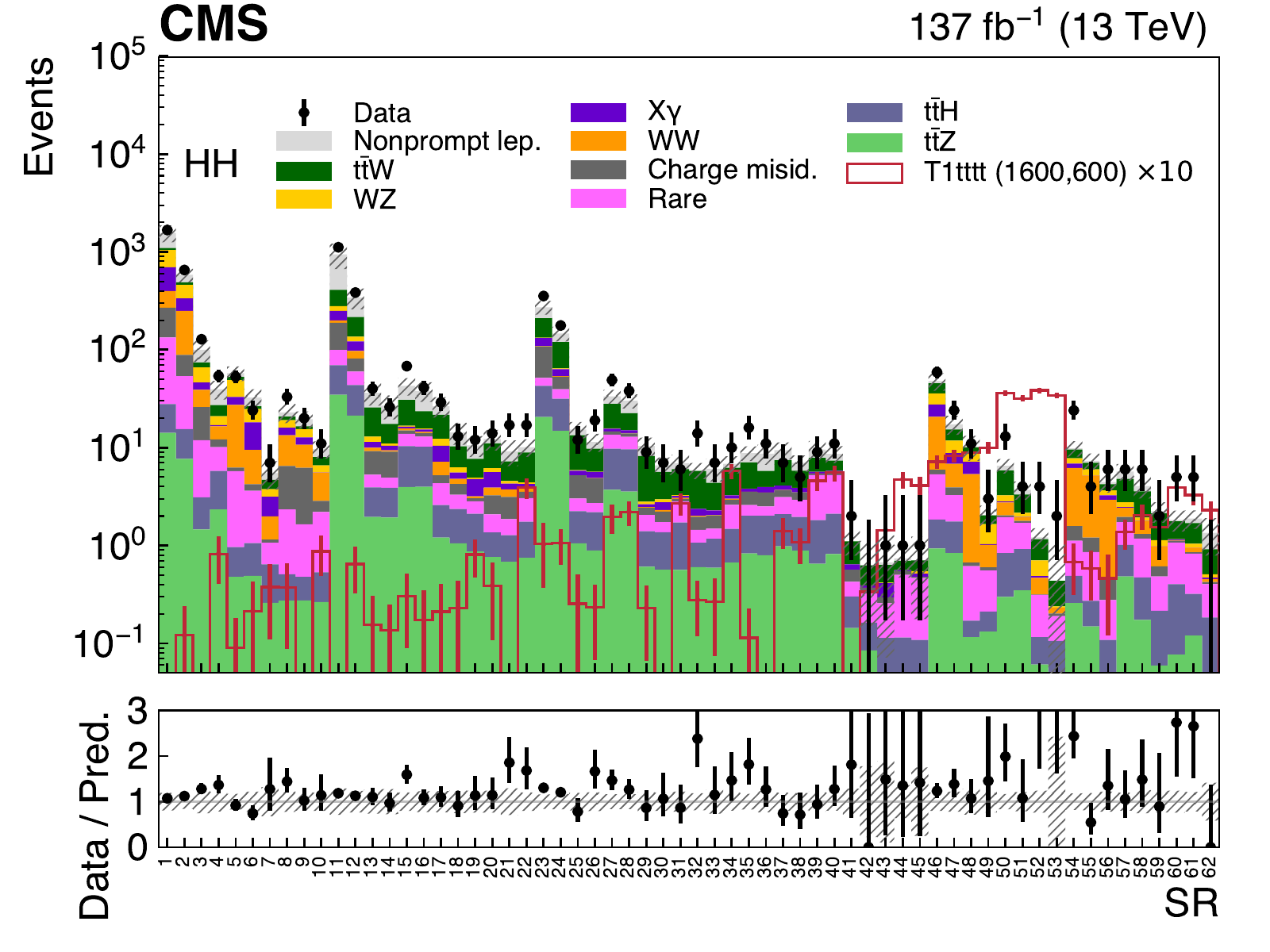}
\includegraphics[width=.60\textwidth]{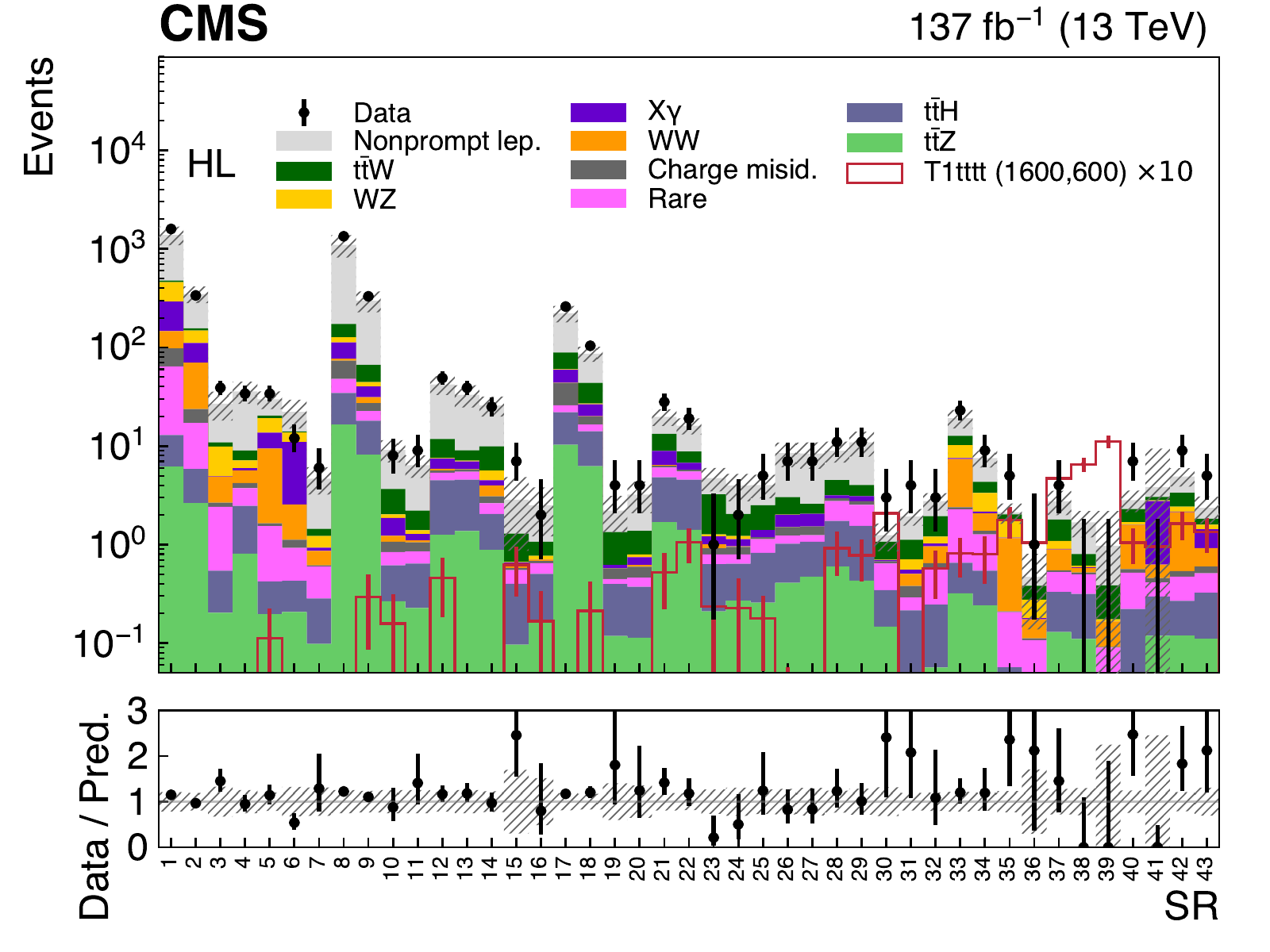}
\includegraphics[width=.60\textwidth]{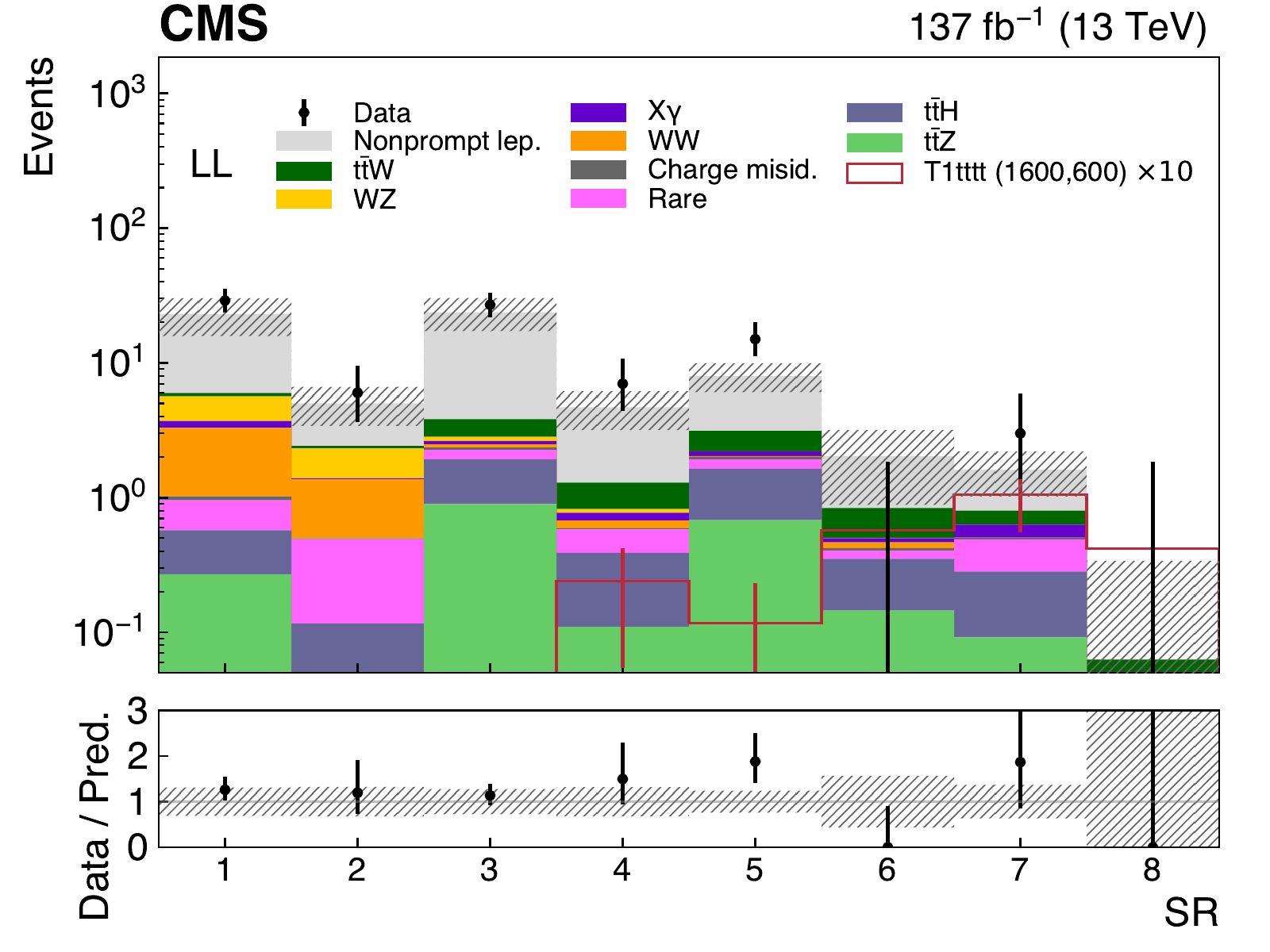}
\\
\caption{Expected and observed SR yields for the HH, HL, LL signal categories.
The hatched area represents the total statistical and systematic uncertainty in the background prediction.
}
\label{fig:SRrun2}
\end{figure*}

\begin{figure*}[!hbtp]
\centering
\includegraphics[width=.60\textwidth]{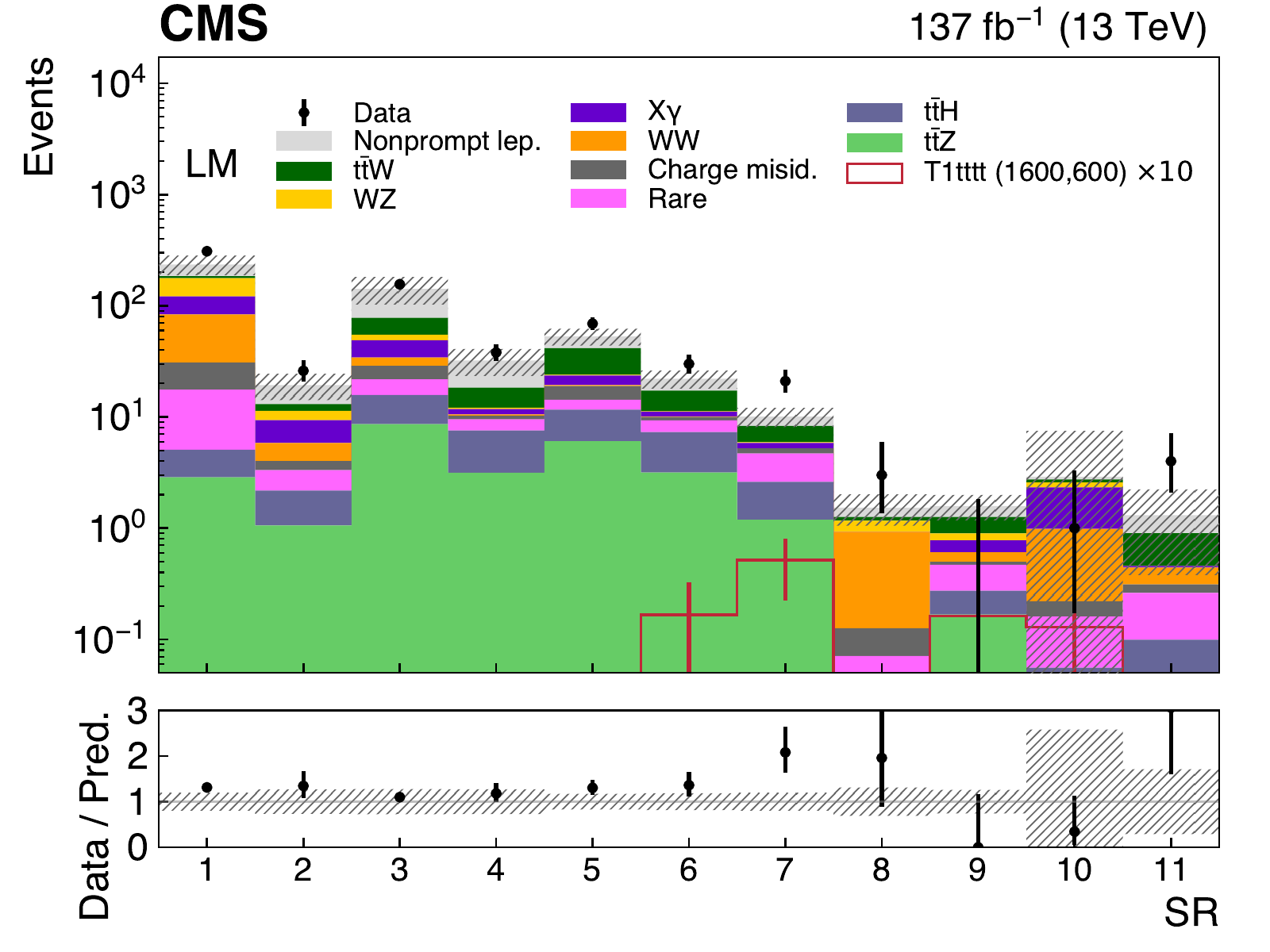}
\includegraphics[width=.60\textwidth]{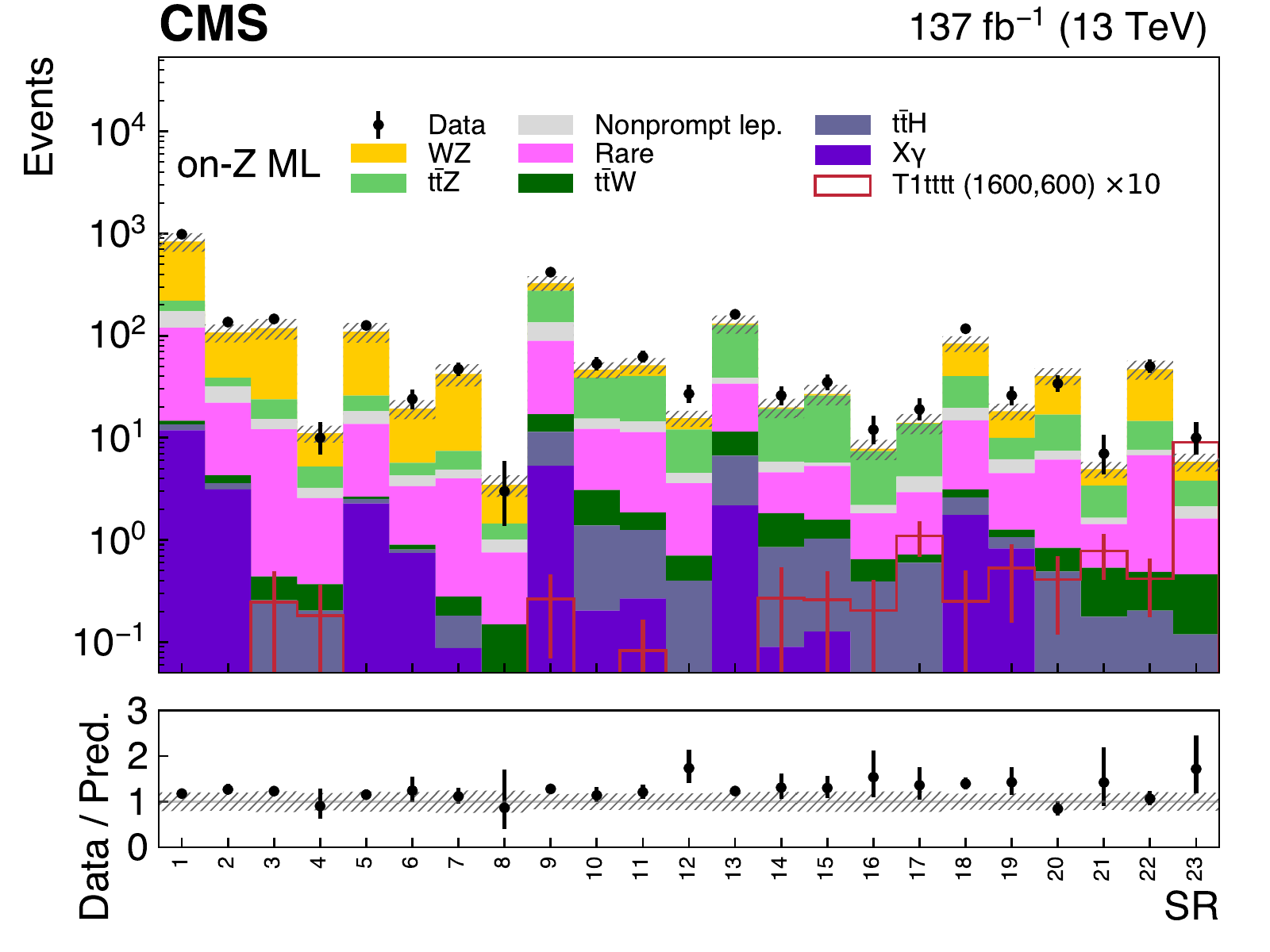}
\includegraphics[width=.60\textwidth]{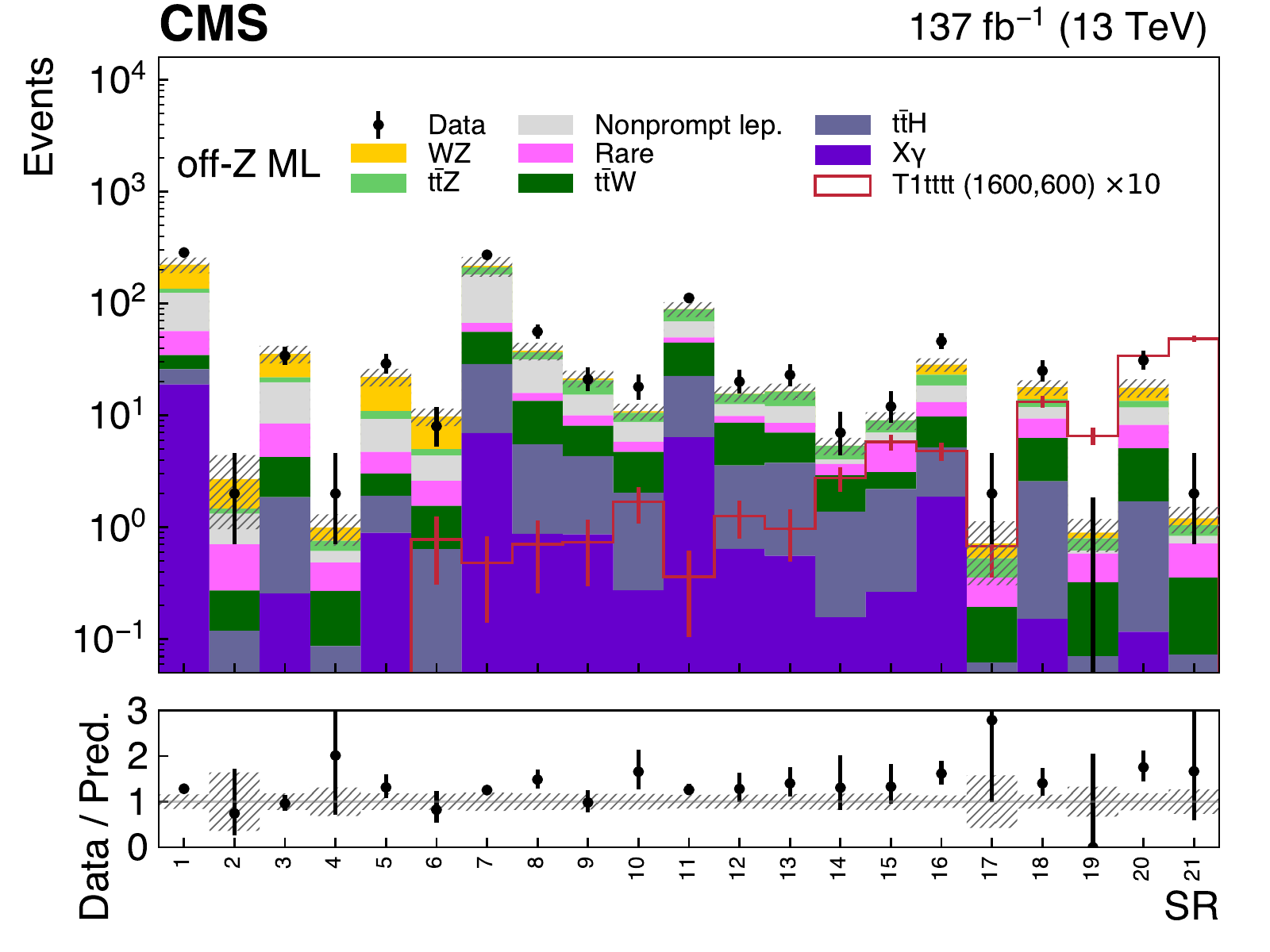}
\\
\caption{Expected and observed SR yields for the LM, on-\PZ ML, off-\PZ ML signal categories. 
The hatched area represents the total statistical and systematic uncertainty in the background prediction.
}
\label{fig:SRrun2b}
\end{figure*}

The distributions of the variables used to define the SRs after the event selection are
shown in Fig.~\ref{fig:kinem}.
Background yields shown as stacked histograms in Figs.~\ref{fig:kinem}, \ref{fig:SRrun2}, and \ref{fig:SRrun2b} are
those determined following the prescriptions detailed in Section~\ref{sec:backgrounds}.
The overall data yields exceed expectation by an amount close to the systematic uncertainty.
However, no particular trend that is not covered by the uncertainties discussed in the previous sections,
is seen in the distributions.
The significance of the excess is of similar magnitude in all categories, with a maximum of around 2 standard deviations (s.d.) in the off-\PZ ML category. 

The results of the search, broken down by SR,
are presented in Figs.~\ref{fig:SRrun2} and~\ref{fig:SRrun2b}, and
are summarized in Table~\ref{tab:slimyields}.
No significant deviation with respect to the SM background prediction is observed.
The largest excess of events found by fitting the data with the background-only
hypothesis is in HH SR54,
corresponding to a local significance of 2.6 s.d.
Its neighboring bin, HH SR55, which is adjacent along the \HT dimension, has a
deficit of events in the data corresponding to a
significance of $1.8$ s.d.

These results are then interpreted as experimental constraints on the cross sections for
the signal models discussed in Section~\ref{sec:samples}. For each model,
event yields in all SRs are used to obtain exclusion limits
on the production cross section at 95\% confidence level (\CL) with
an asymptotic formulation of the modified frequentist \CLs
criterion~\cite{Junk:1999kv,Read:2002hq,ATL-PHYS-PUB-2011-011,Cowan:2010js},
where uncertainties are incorporated as nuisance parameters and profiled~\cite{ATL-PHYS-PUB-2011-011}.
This procedure takes advantage of the differences in the distribution of events amongst the SR between the various SM backgrounds and the signal considered.
The normalizations of the various backgrounds are in particular allowed to float within their uncertainties in the global fit, resulting in several backgrounds (nonprompt lepton, $\ttbar \PW/\PZ/\PH$ and rare processes) being pulled up by around $1$ s.d. for most of the signal points considered, which are often characterized by a distinctive distribution of events across the SRs. 
This observation is consistent with the current measurements of $\ttbar \PW$ and $\ttbar \PZ$ processes performed by the ATLAS and CMS Collaborations~\cite{ATLAS:ttV,CMS:ttV}.
The limits obtained are then used together with the theoretical cross section
calculations to exclude regions of SUSY parameter space.

\begin{table*}[!hbtp]
\centering
\topcaption{Expected background event yields, total uncertainties, and observed event yields in the SRs used in this search.}
\label{tab:slimyields}
\scriptsize
\begin{tabular}{ccc|ccc|ccc}
\hline
\multicolumn{3}{c|}{HH regions}    &    \multicolumn{3}{c|}{HL regions}    &    \multicolumn{3}{c}{LM regions} \\ \hline                                    
 SR    &    Expected SM    &    Obs.    &                                 
 SR    &    Expected SM    &    Obs.    &                                 
 SR    &    Expected SM    &    Obs. \\                                    
 \hline                                                
1   &    $1560\pm300 $    &   1673   &   1   &   $1390\pm300 $    &   1593   &   1   &    $235\pm47 $    &    309 \\
2   &    $582\pm93 $    &   653   &   2   &   $348\pm67 $    &   337   &   2   &    $19.3\pm5.2 $    &    26 \\
3   &    $100\pm25 $    &   128   &   3   &   $26.9\pm8.8 $    &   39   &   3   &    $142\pm39 $    &    156 \\
4   &    $39.5\pm8.5 $    &   54   &   4   &   $35.9\pm9.1 $    &   34   &   4   &    $32.2\pm8.8 $    &    38 \\
5   &    $57.7\pm9.9 $    &   53   &   5   &   $29.8\pm6.0 $    &   34   &   5   &    $53.0\pm9.1 $    &    69 \\
6   &    $32.5\pm7.1 $    &   24   &   6   &   $22.2\pm7.2 $    &   12   &   6   &    $22.0\pm4.0 $    &    30 \\
7   &    $5.5\pm1.8 $    &   7   &   7   &   $4.7\pm1.4 $    &   6   &   7   &    $10.1\pm2.0 $    &    21 \\
8   &    $22.9\pm5.1 $    &   33   &   8   &   $1100\pm280 $    &   1342   &   8   &    $1.53\pm0.48 $    &    3 \\
9   &    $19.5\pm3.9 $    &   20   &   9   &   $299\pm71 $    &   330   &   9   &    $1.58\pm0.41 $    &    0 \\
10   &    $9.6\pm1.9 $    &   11   &   10   &   $9.1\pm2.3 $    &   8   &   10   &    $2.9\pm2.9 $    &    1 \\
11   &    $940\pm270 $    &   1115   &   11   &   $6.4\pm1.6 $    &   9   &   11   &    $1.31\pm0.93 $    &    4 \\
12   &    $340\pm81 $    &   384   &   12   &   $42.1\pm9.2 $    &   49   &       &      &    \\\cline{7-9}
13   &    $36.3\pm9.5 $    &   40   &   13   &   $33.0\pm8.4 $    &   39   &    \multicolumn{3}{c}{on-\PZ ML regions} \\\cline{7-9}            
14   &    $26.8\pm7.4 $    &   26   &   14   &   $25.8\pm5.9 $    &   25   &    SR    &    Expected SM    &    Obs. \\\cline{7-9} 
15   &    $42.7\pm8.6 $    &   68   &   15   &   $2.8\pm2.0 $    &   7   &   1   &    $840\pm170 $    &    985 \\
16   &    $37.9\pm8.6 $    &   41   &   16   &   $2.5\pm1.3 $    &   2   &   2   &    $107\pm21 $    &    136 \\
17   &    $26.5\pm6.2 $    &   29   &   17   &   $222\pm42 $    &   260   &   3   &    $119\pm27 $    &    146 \\
18   &    $14.3\pm3.6 $    &   13   &   18   &   $86\pm15 $    &   104   &   4   &    $11.1\pm2.1 $    &    10 \\ 
19   &    $10.6\pm2.5 $    &   12   &   19   &   $2.22\pm0.90 $    &   4   &   5   &    $109\pm24 $    &    126 \\ 
20   &    $12.3\pm2.9 $    &   14   &   20   &   $3.2\pm1.1 $    &   4   &   6   &    $19.3\pm4.1 $    &    24 \\
21   &    $9.2\pm2.7 $    &   17   &   21   &   $19.8\pm3.8 $    &   28   &   7   &    $42\pm10 $    &    47 \\
22   &    $10.1\pm2.1 $    &   17   &   22   &   $16.1\pm3.0 $    &   19   &   8   &    $3.47\pm0.84 $    &    3 \\
23   &    $272\pm43 $    &   354   &   23   &   $4.7\pm1.3 $    &   1   &   9   &    $327\pm54 $    &    419 \\
24   &    $147\pm25 $    &   177   &   24   &   $4.0\pm1.2 $    &   2   &   10   &    $46.5\pm8.4 $    &    53 \\
25   &    $15.3\pm2.9 $    &   12   &   25   &   $4.0\pm1.1 $    &   5   &   11   &    $51.3\pm9.1 $    &    62 \\
26   &    $11.4\pm2.4 $    &   19   &   26   &   $8.5\pm2.4 $    &   7   &   12   &    $15.6\pm2.8 $    &    27 \\
27   &    $33.4\pm5.4 $    &   49   &   27   &   $8.4\pm2.5 $    &   7   &   13   &    $131\pm27 $    &    162 \\
28   &    $30.1\pm4.9 $    &   38   &   28   &   $8.9\pm2.2 $    &   11   &   14   &    $19.9\pm4.3 $    &    26 \\
29   &    $10.4\pm2.2 $    &   9   &   29   &   $10.9\pm3.1 $    &   11   &   15   &    $26.9\pm6.1 $    &    35 \\
30   &    $6.6\pm1.3 $    &   7   &   30   &   $1.25\pm0.39 $    &   3   &   16   &    $7.8\pm1.8 $    &    12 \\
31   &    $6.9\pm1.5 $    &   6   &   31   &   $1.92\pm0.37 $    &   4   &   17   &    $14.0\pm3.1 $    &    19 \\
32   &    $5.9\pm1.1 $    &   14   &   32   &   $2.77\pm0.56 $    &   3   &   18   &    $84\pm15 $    &    117 \\
33   &    $6.1\pm1.6 $    &   7   &   33   &   $19.1\pm4.1 $    &   23   &   19   &    $18.2\pm3.3 $    &    26 \\
34   &    $6.8\pm1.3 $    &   10   &   34   &   $7.5\pm1.5 $    &   9   &   20   &    $40.4\pm7.6 $    &    34 \\
35   &    $8.8\pm1.5 $    &   16   &   35   &   $2.12\pm0.49 $    &   5   &   21   &    $4.92\pm0.88 $    &    7 \\
36   &    $8.7\pm2.0 $    &   11   &   36   &   $0.47\pm0.33 $    &   1   &   22   &    $46.9\pm9.9 $    &    50 \\
37   &    $9.4\pm1.9 $    &   7   &   37   &   $2.75\pm0.77 $    &   4   &   23   &    $5.8\pm1.2 $    &    10 \\ 
38   &    $7.0\pm1.3 $    &   5   &   38   &   $1.68\pm0.50 $    &   0   &      &      &    \\\cline{7-9}
39   &    $9.6\pm2.1 $    &   9   &   39   &   $0.97\pm0.97 $    &   0   &    \multicolumn{3}{c}{off-\PZ ML regions} \\\cline{7-9}            
40   &    $8.6\pm1.7 $    &   11   &   40   &   $2.83\pm0.70 $    &   7   &    SR    &    Expected SM    &    Obs. \\\cline{7-9} 
41   &    $1.10\pm0.32 $    &   2   &   41   &   $3.8\pm3.8 $    &   0   &   1   &    $222\pm36 $    &    285 \\
42   &    $0.63\pm0.49 $    &   0   &   42   &   $4.9\pm1.0 $    &   9   &   2   &    $2.7\pm1.7 $    &    2 \\
43   &    $0.67\pm0.60 $    &   1   &   43   &   $2.36\pm0.72 $    &   5   &   3   &    $35.5\pm6.4 $    &    34 \\
44   &    $0.74\pm0.27 $    &   1   &       &       &      &   4   &    $0.99\pm0.31 $    &    2 \\\cline{4-6}
45   &    $0.71\pm0.53 $    &   1   &    \multicolumn{3}{c|}{LL regions}                &   5   &    $22.1\pm4.0 $    &    29 \\\cline{4-6}
46   &    $47.8\pm9.7 $    &   59   &    SR    &    Expected SM    &    Obs.    &   6   &    $9.7\pm1.7 $    &    8 \\\cline{4-6}
47   &    $17.3\pm3.8 $    &   24   &   1   &    $23.0\pm7.2 $    &   29   &   7   &    $217\pm44 $    &    272 \\
48   &    $10.3\pm2.9 $    &   11   &   2   &    $5.0\pm1.6 $    &   6   &   8   &    $37.7\pm6.8 $    &    56 \\
49   &    $2.06\pm0.49 $    &   3   &   3   &    $23.8\pm6.6 $    &   27   &   9   &    $21.4\pm3.7 $    &    21 \\
50   &    $6.5\pm1.1 $    &   13   &   4   &    $4.7\pm1.5 $    &   7   &   10   &    $10.9\pm1.9 $    &    18 \\
51   &    $3.72\pm0.79 $    &   4   &   5   &    $8.0\pm1.9 $    &   15   &   11   &    $89\pm14 $    &    112 \\
52   &    $1.21\pm0.29 $    &   4   &   6   &    $2.0\pm1.1 $    &   0   &   12   &    $15.6\pm2.4 $    &    20 \\
53   &    $0.44\pm0.44 $    &   2   &   7   &    $1.61\pm0.59 $    &   3   &   13   &    $16.4\pm2.7 $    &    23 \\
54   &    $9.8\pm1.8 $    &   24   &   8   &    $0.06\pm0.06 $    &   0   &   14   &    $5.36\pm0.95 $    &    7 \\
55   &    $7.3\pm1.4 $    &   4   &       &       &      &   15   &    $9.0\pm1.6 $    &    12 \\
56   &    $4.44\pm0.98 $    &   6   &       &       &      &   16   &    $28.4\pm3.9 $    &    46 \\
57   &    $5.7\pm1.1 $    &   6   &       &       &      &   17   &    $0.72\pm0.41 $    &    2 \\
58   &    $4.0\pm1.0 $    &   6   &       &       &      &   18   &    $17.8\pm2.8 $    &    25 \\
59   &    $2.24\pm0.53 $    &   2   &       &       &      &   19   &    $0.89\pm0.29 $    &    0 \\
60   &    $1.83\pm0.44 $    &   5   &       &       &      &   20   &    $17.7\pm3.3 $    &    31 \\
61   &    $1.88\pm0.40 $    &   5   &       &       &      &   21   &    $1.20\pm0.32 $    &    2 \\
62   &    $1.35\pm0.56 $    &   0   &       &       &      &      &      &   \\
\hline                                                
\end{tabular}                                              
\end{table*}

\begin{figure*}[!hbtp]
\centering
\includegraphics[width=0.45\textwidth]{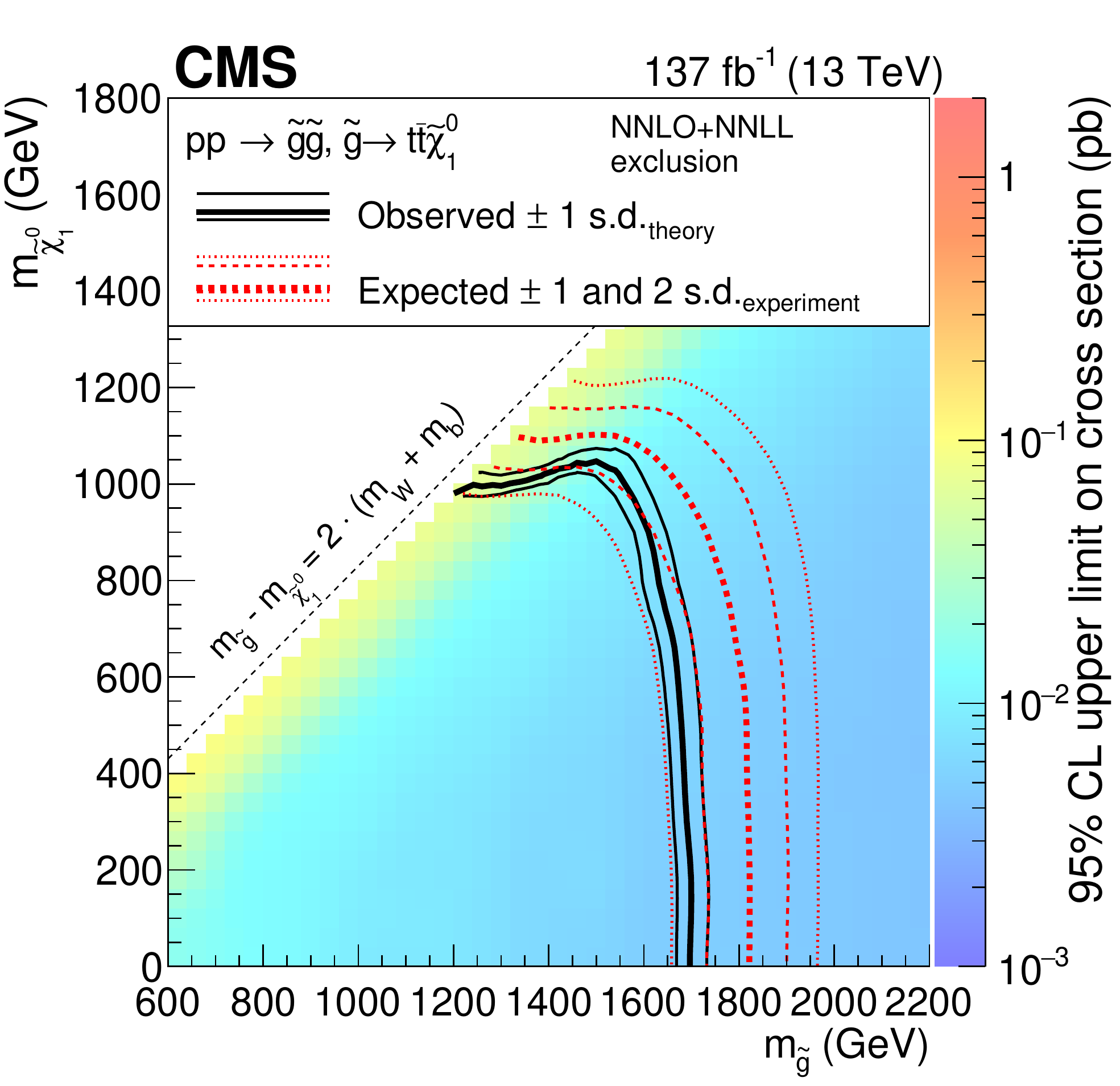}
\includegraphics[width=0.45\textwidth]{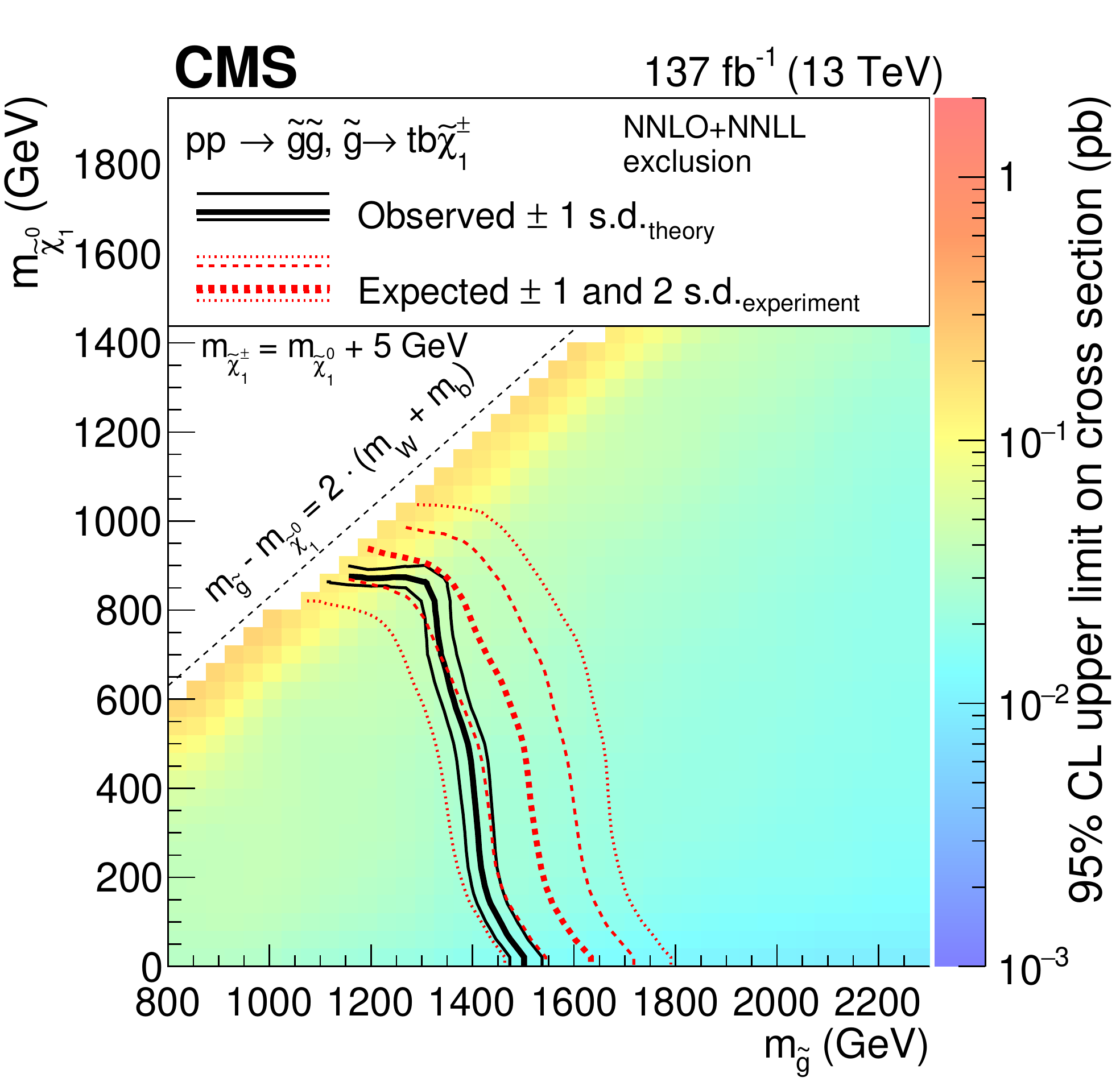}
\includegraphics[width=0.45\textwidth]{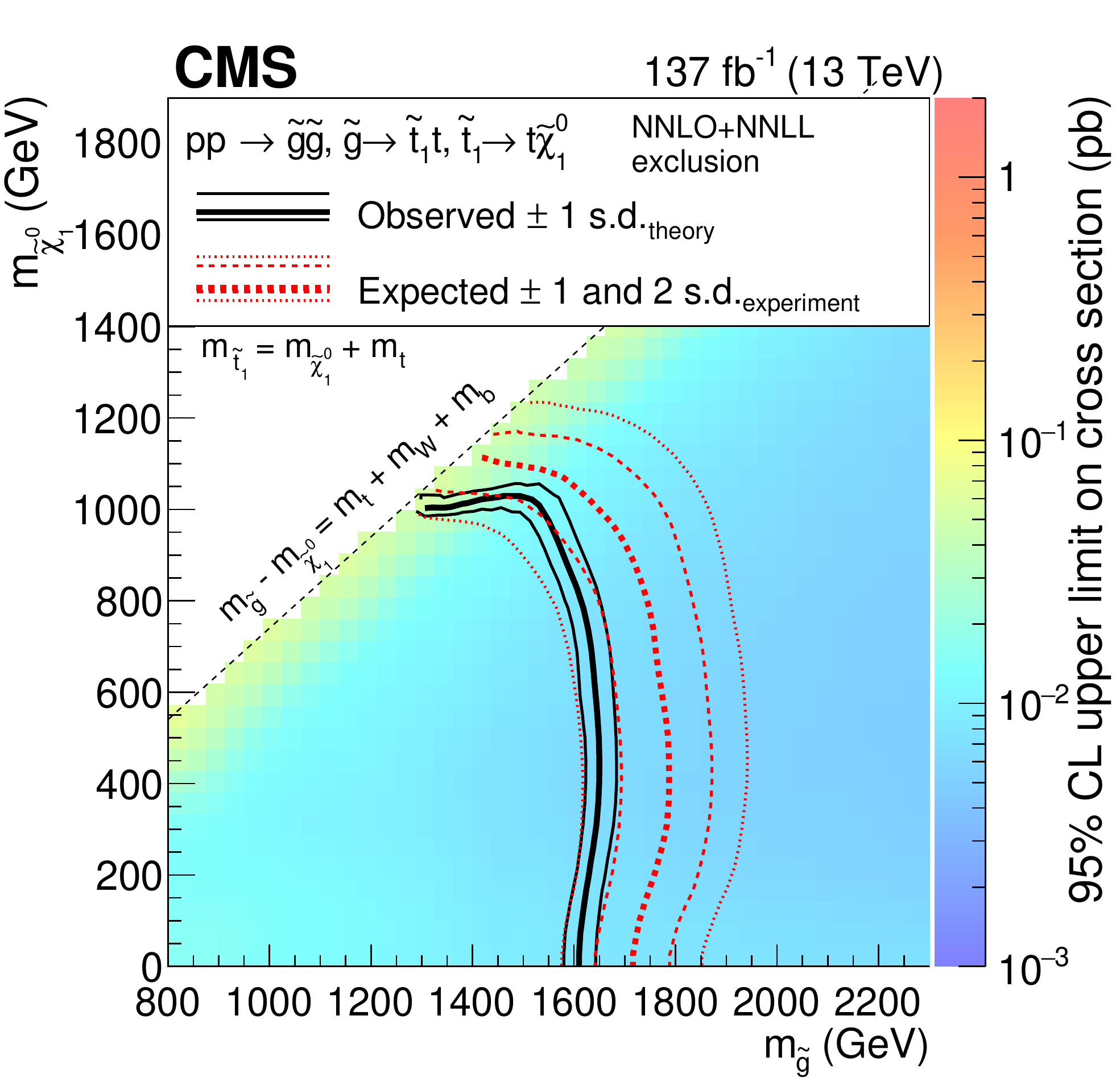}
\includegraphics[width=0.45\textwidth]{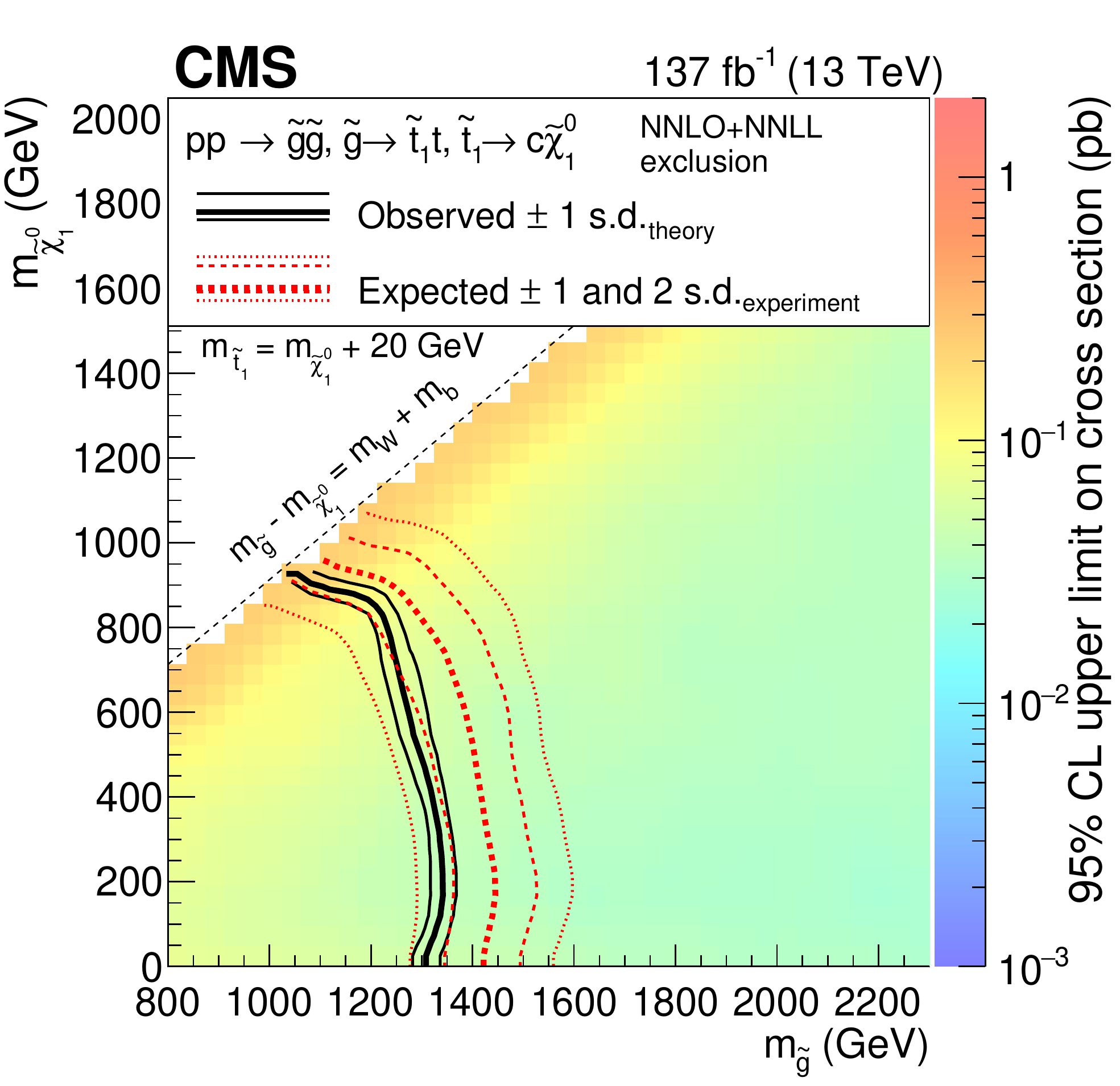}
\caption{ Exclusion regions at 95\% \CL in the $m_{\lsp}$ versus
  $m_{\gluino}$ plane for the \Totttt~(upper left) and \TfttbbWW~(upper right) models, with off-shell third-generation squarks, and the
    \Tftttt~(lower left) and \Tfttcc (lower right) models, with on-shell third-generation squarks.
For the \TfttbbWW model, $m_{\chiplmin} = m_{\lsp} + 5\GeV$, for the \Tftttt model, $m_{\susytop} - m_{\lsp} = m_{\PQt}$, and
for the \Tfttcc model, $m_{\susytop} - m_{\lsp} = 20\GeV$ and the decay proceeds through $\susytop \to \PQc \lsp$.
The right-hand side color scale indicates the excluded cross section values for a given point in the SUSY particle mass plane.
The solid black curves represent the observed exclusion limits
assuming the approximate-NNLO+NNLL cross
    sections~\protect\cite{bib-nnll,bib-nlo-nll-01,bib-nlo-nll-02,bib-nlo-nll-03,bib-nlo-nll-04,bib-nlo-nll-05,Borschensky:2014cia} (thick line), or their variations of $\pm 1$ standard deviations (s.d.) (thin lines).
The dashed red curves show the expected limits with the corresponding $\pm 1$ s.d. and $\pm 2$ s.d. uncertainties.
Excluded regions are to the left and below the limit curves.
}
\label{fig:t1ttxx_scan_xsec}
\end{figure*}

\begin{figure*}[!hbtp]
\centering
\includegraphics[width=0.45\textwidth]{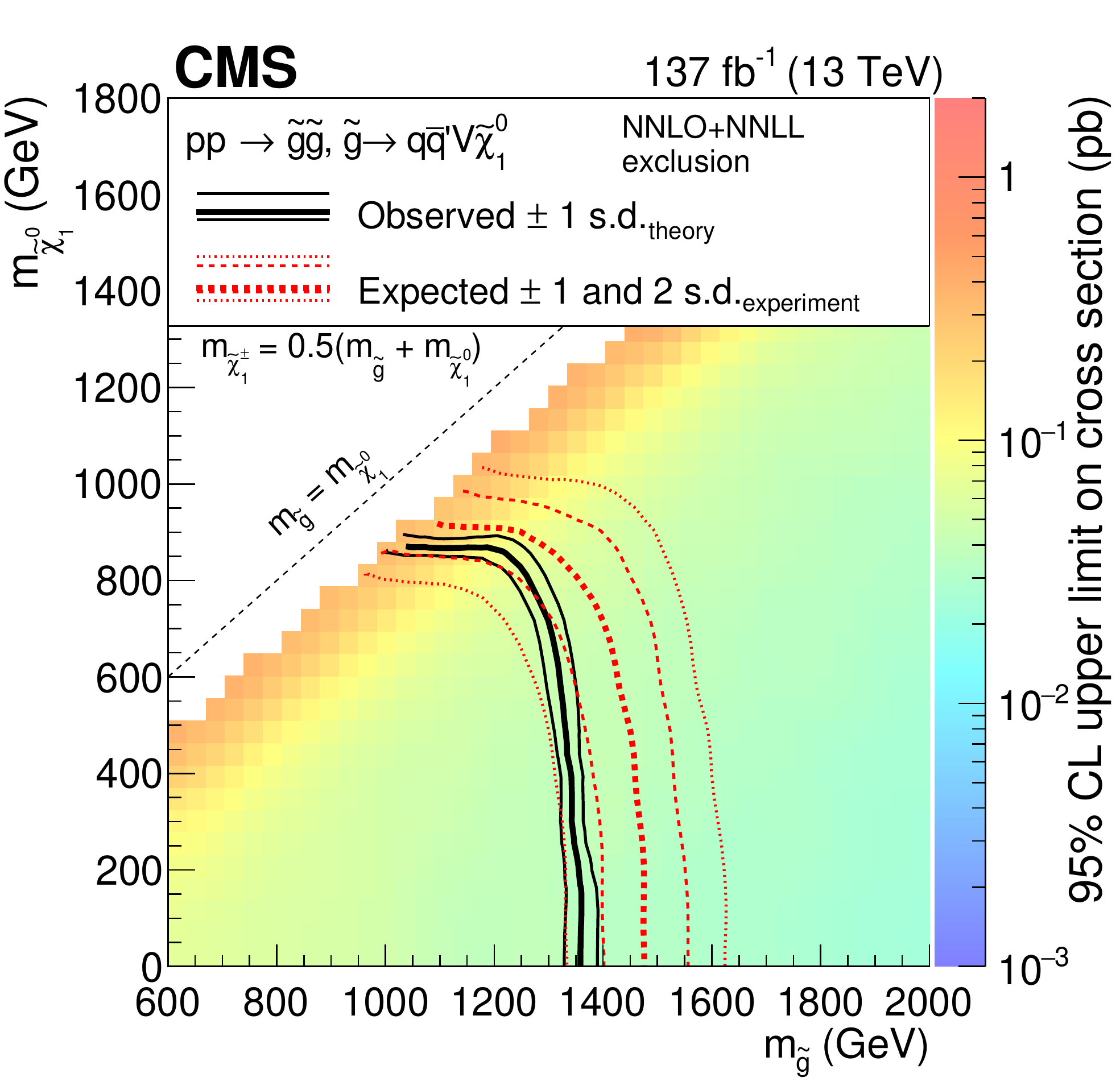}
\includegraphics[width=0.45\textwidth]{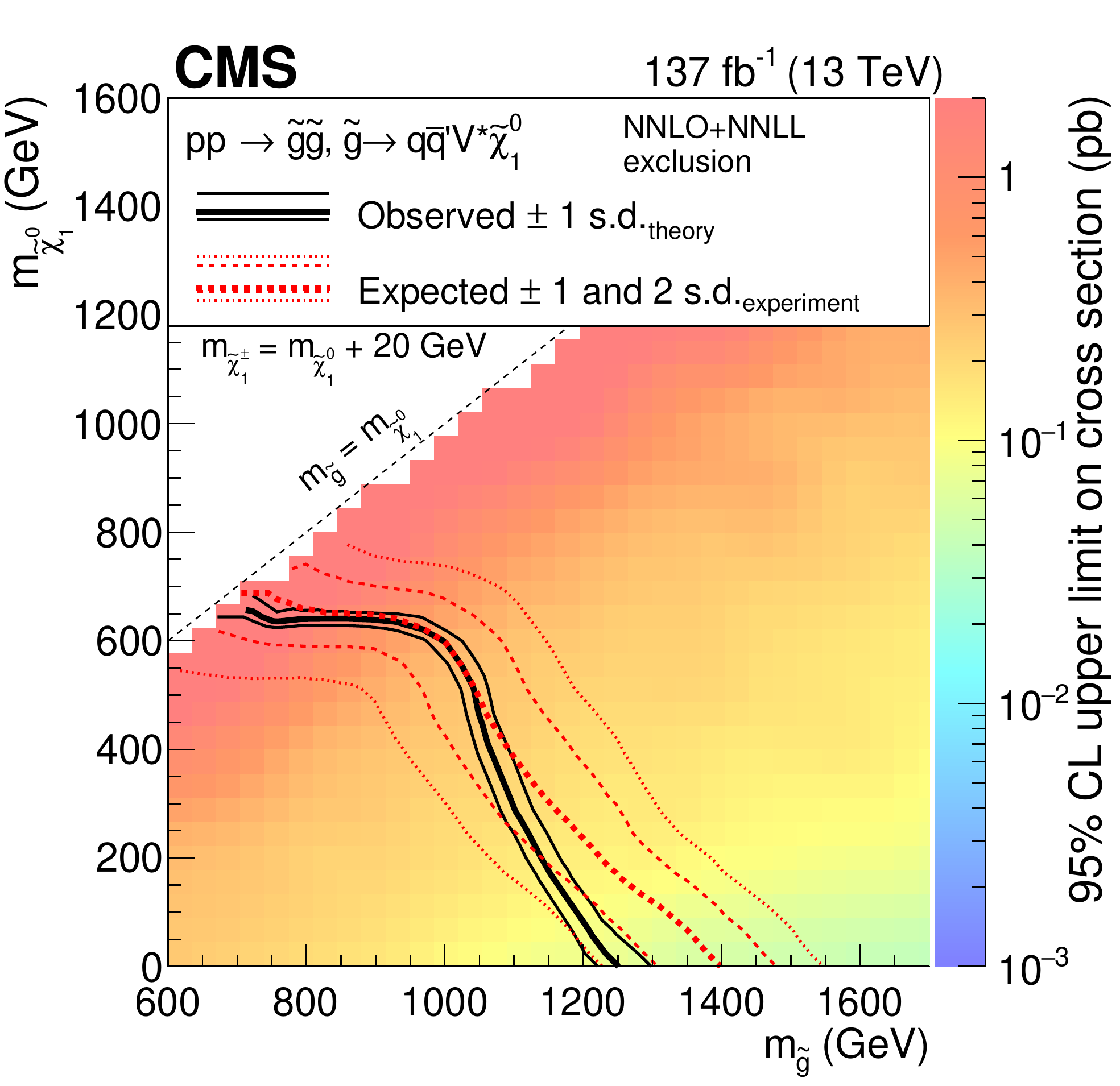}
\caption{Exclusion regions at 95\% \CL in the plane of $m_{\lsp}$ versus $m_{\gluino}$ for the \TfqqqqWZ model
with $m_{\chiplmin}=0.5(m_{\gluino} + m_{\lsp})$~(left) and with $m_{\chiplmin} = m_{\lsp} + 20\GeV$~(right).
The notations are as in Fig.~\protect\ref{fig:t1ttxx_scan_xsec}.  }
\label{fig:t5qqqqvv_scan_xsec}
\end{figure*}

\begin{figure*}[!hbtp]
\centering
\includegraphics[width=0.45\textwidth]{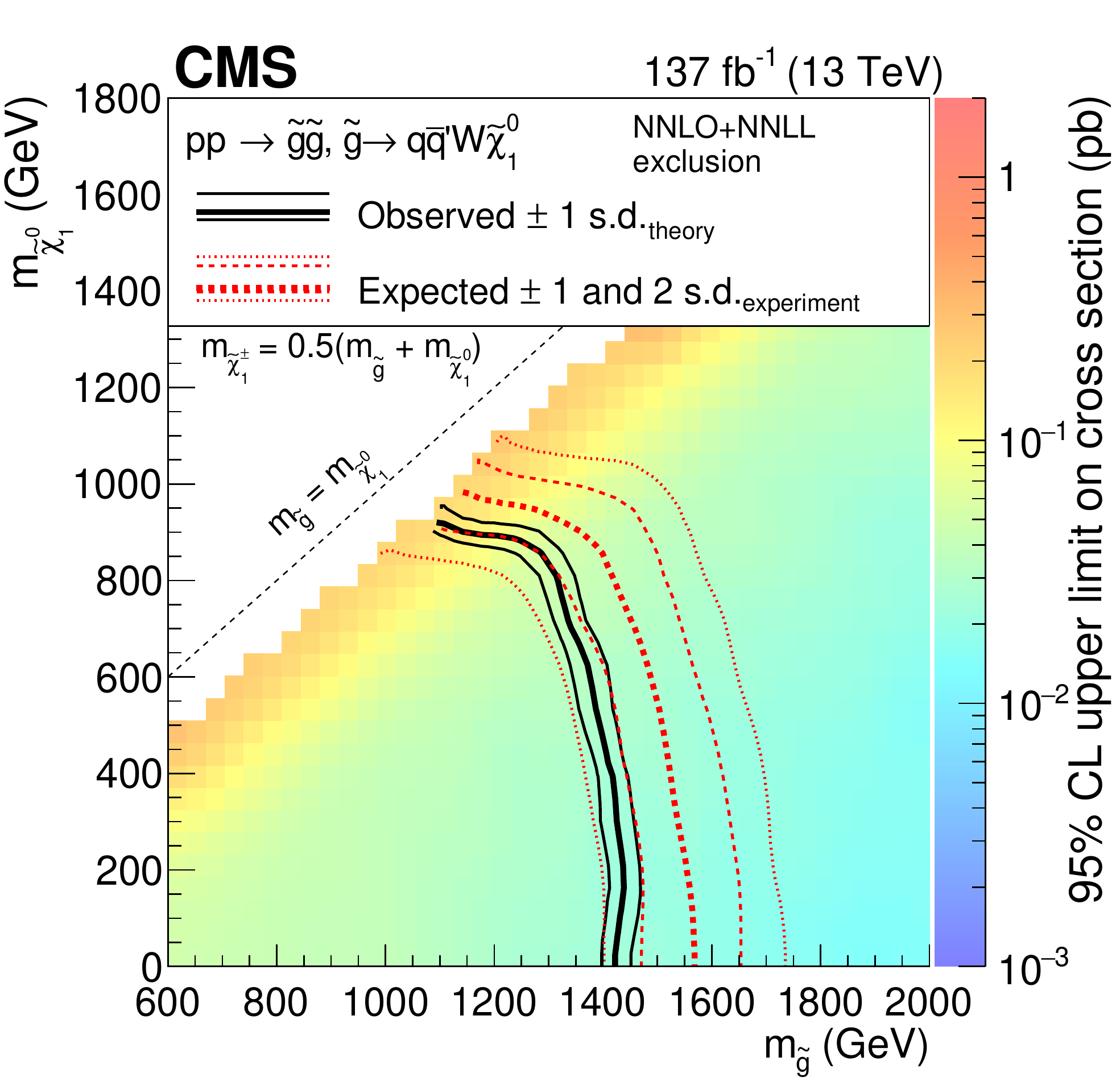}
\includegraphics[width=0.45\textwidth]{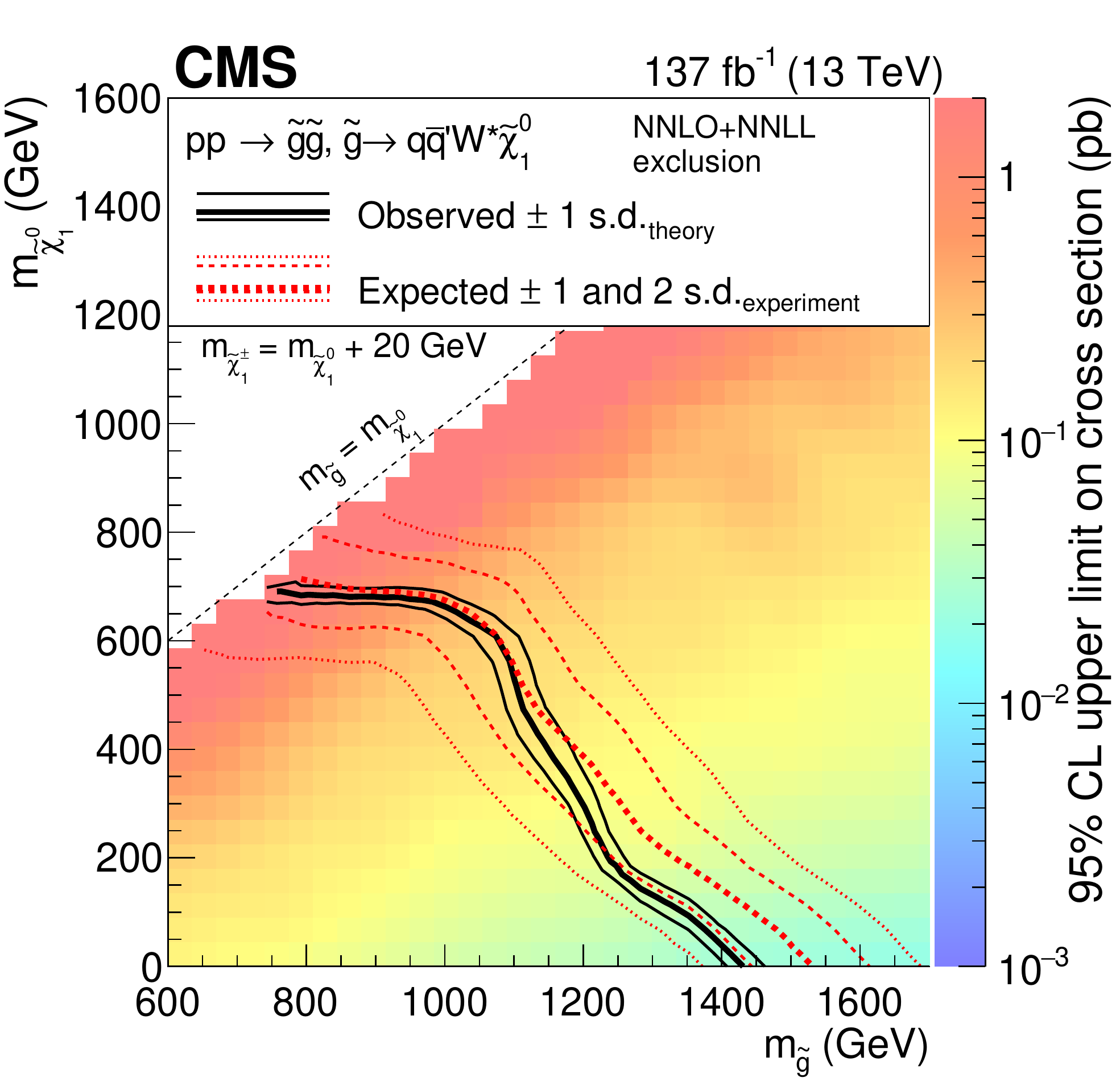}
\caption{Exclusion regions at 95\% \CL in the plane of $m_{\lsp}$ versus $m_{\gluino}$ for the \TfqqqqWW model
with $m_{\chiplmin}=0.5(m_{\gluino} + m_{\lsp})$~(left) and with $m_{\chiplmin} = m_{\lsp} + 20\GeV$~(right).
The notations are as in Fig.~\protect\ref{fig:t1ttxx_scan_xsec}.  }
\label{fig:t5qqqqww_scan_xsec}
\end{figure*}

\begin{figure*}[!hbtp]
\centering
\includegraphics[width=0.45\textwidth]{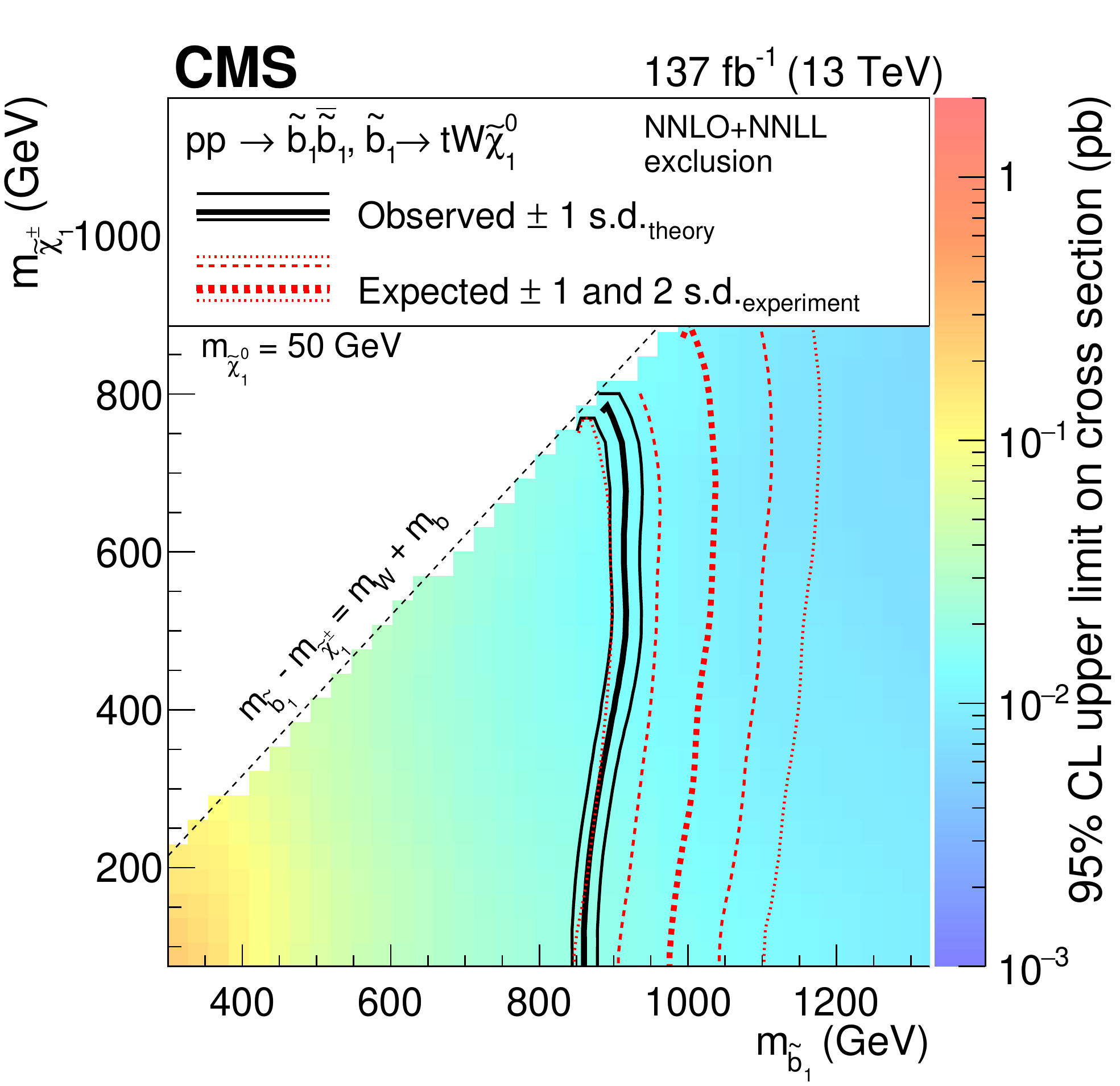}\\
\caption{ Exclusion regions at 95\% \CL in the plane of $m_{\chiplmin}$ versus $m_{\sbottomone}$ for the \TsttWW model with $m_{\lsp}=50\GeV$.
    The notations are as in Fig.~\ref{fig:t1ttxx_scan_xsec}. }
\label{fig:t6ttww_scan_xsec}
\end{figure*}

\begin{figure*}[!hbtp]
\centering
\includegraphics[width=.45\textwidth]{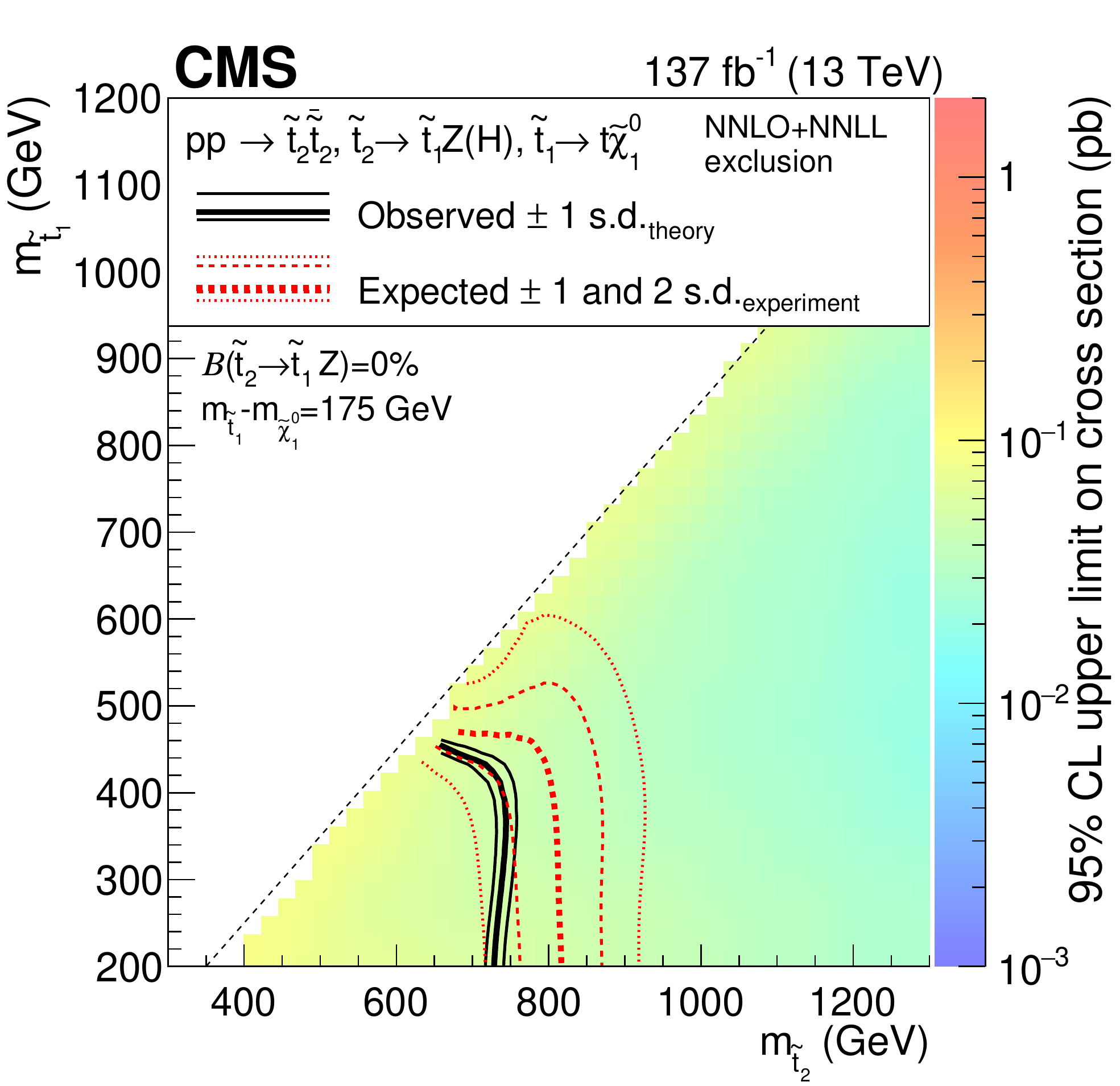}
\includegraphics[width=.45\textwidth]{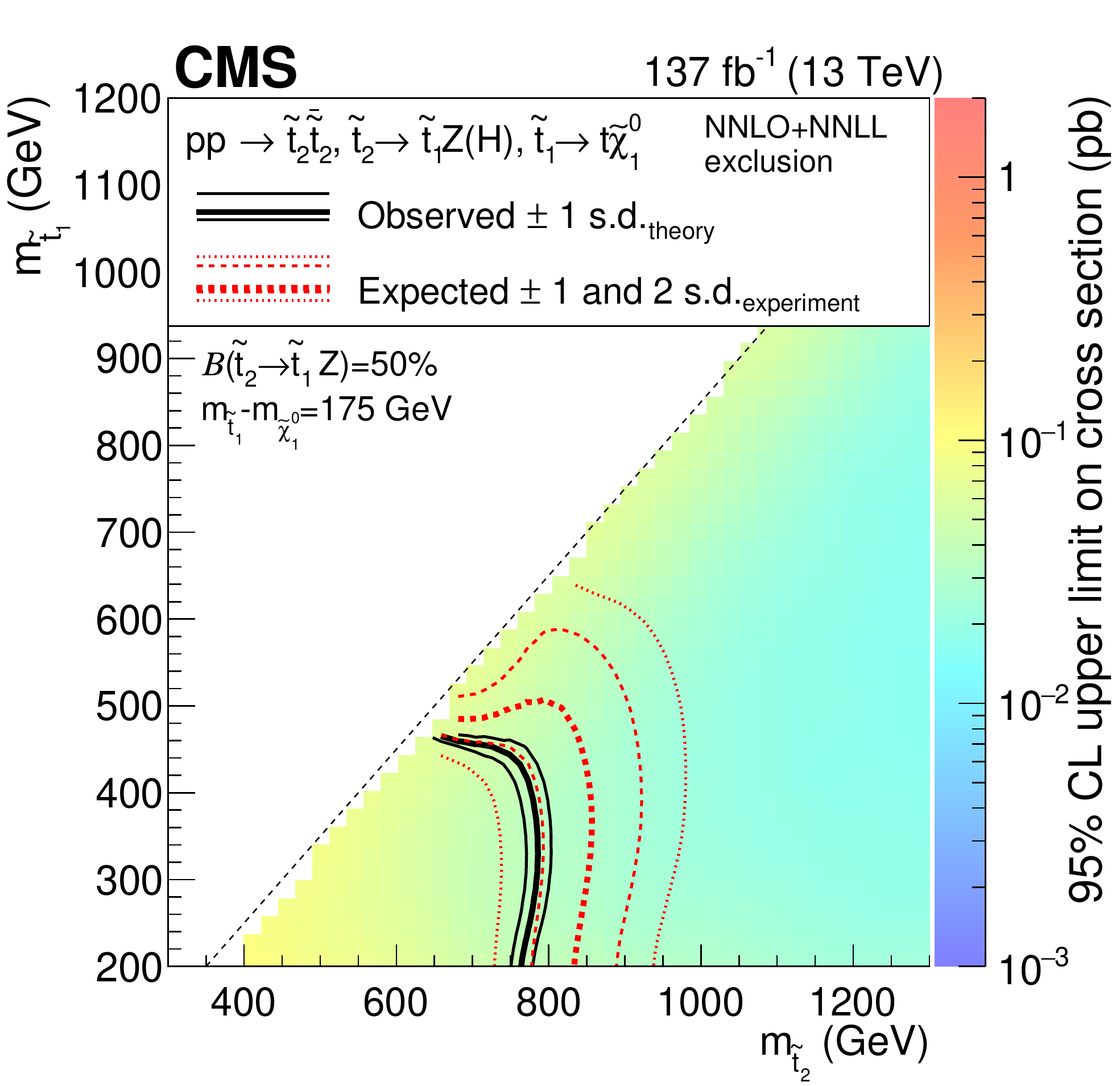} \\
\includegraphics[width=.45\textwidth]{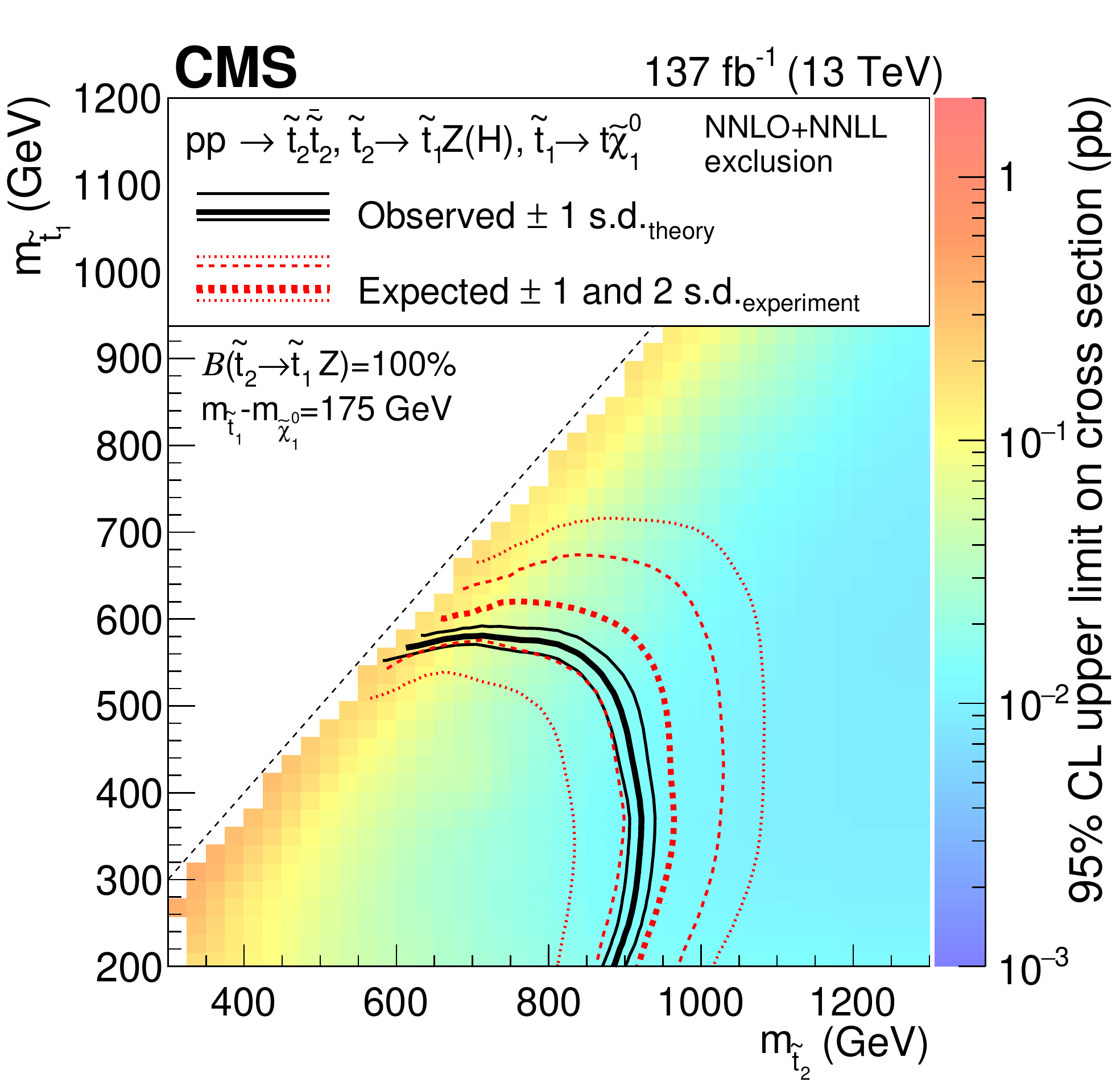}
\\
\caption{
Exclusion regions at 95\% \CL in the plane of $m(\susytopone)$ versus $m(\susytoptwo)$ for the \TsttHZ model
    with $m(\susytopone)-m(\lsp)=175\gev$. The three exclusions represent $\mathcal{B}(\susytoptwo\to\susytopone\PZ)$ of 0, 50, and 100\%, respectively.
The notations are as in Fig.~\ref{fig:t1ttxx_scan_xsec}.
}
\label{fig:t6tthz_scan_xsec}
\end{figure*}

{\tolerance=5000
Figure~\ref{fig:t1ttxx_scan_xsec} shows observed and expected exclusion limits
for simplified models of gluino pair production with each gluino decaying
to off- or on-shell third-generation squarks. These models were introduced in Section~\ref{sec:samples} and denoted as
\Totttt, \TfttbbWW, \Tftttt, and \Tfttcc.
Similarly, Figs.~\ref{fig:t5qqqqvv_scan_xsec} and~\ref{fig:t5qqqqww_scan_xsec} show the corresponding limits for \TfqqqqWZ and \TfqqqqWW, with
two different assumptions on the chargino mass.
Note that the \TfqqqqWZ model assumes equal probabilities for
the decay of the gluino into \chiplus, \chiminus, and \neutralinotwo.
The exclusion limits for \TsttWW and \TsttHZ are displayed in
Figs.~\ref{fig:t6ttww_scan_xsec} and~\ref{fig:t6tthz_scan_xsec},
respectively.
In the \TsttHZ model, the heavier top squark decays
into a lighter top squark and a \PZ\ or \PH\ boson. The three
sets of exclusion limits shown in Fig.~\ref{fig:t6tthz_scan_xsec}
correspond to the branching fraction $\mathcal{B}(\susytoptwo\to\susytopone\PZ)$ having values of
0, 50, and 100\%.
\par}

\begin{figure*}[!hbtp]
\centering
    \includegraphics[width=0.45\textwidth]{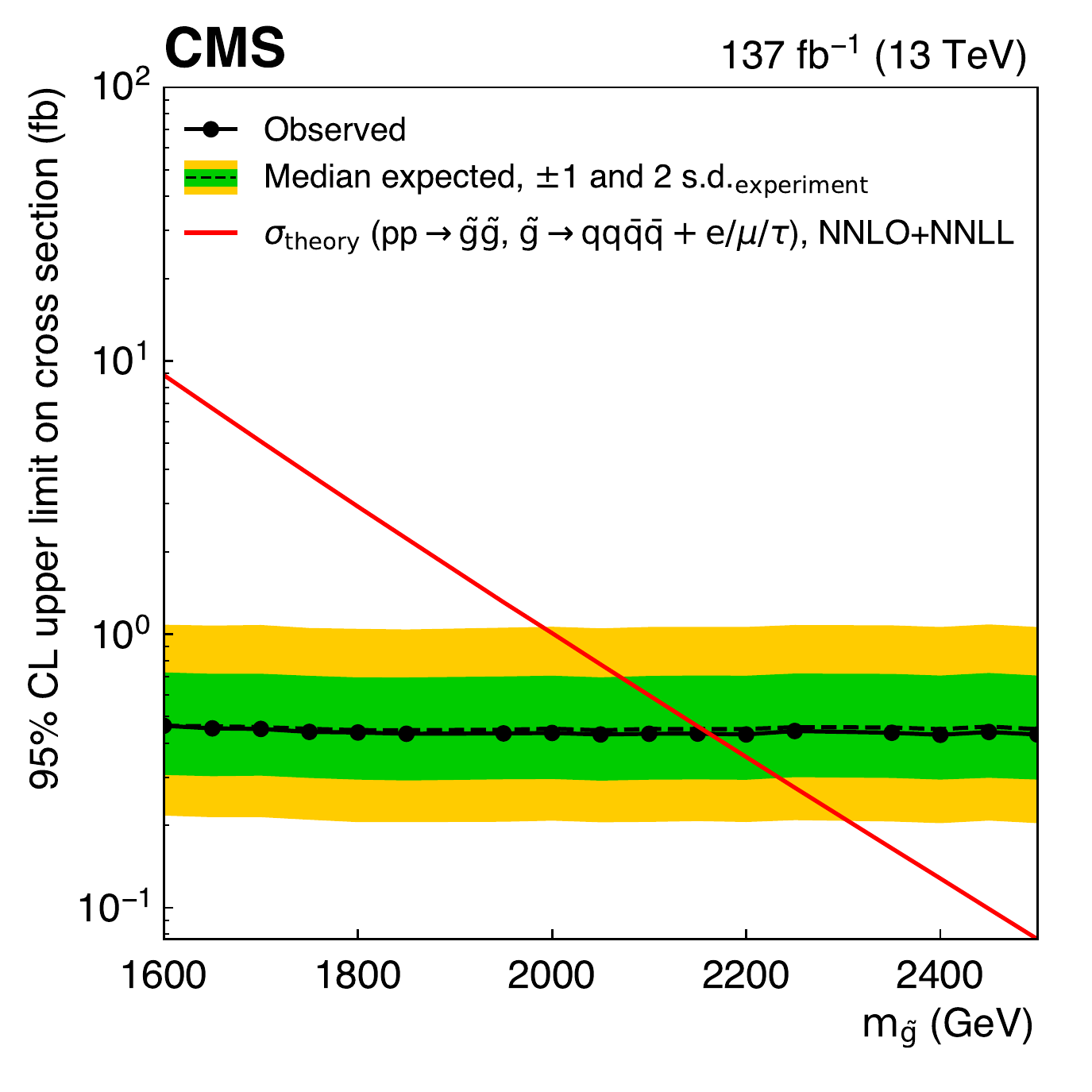}
\includegraphics[width=0.45\textwidth]{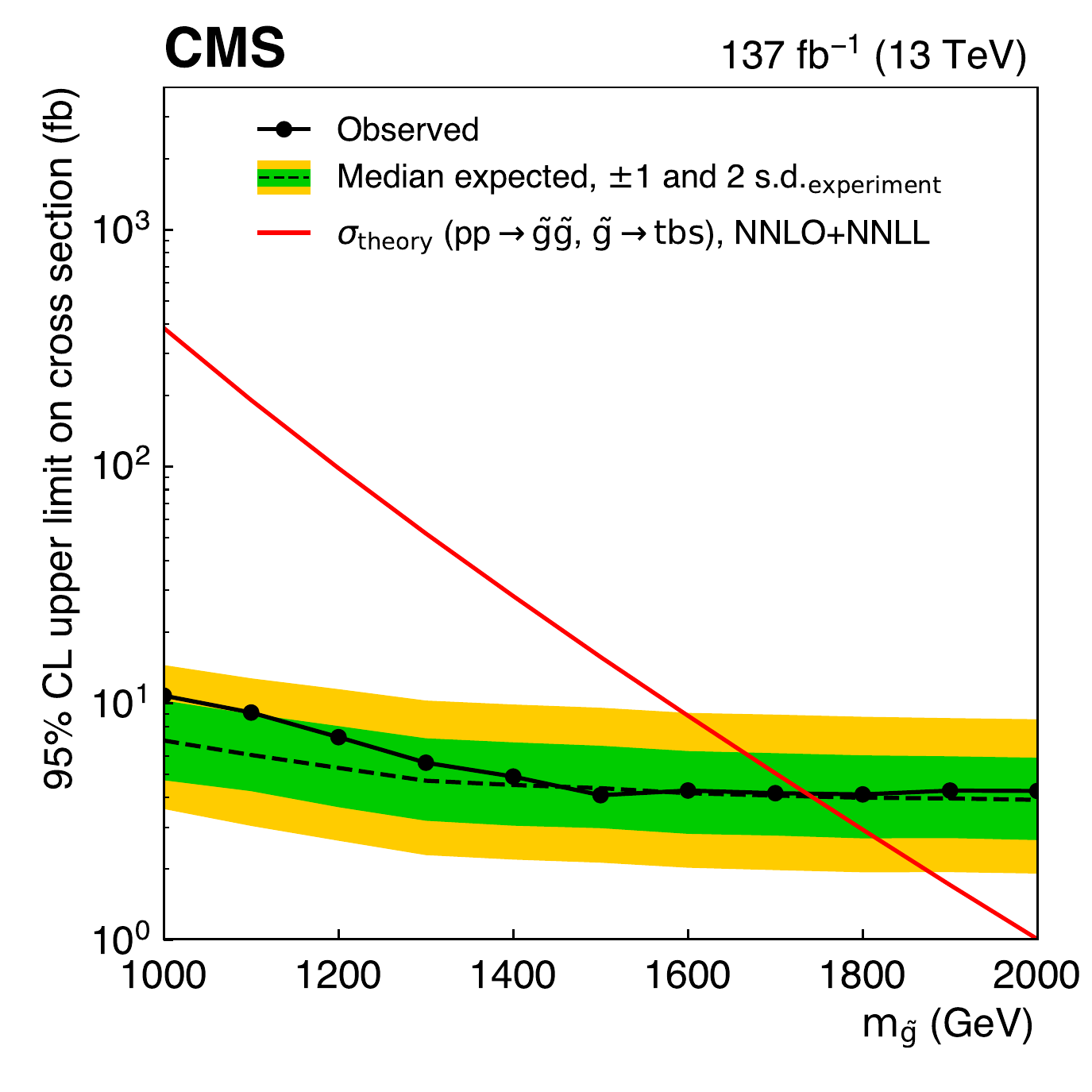}\\
\caption{
          Upper limits at 95\% \CL on the cross section for RPV gluino pair production with each gluino decaying into four quarks and one lepton (\ToqqqqL, left), and
    each gluino decaying into a top, bottom, and strange quarks (\Totbs, right).
    }
\label{fig:rpvlimits}
\end{figure*}

Finally, Fig.~\ref{fig:rpvlimits} shows observed and expected limits on
the cross section of gluino pair production as a function of the gluino masses in
the two RPV models described in Section~\ref{sec:samples}.
The observed and expected exclusions on the gluino mass are similar and reach 2.1 and 1.7 \TeV for the \ToqqqqL and \Totbs models, respectively.

The analysis sensitivity for the various models studied in Figs.~\ref{fig:t1ttxx_scan_xsec}--\ref{fig:t6tthz_scan_xsec} is often driven by the event yields in a few SRs (off-\Z ML21, HH53 and HH52), where a slight excess of data is observed.
This in particular applies to the uncompressed mass regime, resulting in an observed limit weaker than the expected one by one or two s.d.
In the compressed mass regime, however, other SRs can become dominant, for example when the hadronic activity becomes limited. 
This happens in the \TfqqqqWZ and \TfqqqqWW models where the gluino and the lightest neutralino present a limited mass splitting (the region close to the diagonal in the left plots of Figs.~\ref{fig:t5qqqqvv_scan_xsec} and ~\ref{fig:t5qqqqww_scan_xsec}). In those scenarios the on-\PZ ML4 and HH3 SRs provide the best sensitivity, respectively. 
Additionally, if the intermediate chargino is nearly degenerate in mass with the lightest neutralino, both leptons become soft and LL SRs such as LL2 become relevant. Such a situation is encountered in the phase space region close to the diagonal in the right plots of Figs.~\ref{fig:t5qqqqvv_scan_xsec} and ~\ref{fig:t5qqqqww_scan_xsec}.
On-\PZ SRs (especially on-\PZ ML23) become important for models where an on-shell \PZ boson is produced (bottom plot in Fig.~\ref{fig:t6tthz_scan_xsec}).
The limits on the RPV models presented in Fig.~\ref{fig:rpvlimits} are mostly driven by another set of SRs (HH62 and LM11, the latter becoming more relevant for lower masses).

Compared to the previous versions of the analysis~\cite{SUS-16-035,SUS-16-041}, the limits for the RPC models extend the gluino and squark mass observed and expected exclusions by up to 200\GeV because of the increase in the integrated luminosity and the corresponding re-optimization of SR definitions. These results also complement searches for gluino pair production conducted by CMS in final states with 0 or 1 lepton~\cite{CMS:2019tlp,Sirunyan:2019ctn,Sirunyan:2019xwh}. For the \Totttt scenario, the expected sensitivity of this analysis suffers from a lower branching fraction that makes it uncompetitive in the uncompressed mass regime. However, for a nearly degenerate mass spectrum, the SM background becomes of higher importance and the presence of an SS lepton pair significantly reduces it, leading to a similar sensitivity. The constraints on the two RPV models that were not previously included demonstrate the sensitivity of the analysis to RPV scenarios. 
The final state is particularly well suited to study the \ToqqqqL model since no leptonic branching fraction penalty applies, resulting in exclusion limits on the gluino mass beyond 2.1 TeV, comparable to other results in fully hadronic final states~\cite{Sirunyan:2019ctn,Sirunyan:2019xwh}. The limits obtained on the \Totbs model are stronger than those previously obtained in the one-lepton channel based on the analysis of the 2016 dataset~\cite{Sirunyan:2017dhe}. They are expected to remain competitive after an update with the full Run 2 dataset.

\begin{figure*}[!hbtp]
\centering
\includegraphics[width=0.45\textwidth]{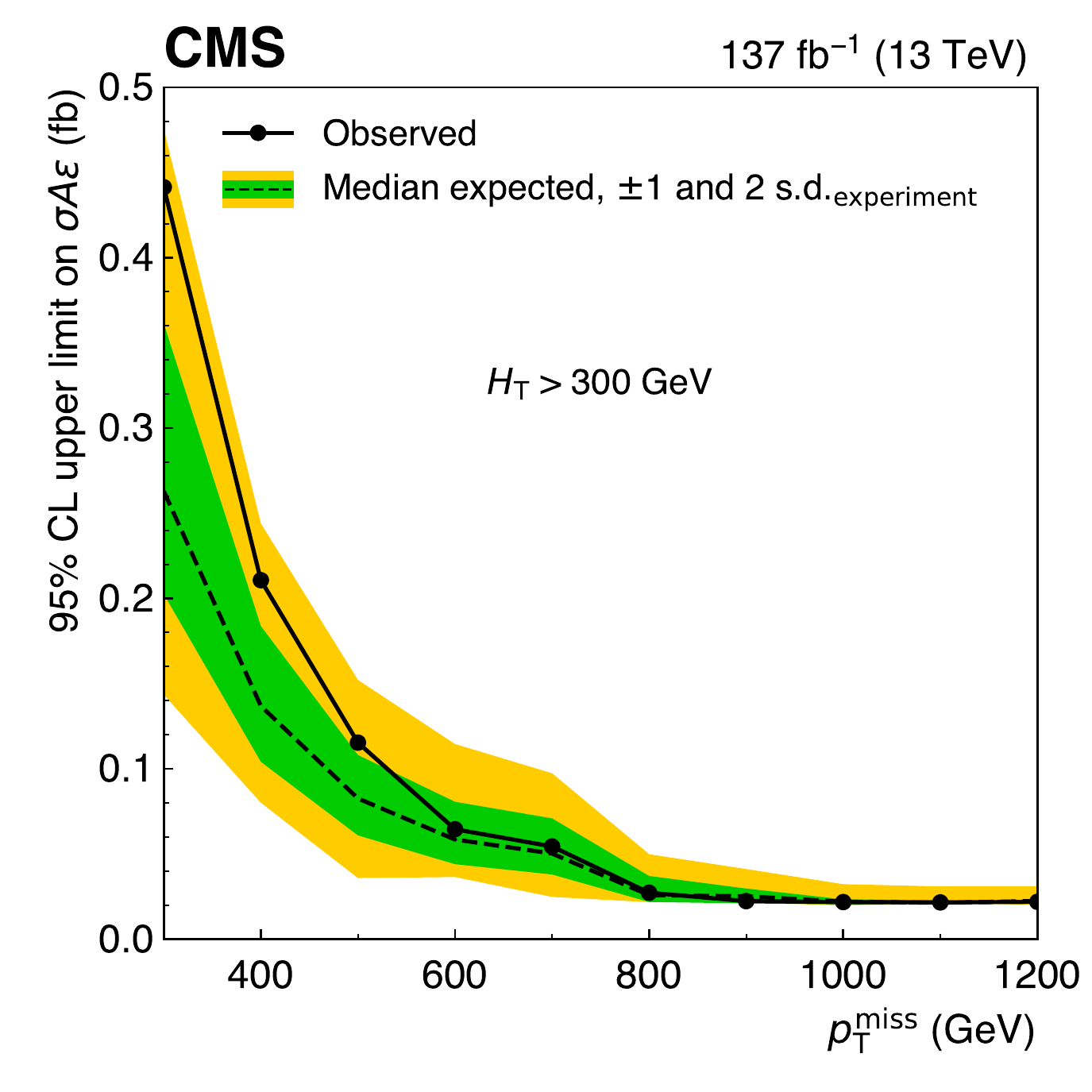}
\includegraphics[width=0.45\textwidth]{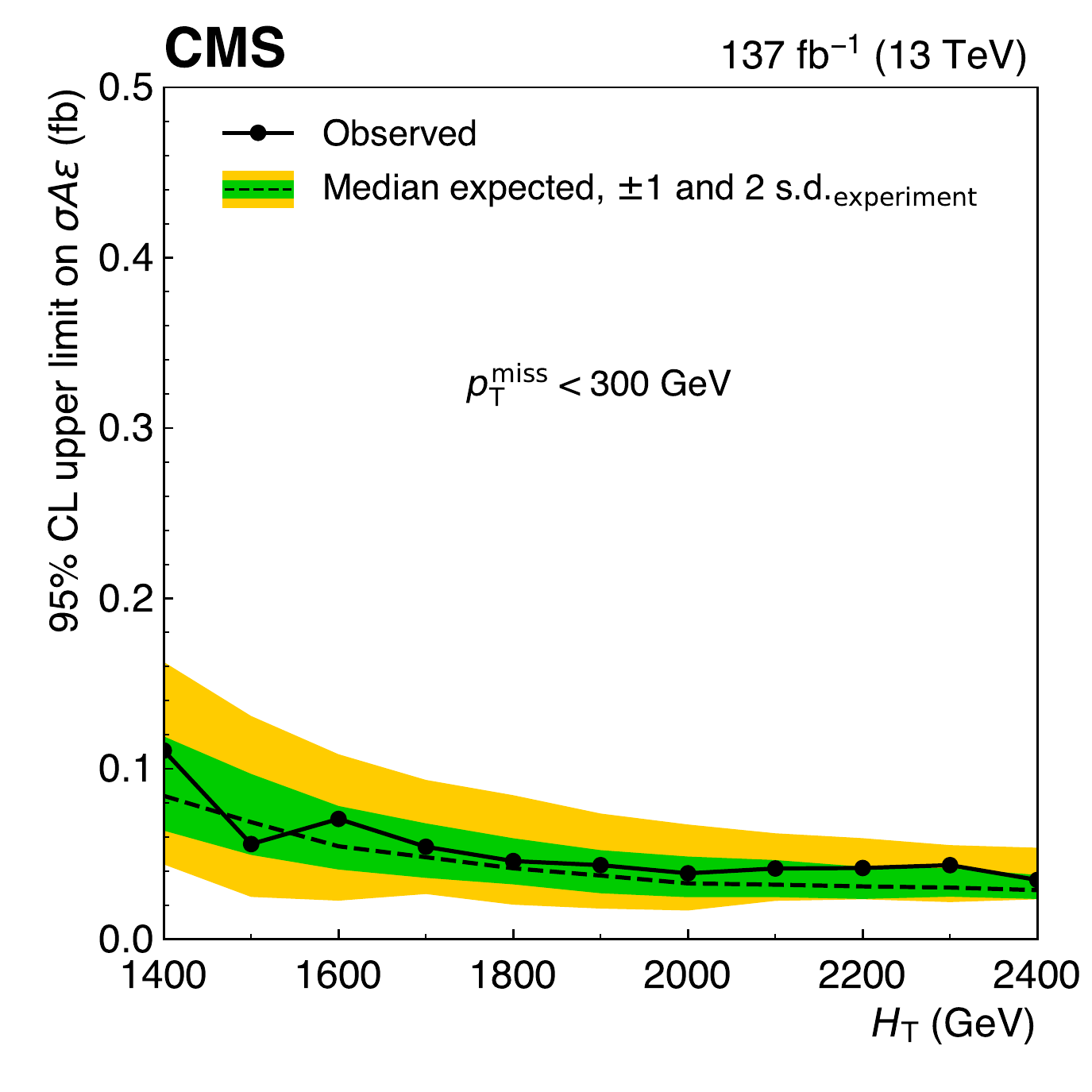}
\caption{Upper limits at 95\% \CL on the product of cross section, detector acceptance, and selection efficiency, $\sigma \! \mathcal{A} \epsilon$,
for the production of an SS lepton pair with at least two jets, as a function of the minimum \ptmiss threshold, when $\HT>300\GeV$ (left), or the minimum \HT threshold, when $\ptmiss<300\GeV$ (right).
    }
\label{fig:milimits}
\end{figure*}

Model-independent limits are also set on the product of cross section, branching fraction, detector acceptance, and reconstruction efficiency,
for the production of an SS lepton pair with at least two extra jets and $\HT>300\GeV$.
For this purpose, we select events from the HH and LM categories
and calculate limits as a function of minimum \ptmiss or \HT requirements
starting at 300 and 1400\GeV, respectively.
In order to remove the overlap between the two conditions, events selected for the \HT scan must also satisfy $\ptmiss<300\GeV$.
The corresponding limits are presented in Fig.~\ref{fig:milimits}.

Finally, in order to facilitate reinterpretations of our results, we present in Table~\ref{tab:inclusive_aggregate} the expected and observed yields for a  number of inclusive SRs.
This procedure focuses on events with large \HT, \ptmiss, \Nbjets, and/or
\Njets, and the SRs are defined such that they typically lead to 5 to 10 expected background events.
The last column in the table indicates the upper limit at 95\% \CL on the number of BSM events in each SR.

\begin{table*}[!hbtp]
\centering
\topcaption{Inclusive SR definitions, expected background yields and uncertainties, and observed yields, as well as the observed 95\% \CL upper limits on the number of BSM events contributing to each region.
No uncertainty in the signal acceptance is assumed in calculating these limits. A dash (\NA) indicates that a particular selection is not required.
}
\label{tab:inclusive_aggregate}
    \cmsTable{
  \begin{tabular}{cccccccccc}
  \hline
         SR     & Category              & \Njets   & \Nbjets  & \HT (\GeVns{})  & \ptmiss (\GeVns{}) & \MTmin (\GeVns{}) & SM expected       & Obs.  &  $N_{\mathstrut{\text{BSM}}}^{\mathstrut{\text{max}}}(95\%\ \CL)$ \\
         \hline
        ISR1  & \multirow{11}{*}{HH} & $\geq$2  & 0        & $\geq$1000  & $\geq$250  & \NA        & $12.7 \pm 7.4$ & 16 & 12.32  \\
        ISR2  &                      & $\geq$2  & $\geq$2  & $\geq$1100  & \NA        & \NA        & $11.0 \pm 3.8$ & 14 & 11.33  \\
        ISR3  &                      & $\geq$2  & 0        & \NA         & $\geq$500  & \NA        & $10.4 \pm 9.7$ & 13 & 11.26  \\
        ISR4  &                      & $\geq$2  & $\geq$2  & \NA         & $\geq$300  & \NA        & $11.4 \pm 3.8$ & 17 & 14.22  \\
        ISR5  &                      & $\geq$2  & 0        & \NA         & $\geq$250  & $\geq$120  & $6.6 \pm 5.7$  & 10 & 10.77  \\
        ISR6  &                      & $\geq$2  & $\geq$2  & \NA         & $\geq$200  & $\geq$120  & $6.3 \pm 1.3$  & 8  & 8.22   \\
        ISR7  &                      & $\geq$8  & \NA      & \NA         & \NA        & \NA        & $7.0 \pm 2.8$  & 12 & 12.17  \\
        ISR8  &                      & $\geq$6  & \NA      & \NA         & \NA        & $\geq$120  & $6.2 \pm 1.4$  & 10 & 10.45  \\
        ISR9  &                      & $\geq$2  & $\geq$3  & $\geq$800   & \NA        & \NA        & $7.8 \pm 3.5$  & 8  & 7.53   \\
[\cmsTabSkip]
        ISR10 & \multirow{4}{*}{LL}  & $\geq$2  & \NA      & $\geq$700   & \NA        & \NA        & $10.4 \pm 9.0$ & 12 & 10.37  \\
        ISR11 &                      & $\geq$2  & \NA      & \NA         & $\geq$200  & \NA        & $12.1 \pm 5.6$ & 13 & 9.94   \\
        ISR12 &                      & $\geq$6  & \NA      & \NA         & \NA        & \NA        & $7.1 \pm 4.3$   & 7  & 7.10   \\
        ISR13 &                      & $\geq$2  & $\geq$3  & \NA         & \NA        & \NA        & $1.61 \pm 0.39$ & 3  & 5.70   \\
[\cmsTabSkip]
        ISR14 & \multirow{2}{*}{LM}  & $\geq$2  & 0        & $\geq$1200  & $<$50      & \NA        & $3.6 \pm 3.6$   & 3  & 5.10   \\
        ISR15 &                      & $\geq$2  & $\geq$2  & $\geq$1000  & $<$50      & \NA        & $2.34 \pm 0.51$ & 4  & 6.41   \\
[\cmsTabSkip]
        ISR16 & \multirow{2}{*}{ML}  & $\geq$2  & 0        & $\geq$1000  & $\geq$300  & \NA        & $5.6 \pm 1.6$   & 7  & 7.78   \\
        ISR17 &                      & $\geq$2  & $\geq$2  & $\geq$1000  & \NA        & \NA        & $5.7 \pm 1.9$   & 7  & 7.62   \\ \hline
\end{tabular}
    }

\end{table*}

\clearpage

\section{Summary}
\label{sec:summary}

A sample of events with two same-sign or at least three charged leptons (electrons or muons) produced in association with several jets in proton-proton collisions at 13\TeV,
corresponding to an integrated luminosity of \sslumiruntwo, has been studied to search for
manifestations of physics beyond the standard model.
The data are found to be consistent with the standard model expectations.
The results are interpreted as limits on cross sections at 95\% confidence level for the
production of new particles in simplified supersymmetric models, considering both R parity conserving and violating scenarios.
Using calculations for these cross sections as functions of particle
masses, the limits are translated into lower mass limits
that are as large as 2.1\TeV for gluinos and 0.9\TeV for top and bottom squarks,
depending on the details of the model.
The results extend the gluino and squark mass observed and expected exclusions by up to 200\GeV,
compared to the previous versions of this analysis.
Finally, to facilitate further interpretations of the search, model-independent
limits are provided as a function of the missing transverse momentum and the scalar sum of jet transverse momenta in an event,
together with the background prediction and data yields in a set of
simplified signal regions.

\begin{acknowledgments}
We congratulate our colleagues in the CERN accelerator departments for the excellent performance of the LHC and thank the technical and administrative staffs at CERN and at other CMS institutes for their contributions to the success of the CMS effort. In addition, we gratefully acknowledge the computing centers and personnel of the Worldwide LHC Computing Grid for delivering so effectively the computing infrastructure essential to our analyses. Finally, we acknowledge the enduring support for the construction and operation of the LHC and the CMS detector provided by the following funding agencies: BMBWF and FWF (Austria); FNRS and FWO (Belgium); CNPq, CAPES, FAPERJ, FAPERGS, and FAPESP (Brazil); MES (Bulgaria); CERN; CAS, MoST, and NSFC (China); COLCIENCIAS (Colombia); MSES and CSF (Croatia); RPF (Cyprus); SENESCYT (Ecuador); MoER, ERC IUT, PUT and ERDF (Estonia); Academy of Finland, MEC, and HIP (Finland); CEA and CNRS/IN2P3 (France); BMBF, DFG, and HGF (Germany); GSRT (Greece); NKFIA (Hungary); DAE and DST (India); IPM (Iran); SFI (Ireland); INFN (Italy); MSIP and NRF (Republic of Korea); MES (Latvia); LAS (Lithuania); MOE and UM (Malaysia); BUAP, CINVESTAV, CONACYT, LNS, SEP, and UASLP-FAI (Mexico); MOS (Montenegro); MBIE (New Zealand); PAEC (Pakistan); MSHE and NSC (Poland); FCT (Portugal); JINR (Dubna); MON, RosAtom, RAS, RFBR, and NRC KI (Russia); MESTD (Serbia); SEIDI, CPAN, PCTI, and FEDER (Spain); MOSTR (Sri Lanka); Swiss Funding Agencies (Switzerland); MST (Taipei); ThEPCenter, IPST, STAR, and NSTDA (Thailand); TUBITAK and TAEK (Turkey); NASU (Ukraine); STFC (United Kingdom); DOE and NSF (USA).

\hyphenation{Rachada-pisek} Individuals have received support from the Marie-Curie program and the European Research Council and Horizon 2020 Grant, contract Nos.\ 675440, 752730, and 765710 (European Union); the Leventis Foundation; the A.P.\ Sloan Foundation; the Alexander von Humboldt Foundation; the Belgian Federal Science Policy Office; the Fonds pour la Formation \`a la Recherche dans l'Industrie et dans l'Agriculture (FRIA-Belgium); the Agentschap voor Innovatie door Wetenschap en Technologie (IWT-Belgium); the F.R.S.-FNRS and FWO (Belgium) under the ``Excellence of Science -- EOS" -- be.h project n.\ 30820817; the Beijing Municipal Science \& Technology Commission, No. Z191100007219010; the Ministry of Education, Youth and Sports (MEYS) of the Czech Republic; the Deutsche Forschungsgemeinschaft (DFG) under Germany’s Excellence Strategy -- EXC 2121 ``Quantum Universe" -- 390833306; the Lend\"ulet (``Momentum") Program and the J\'anos Bolyai Research Scholarship of the Hungarian Academy of Sciences, the New National Excellence Program \'UNKP, the NKFIA research grants 123842, 123959, 124845, 124850, 125105, 128713, 128786, and 129058 (Hungary); the Council of Science and Industrial Research, India; the HOMING PLUS program of the Foundation for Polish Science, cofinanced from European Union, Regional Development Fund, the Mobility Plus program of the Ministry of Science and Higher Education, the National Science Center (Poland), contracts Harmonia 2014/14/M/ST2/00428, Opus 2014/13/B/ST2/02543, 2014/15/B/ST2/03998, and 2015/19/B/ST2/02861, Sonata-bis 2012/07/E/ST2/01406; the National Priorities Research Program by Qatar National Research Fund; the Ministry of Science and Education, grant no. 14.W03.31.0026 (Russia); the Programa Estatal de Fomento de la Investigaci{\'o}n Cient{\'i}fica y T{\'e}cnica de Excelencia Mar\'{\i}a de Maeztu, grant MDM-2015-0509 and the Programa Severo Ochoa del Principado de Asturias; the Thalis and Aristeia programs cofinanced by EU-ESF and the Greek NSRF; the Rachadapisek Sompot Fund for Postdoctoral Fellowship, Chulalongkorn University and the Chulalongkorn Academic into Its 2nd Century Project Advancement Project (Thailand); the Kavli Foundation; the Nvidia Corporation; the SuperMicro Corporation; the Welch Foundation, contract C-1845; and the Weston Havens Foundation (USA).
\end{acknowledgments}

\bibliography{auto_generated}

\appendix

\section{Extended results}
Tables~\ref{tab:yieldsHHrun2}--\ref{tab:yieldsLMrun2}, corresponding to Figures~\ref{fig:SRrun2}--\ref{fig:SRrun2b}, show
background predictions per process within each signal region.

\begin{table*}[htb]
\centering
\topcaption{Event yields in HH regions. Yields shown as ``-'' have a contribution smaller than 0.01, or do not contribute to a particular region. }
\label{tab:yieldsHHrun2}
\cmsTable{
\begin{tabular}{c|ccccccccc|cc}\hline
&$\ttbar \PW$      &$\ttbar \PZ$      &$\ttbar \PH$      &\PW\PZ       &WW          &X+$\gamma$  &Rare       &Charge misid.&Nonprompt lep.&SM expected &Data         \\
\hline
SR1 & 53$\pm$15 & 14.1$\pm$4.1 & 13.7$\pm$3.5 & 349$\pm$97 & 129$\pm$37 & 300$\pm$120 & 105$\pm$43 & 136$\pm$16 & 460$\pm$260 & 1560$\pm$300 & 1673 \\ 
SR2 & 28.5$\pm$8.1 & 7.7$\pm$2.2 & 7.7$\pm$2.0 & 124$\pm$34 & 161$\pm$46 & 86$\pm$40 & 38$\pm$16 & 34.6$\pm$3.9 & 94$\pm$60 & 582$\pm$93 & 653 \\ 
SR3 & 8.1$\pm$2.3 & 1.46$\pm$0.41 & 1.64$\pm$0.42 & 19.8$\pm$5.6 & 13.0$\pm$3.8 & 7.3$\pm$7.3 & 8.8$\pm$3.7 & 14.0$\pm$1.6 & 26$\pm$21 & 100$\pm$25 & 128 \\ 
SR4 & 6.2$\pm$1.8 & 2.34$\pm$0.67 & 3.45$\pm$0.86 & 4.0$\pm$1.1 & 4.4$\pm$1.3 & 0.50$\pm$0.23 & 4.3$\pm$1.8 & 2.01$\pm$0.22 & 12.4$\pm$7.7 & 39.5$\pm$8.5 & 54 \\ 
SR5 & 4.0$\pm$1.1 & 0.48$\pm$0.14 & 0.49$\pm$0.12 & 15.9$\pm$4.4 & 21.1$\pm$6.1 & 5.5$\pm$4.7 & 4.8$\pm$2.0 & 0.50$\pm$0.06 & 4.9$\pm$2.4 & 57.7$\pm$9.9 & 53 \\ 
SR6 & 1.64$\pm$0.49 & 0.49$\pm$0.14 & 0.57$\pm$0.15 & 6.7$\pm$2.0 & 5.2$\pm$1.6 & 8.6$\pm$5.3 & 2.6$\pm$1.0 & 0.61$\pm$0.07 & 6.0$\pm$3.3 & 32.5$\pm$7.1 & 24 \\ 
SR7 & 0.91$\pm$0.26 & 0.26$\pm$0.07 & 0.38$\pm$0.09 & 0.62$\pm$0.19 & 0.84$\pm$0.23 & 1.2$\pm$1.2 & 0.43$\pm$0.18 & 0.08$\pm$0.01 & 0.81$\pm$0.51 & 5.5$\pm$1.8 & 7 \\
\hline
\end{tabular}
}
\end{table*}

\begin{table*}
\centering
\topcaption{Event yields in HL regions. Yields shown as ``-'' have a contribution smaller than 0.01, or do not contribute to a particular region. }
\label{tab:yieldsHLrun2}
\cmsTable{
\begin{tabular}{c|ccccccccc|cc}\hline
&$\ttbar \PW$      &$\ttbar \PZ$      &$\ttbar \PH$      &\PW\PZ       &WW          &X+$\gamma$  &Rare       &Charge misid.&Nonprompt lep.&SM expected &Data         \\
\hline
SR1 & 17.5$\pm$4.9 & 6.2$\pm$1.8 & 6.8$\pm$1.8 & 168$\pm$47 & 49$\pm$14 & 145$\pm$67 & 51$\pm$21 & 33.6$\pm$3.8 & 910$\pm$290 & 1390$\pm$300 & 1593 \\ 
SR2 & 7.2$\pm$2.1 & 2.63$\pm$0.76 & 3.21$\pm$0.82 & 38$\pm$10 & 46$\pm$13 & 41$\pm$25 & 11.3$\pm$4.8 & 6.54$\pm$0.73 & 192$\pm$62 & 348$\pm$67 & 337 \\ 
SR3 & 0.97$\pm$0.28 & 0.20$\pm$0.07 & 0.34$\pm$0.09 & 5.0$\pm$1.4 & 2.24$\pm$0.68 & 0.04$\pm$0.04 & 1.88$\pm$0.79 & 0.24$\pm$0.03 & 16.0$\pm$5.3 & 26.9$\pm$8.8 & 39 \\ 
SR4 & 1.85$\pm$0.54 & 0.81$\pm$0.23 & 1.67$\pm$0.42 & 1.19$\pm$0.34 & 1.45$\pm$0.42 & 0.30$\pm$0.13 & 1.25$\pm$0.50 & 0.48$\pm$0.06 & 26.9$\pm$9.1 & 35.9$\pm$9.1 & 34 \\ 
SR5 & 1.15$\pm$0.33 & 0.20$\pm$0.06 & 0.23$\pm$0.06 & 5.5$\pm$1.5 & 7.8$\pm$2.3 & 4.2$\pm$3.7 & 1.12$\pm$0.46 & 0.10$\pm$0.01 & 9.4$\pm$3.0 & 29.8$\pm$6.0 & 34 \\ 
SR6 & 0.36$\pm$0.10 & 0.21$\pm$0.07 & 0.22$\pm$0.06 & 2.73$\pm$0.76 & 1.43$\pm$0.43 & 8.4$\pm$6.7 & 0.49$\pm$0.22 & 0.19$\pm$0.02 & 8.2$\pm$2.5 & 22.2$\pm$7.2 & 12 \\ 
SR7 & 0.22$\pm$0.07 & 0.10$\pm$0.05 & 0.19$\pm$0.05 & 0.29$\pm$0.10 & 0.25$\pm$0.09 & 0.07$\pm$0.04 & 0.32$\pm$0.14 & 0.02$\pm$0.00 & 3.2$\pm$1.4 & 4.7$\pm$1.4 & 6 \\
\hline 
\end{tabular}
}
\end{table*}

\begin{table*}
\centering
\topcaption{Event yields in LL regions. Yields shown as ``-'' have a contribution smaller than 0.01, or do not contribute to a particular region. }
\label{tab:yieldsLLrun2}
\cmsTable{
\begin{tabular}{c|ccccccccc|cc}\hline
&$\ttbar \PW$      &$\ttbar \PZ$      &$\ttbar \PH$      &\PW\PZ       &WW          &X+$\gamma$  &Rare       &Charge misid.&Nonprompt lep.&SM expected &Data         \\
\hline
SR1 & 0.34$\pm$0.10 & 0.27$\pm$0.08 & 0.30$\pm$0.08 & 1.93$\pm$0.57 & 2.27$\pm$0.66 & 0.41$\pm$0.28 & 0.38$\pm$0.17 & 0.06$\pm$0.01 & 17.0$\pm$7.2 & 23.0$\pm$7.2 & 29 \\ 
SR2 & 0.10$\pm$0.04 & 0.05$\pm$0.03 & 0.07$\pm$0.02 & 0.93$\pm$0.31 & 0.86$\pm$0.25 & 0.03$\pm$0.01 & 0.37$\pm$0.16 & - & 2.6$\pm$1.5 & 5.0$\pm$1.6 & 6 \\ 
SR3 & 1.00$\pm$0.27 & 0.90$\pm$0.26 & 1.03$\pm$0.26 & 0.19$\pm$0.06 & 0.14$\pm$0.04 & 0.14$\pm$0.14 & 0.33$\pm$0.14 & 0.09$\pm$0.01 & 19.9$\pm$6.5 & 23.8$\pm$6.6 & 27 \\ 
SR4 & 0.47$\pm$0.14 & 0.11$\pm$0.03 & 0.28$\pm$0.07 & 0.05$\pm$0.02 & 0.08$\pm$0.02 & 0.09$\pm$0.04 & 0.19$\pm$0.08 & 0.01$\pm$0.00 & 3.4$\pm$1.5 & 4.7$\pm$1.5 & 7 \\ 
SR5 & 0.92$\pm$0.26 & 0.68$\pm$0.20 & 0.96$\pm$0.25 & 0.01$\pm$0.00 & 0.02$\pm$0.02 & 0.16$\pm$0.14 & 0.28$\pm$0.12 & 0.09$\pm$0.01 & 4.8$\pm$1.8 & 8.0$\pm$1.9 & 15 \\ 
SR6 & 0.33$\pm$0.13 & 0.14$\pm$0.07 & 0.21$\pm$0.06 & - & 0.04$\pm$0.04 & 0.03$\pm$0.03 & 0.05$\pm$0.02 & 0.02$\pm$0.01 & 1.2$\pm$1.1 & 2.0$\pm$1.1 & 0 \\ 
SR7 & 0.17$\pm$0.06 & 0.09$\pm$0.03 & 0.19$\pm$0.06 & - & - & 0.12$\pm$0.05 & 0.21$\pm$0.09 & 0.02$\pm$0.00 & 0.81$\pm$0.56 & 1.61$\pm$0.59 & 3 \\ 
SR8 & 0.02$\pm$0.02 & 0.01$\pm$0.01 & 0.01$\pm$0.01 & - & - & - & - & - & - & 0.06$\pm$0.06 & 0 \\ \hline 
Total & 3.36$\pm$0.98 & 2.25$\pm$0.60 & 3.04$\pm$0.80 & 3.09$\pm$0.82 & 3.42$\pm$0.96 & 0.79$\pm$0.42 & 1.84$\pm$0.97 & 0.30$\pm$0.03 & 50$\pm$13 & 68$\pm$13 & 87 \\ 
\hline 
\end{tabular}
}
\end{table*}

\begin{table*}
\centering
\topcaption{Event yields in on-\PZ ML regions. Yields shown as ``-'' have a contribution smaller than 0.01, or do not contribute to a particular region. }
\label{tab:yieldsMLONZrun2}
\cmsTable{
\begin{tabular}{c|ccccccc|cc}\hline
&$\ttbar \PW$      &$\ttbar \PZ$      &$\ttbar \PH$      &\PW\PZ       &X+$\gamma$  &Rare       &Nonprompt lep.&SM expected &Data         \\
\hline
SR1 & 1.12$\pm$0.30 & 46$\pm$13 & 1.67$\pm$0.43 & 620$\pm$170 & 11.9$\pm$5.7 & 105$\pm$42 & 54$\pm$18 & 840$\pm$170 & 985 \\ 
SR2 & 0.71$\pm$0.21 & 6.8$\pm$1.9 & 0.46$\pm$0.12 & 68$\pm$19 & 3.1$\pm$2.6 & 17.8$\pm$7.3 & 9.9$\pm$4.6 & 107$\pm$21 & 136 \\ 
SR3 & 0.18$\pm$0.05 & 8.5$\pm$2.4 & 0.23$\pm$0.06 & 95$\pm$26 & 0.02$\pm$0.01 & 11.7$\pm$4.8 & 3.2$\pm$1.8 & 119$\pm$27 & 146 \\ 
SR4 & 0.17$\pm$0.06 & 1.99$\pm$0.54 & 0.17$\pm$0.04 & 5.8$\pm$1.6 & 0.03$\pm$0.01 & 2.19$\pm$0.87 & 0.68$\pm$0.46 & 11.1$\pm$2.1 & 10 \\ 
SR5 & 0.13$\pm$0.04 & 7.7$\pm$2.2 & 0.25$\pm$0.06 & 83$\pm$23 & 2.3$\pm$1.2 & 11.0$\pm$4.5 & 4.5$\pm$2.1 & 109$\pm$24 & 126 \\ 
SR6 & 0.09$\pm$0.03 & 1.42$\pm$0.39 & 0.06$\pm$0.02 & 13.6$\pm$3.8 & 0.74$\pm$0.61 & 2.5$\pm$1.0 & 0.93$\pm$0.69 & 19.3$\pm$4.1 & 24 \\ 
SR7 & 0.10$\pm$0.03 & 2.55$\pm$0.73 & 0.09$\pm$0.02 & 35$\pm$10 & 0.09$\pm$0.04 & 3.7$\pm$1.6 & 0.89$\pm$0.53 & 42$\pm$10 & 47 \\
\hline 
\end{tabular}
}
\end{table*}

\begin{table*}
\centering
\topcaption{Event yields in off-\PZ ML regions. Yields shown as ``-'' have a contribution smaller than 0.01, or do not contribute to a particular region. }
\label{tab:yieldsMLOFFZrun2}
\cmsTable{
\begin{tabular}{c|ccccccc|cc}\hline
&$\ttbar \PW$      &$\ttbar \PZ$      &$\ttbar \PH$      &\PW\PZ       &X+$\gamma$  &Rare       &Nonprompt lep.&SM expected &Data         \\
\hline
SR1 & 8.8$\pm$2.5 & 11.2$\pm$3.2 & 7.0$\pm$1.8 & 87$\pm$24 & 18.8$\pm$9.5 & 22.0$\pm$9.0 & 68$\pm$25 & 222$\pm$36 & 285 \\ 
SR2 & 0.15$\pm$0.07 & 0.15$\pm$0.08 & 0.10$\pm$0.03 & 1.23$\pm$0.50 & 0.02$\pm$0.02 & 0.43$\pm$0.22 & 0.61$\pm$0.61 & 2.7$\pm$1.7 & 2 \\ 
SR3 & 2.38$\pm$0.69 & 2.07$\pm$0.59 & 1.60$\pm$0.42 & 13.6$\pm$3.8 & 0.26$\pm$0.11 & 4.2$\pm$1.7 & 11.3$\pm$4.9 & 35.5$\pm$6.4 & 34 \\ 
SR4 & 0.18$\pm$0.06 & 0.14$\pm$0.04 & 0.08$\pm$0.02 & 0.24$\pm$0.11 & - & 0.22$\pm$0.12 & 0.13$\pm$0.13 & 0.99$\pm$0.31 & 2 \\ 
SR5 & 1.11$\pm$0.31 & 1.64$\pm$0.47 & 1.02$\pm$0.26 & 11.2$\pm$3.1 & 0.89$\pm$0.84 & 1.68$\pm$0.68 & 4.6$\pm$2.0 & 22.1$\pm$4.0 & 29 \\ 
SR6 & 0.91$\pm$0.25 & 0.64$\pm$0.19 & 0.59$\pm$0.15 & 4.7$\pm$1.4 & 0.05$\pm$0.02 & 1.05$\pm$0.44 & 1.76$\pm$0.80 & 9.7$\pm$1.7 & 8 \\ 
SR7 & 26.8$\pm$7.5 & 29.7$\pm$8.5 & 21.9$\pm$5.7 & 7.0$\pm$2.0 & 6.9$\pm$3.3 & 11.3$\pm$4.7 & 113$\pm$41 & 217$\pm$44 & 272 \\
\hline 
\end{tabular}
}
\end{table*}

\begin{table*}
\centering
\topcaption{Event yields in LM regions. Yields shown as ``-'' have a contribution smaller than 0.01, or do not contribute to a particular region. }
\label{tab:yieldsLMrun2}
\cmsTable{
\begin{tabular}{c|ccccccccc|cc}\hline
&$\ttbar \PW$      &$\ttbar \PZ$      &$\ttbar \PH$      &\PW\PZ       &WW          &X+$\gamma$  &Rare       &Charge misid.&Nonprompt lep.&SM expected &Data         \\
\hline
SR1 & 7.7$\pm$2.1 & 2.86$\pm$0.84 & 2.22$\pm$0.57 & 56$\pm$15 & 53$\pm$15 & 38$\pm$19 & 12.5$\pm$5.3 & 13.4$\pm$1.5 & 51$\pm$36 & 235$\pm$47 & 309 \\ 
SR2 & 1.69$\pm$0.50 & 1.06$\pm$0.30 & 1.12$\pm$0.28 & 1.99$\pm$0.56 & 1.80$\pm$0.53 & 3.5$\pm$2.3 & 1.14$\pm$0.53 & 0.70$\pm$0.08 & 6.3$\pm$4.3 & 19.3$\pm$5.2 & 26 \\ 
SR3 & 23.0$\pm$6.4 & 8.6$\pm$2.5 & 7.1$\pm$1.9 & 5.9$\pm$1.7 & 5.5$\pm$1.6 & 14.8$\pm$7.7 & 6.1$\pm$2.6 & 6.97$\pm$0.79 & 64$\pm$37 & 142$\pm$39 & 156 \\ 
SR4 & 6.3$\pm$1.8 & 3.15$\pm$0.89 & 4.4$\pm$1.1 & 0.34$\pm$0.09 & 0.25$\pm$0.10 & 1.22$\pm$0.57 & 1.96$\pm$0.81 & 0.77$\pm$0.09 & 13.8$\pm$8.2 & 32.2$\pm$8.8 & 38 \\ 
SR5 & 17.6$\pm$5.0 & 6.1$\pm$1.7 & 5.6$\pm$1.5 & 0.48$\pm$0.14 & 0.37$\pm$0.12 & 4.2$\pm$2.1 & 2.6$\pm$1.1 & 4.71$\pm$0.54 & 11.4$\pm$5.8 & 53.0$\pm$9.1 & 69 \\ 
SR6 & 6.0$\pm$1.7 & 3.17$\pm$0.91 & 4.2$\pm$1.1 & 0.10$\pm$0.04 & 0.12$\pm$0.04 & 1.05$\pm$0.45 & 2.00$\pm$0.83 & 0.68$\pm$0.08 & 4.8$\pm$3.0 & 22.0$\pm$4.0 & 30 \\ 
SR7 & 2.37$\pm$0.71 & 1.19$\pm$0.36 & 1.42$\pm$0.40 & 0.08$\pm$0.03 & - & 0.61$\pm$0.36 & 2.08$\pm$0.86 & 0.53$\pm$0.06 & 1.8$\pm$1.4 & 10.1$\pm$2.0 & 21 \\
\hline 
\end{tabular}
}
\end{table*}

\clearpage
\section{Top five SRs for several representative models}
Table~\ref{tab:toprankedSRs} presents the top five SRs for several representative models, ranked based on the largest values of $N_\text{sig.}/\sqrt{\smash[b]{N_\text{bkg.} + N_\text{sig.}}}$,
    where $N_\text{sig.}$ and $N_\text{bkg.}$ are the signal and total background yields in each SR, respectively.
\begin{table*}
\centering
    \topcaption{
        Top five SRs for several representative models, ranked based on the largest values of $N_\text{sig.}/\sqrt{\smash[b]{N_\text{bkg.} + N_\text{sig.}}}$,
    where $N_\text{sig.}$ and $N_\text{bkg.}$ are the signal and total background yields in each SR, respectively. 
    }
\label{tab:toprankedSRs}
\cmsTable{
\begin{tabular}{ccc}\hline
model      & mass point   & top SRs \\
\hline
\Totttt                                               & $m_{\gluino}=1400, m_{\lsp}=400$           & off-Z ML21, HH53, HH52, HH51, HH50 \\
\Totttt                                               & $m_{\gluino}=2000, m_{\lsp}=100$           & HH53, HH52, off-Z ML21, HL39, HH49 \\
\Totttt                                               & $m_{\gluino}=1800, m_{\lsp}=100$           & HH53, off-Z ML21, HH52, HL39, HH51 \\
\Totttt                                               & $m_{\gluino}=1800, m_{\lsp}=1000$          & off-Z ML21, HH53, HH52, HH51, HH50 \\
\Totttt                                               & $m_{\gluino}=1800, m_{\lsp}=1550$          & HH53, HL39, off-Z ML21, HH49, HH52 \\
\TsttWW                                               & $m_{\sbottomone}=1000, m_{\chiplmin}=600$  & off-Z ML21, HH53, HH51, HH50, HH52 \\
\TsttWW                                               & $m_{\sbottomone}=900, m_{\chiplmin}=400$   & off-Z ML21, HH51, HH50, HH53, off-Z ML20 \\
\TsttWW                                               & $m_{\sbottomone}=800, m_{\chiplmin}=400$   & off-Z ML21, HH51, HH50, HH34, off-Z ML20 \\
\TfqqqqWZ                                             & $m_{\gluino}=1400, m_{\lsp}=1$             & on-Z ML23, HH53, HH52, HH51, HH49 \\
\TfqqqqWZ                                             & $m_{\gluino}=900, m_{\lsp}=600$            & on-Z ML4, HH3, HH10, on-Z ML23, HH4 \\
\TfqqqqWW                                             & $m_{\gluino}=1400, m_{\lsp}=1$             & HH53, HH52, HH49, HH51, HH50 \\
\TfqqqqWW                                             & $m_{\gluino}=900, m_{\lsp}=600$            & HH3, HH10, HH4, HH7, HH50 \\
\TfqqqqWZ ($m_{\chiplmin} = m_{\lsp} + 20\GeV$)       & $m_{\gluino}=1400, m_{\lsp}=1$             & HH59, HH53, HH52, HH62, HH51 \\
\TfqqqqWZ ($m_{\chiplmin} = m_{\lsp} + 20\GeV$)       & $m_{\gluino}=900, m_{\lsp}=600$            & LL2, LL1, LL4, HL39, HL37 \\
\TfqqqqWW ($m_{\chiplmin} = m_{\lsp} + 20\GeV$)       & $m_{\gluino}=1400, m_{\lsp}=1$             & HH59, HH53, HH52, HH51, HH62 \\
\TfqqqqWW ($m_{\chiplmin} = m_{\lsp} + 20\GeV$)       & $m_{\gluino}=900, m_{\lsp}=600$            & LL2, LL4, HL39, LL1, HL37 \\
\TsttHZ ($\mathcal{B}(\susytoptwo\to\susytopone\PZ)$=1) & $m_{\susytoptwo}=850, m_{\susytopone}=625$ & on-Z ML23, on-Z ML21, on-Z ML16, on-Z ML14, on-Z ML17 \\
\TsttHZ ($\mathcal{B}(\susytoptwo\to\susytopone\PZ)$=0.5) & $m_{\susytoptwo}=850, m_{\susytopone}=625$ & on-Z ML17, on-Z ML23, on-Z ML21, on-Z ML14, on-Z ML16 \\
\TsttHZ ($\mathcal{B}(\susytoptwo\to\susytopone\PZ)$=0) & $m_{\susytoptwo}=850, m_{\susytopone}=625$ & off-Z ML15, HH40, HH39, HH45, HH44 \\
\ToqqqqL                                              & $m_{\gluino}=1600$                         & HH62, LM11, HH59, HH61, HH51 \\
\ToqqqqL                                              & $m_{\gluino}=2400$                         & HH62, LM11, HH59, HH53, HH52 \\
\Totbs                                                & $m_{\gluino}=1200$                         & HH62, HH50, HH59, HH61, HH58 \\
\Totbs                                                & $m_{\gluino}=1700$                         & HH62, HH59, HH50, HH52, LM11 \\
\hline 
\end{tabular}
}
\end{table*}

\cleardoublepage \section{The CMS Collaboration \label{app:collab}}\begin{sloppypar}\hyphenpenalty=5000\widowpenalty=500\clubpenalty=5000\input{SUS-19-008-authorlist.tex}\end{sloppypar}
\end{document}

%% file: SUS-19-008-authorlist.tex
\vskip\cmsinstskip
\textbf{Yerevan Physics Institute, Yerevan, Armenia}\\*[0pt]
A.M.~Sirunyan$^{\textrm{\dag}}$, A.~Tumasyan
\vskip\cmsinstskip
\textbf{Institut f\"{u}r Hochenergiephysik, Wien, Austria}\\*[0pt]
W.~Adam, F.~Ambrogi, T.~Bergauer, M.~Dragicevic, J.~Er\"{o}, A.~Escalante~Del~Valle, M.~Flechl, R.~Fr\"{u}hwirth\cmsAuthorMark{1}, M.~Jeitler\cmsAuthorMark{1}, N.~Krammer, I.~Kr\"{a}tschmer, D.~Liko, T.~Madlener, I.~Mikulec, N.~Rad, J.~Schieck\cmsAuthorMark{1}, R.~Sch\"{o}fbeck, M.~Spanring, W.~Waltenberger, C.-E.~Wulz\cmsAuthorMark{1}, M.~Zarucki
% ...
\vskip\cmsinstskip
\dag: Deceased\\
1:  Also at Vienna University of Technology, Vienna, Austria\\
% ...
